\documentclass[letter]{aa} 
\usepackage{graphicx}
\usepackage{soul}

\usepackage{txfonts}
\usepackage{natbib}
\begin{document}
\newcommand{\phn}   {\phantom{0}}
\newcommand{\phnnn} {\phantom{0}\phantom{0}\phantom{0}}
\newcommand{\Td}    {T_\mathrm{d}}	
\newcommand{\et}    {et al.}
\newcommand{\mum}   {$\mu$m}
\newcommand{\kms}   {km~s$^{-1}$}
\newcommand{\cmt}   {cm$^{-3}$}
\newcommand{\jpb}   {$\rm Jy~beam^{-1}$}	
\newcommand{\nh}    {NH$_3$}
\newcommand{\nth}   {N$_2$H$^+$}
\newcommand{\eg}    {e.\,g.,}
\newcommand{\Tex}   {T_\mathrm{ex}}
\newcommand{\Trot}  {T_\mathrm{rot}}
\newcommand{\cmg}   {cm$^{2}$~g$^{-1}$}
\newcommand{\chtoh} {CH$_3$OH}
\newcommand{\water} {H$_2$O}
\newcommand{\juc}   {\mbox{$J$=1$\rightarrow$0}}
\newcommand{\J}[2]  {\mbox{#1--#2}}
\newcommand{\uchii} {UCH{\small II}~}
\newcommand{\hchii} {HCH{\small II}~}
\newcommand{\raun}  {$^\mathrm{h~m~s}$}
\newcommand{\deun}  {$\mathrm{\arcdeg~\arcmin~\arcsec}$}
\newcommand{\myemail} {s.trevino@crya.unam.mx} 
\newcommand{\ie}      {i.\,e.,}
\newcommand{\lo}      {$L_{\sun}$}
\newcommand{\mo}      {$M_{\sun}$}    
\newcommand{\Halpha}  {H110$\alpha$} 
\newcommand{\Calpha}  {C110$\alpha$} 
\newcommand{\hii}     {H{\small II}}
\newcommand{\hi}      {H\small{I}}
   \title{Deuteration around the ultracompact \hii\ region Mon~R2}


   \author{S. P. Trevi\~no-Morales
      \inst{1} \and
      P. Pilleri\inst{2,3,4}\and
	  A. Fuente\inst{3}\and
	  C. Kramer\inst{1}\and
	  E. Roueff\inst{5}\and	  
	  M. Gonz\'alez-Garc\'{\i}a\inst{1}\and
 	  J. Cernicharo\inst{4}\and
	  M. Gerin\inst{6}\and
	  J. R. Goicoechea\inst{4}\and
	  J. Pety\inst{7, 8}\and
	  O. Bern\'e\inst{9, 10} \and
	  V. Ossenkopf\inst{11}\and
	  D. Ginard\inst{3}\and
	  S. Garc\'{\i}a-Burillo\inst{3}\and
	  J. R. Rizzo\inst{4} 	  
          \and	  
	  S. Viti\inst{12}
          }
         \institute{Instituto de Radioastronom\'ia Milim\'etrica (IRAM-Spain), Ave. Divina Pastora, 7, Local 20 18012, Granada (Spain).\\
         \email{trevino@iram.es}
	     \and 
	     Los Alamos National Laboratory, P.O. Box 1663, Los Alamos, NM 87545, (USA).
	     \and	     
         Observatorio Astron\'omico Nacional, Apdo. 112, E-28803 Alcal\'a de Henares (Madrid), Spain. 
	     \and
	     Centro de Astrobiolog\'{\i}a (INTA-CSIC), Departamento de Astrof\'{\i}sica, Ctra. M-108, km.~4, E-28850 Torrej\'on de Ardoz, (Spain). 
	     \and	     
         LUTH UMR 8102, CNRS and Observatoire de Paris, Place J. Janssen, 92195, Meudon Cedex, (France).
	     \and
         LERMA, UMR 8112, CNRS and Observatoire de Paris, 61 avenue de l'Observatoire, 75014, Paris, (France).
	     \and
	     IRAM, 300 rue de la Piscine, 38406, Saint-Martin d'H\'eres, France; 
	     \and
	     LERMA - LRA, UMR 8112, Observatoire de Paris and \'Ecole normale Sup\'erieure, 24 rue Lhomond, 75231, Paris, (France).
	     \and
	     Universit\'e de Toulouse, UPS-OMP, IRAP, 31028 Toulouse, (France).
	     \and
	     CNRS, IRAP, 9 Av. colonel Roche, BP 44346, 31028 Toulouse Cedex 4, (France).
	     \and	     
	     I. Physikalisches Institut der Universit\"at zu K\"oln, Z\"ulpicher Stra$\ss$e 77, 50937 K\"oln, (Germany).
	     \and	     
	     Department of Physics and Astronomy, UCL, Gower Place (London).
              }

   \date{Received ????; accepted ????}

\abstract
{The massive star-forming region Monoceros~R2 (Mon~R2) hosts the closest ultra-compact {\sc Hii}~region, where the photon-dominated region (PDR) between the ionized and molecular gas can be spatially resolved with current single-dish telescopes.}
{We aim at studying the chemistry of deuterated molecules toward Mon~R2 to determine the deuterium fractions around a high-UV irradiated PDR and investigate the chemistry of these species.}
{We used the IRAM-30m telescope to carry out an unbiased spectral survey toward two important positions (namely IF and MP2) in Mon~R2 at 1, 2, and 3~mm. This spectral survey is the observational basis of our study of the deuteration in this massive star forming region. Our high spectral resolution observations ($\sim0.25$--0.65 \kms) allowed us to resolve the line profiles of the different species detected.}
{We found a rich chemistry of deuterated species at both positions of Mon~R2, with detections of C$_2$D, DCN, DNC, DCO$^{+}$, D$_{2}$CO, HDCO, NH$_{2}$D, and N$_{2}$D$^{+}$
and their corresponding hydrogenated species and rarer isotopologs. The high spectral resolution of our observations 
allowed us to resolve three velocity components: the component at 10 \kms\ is detected at both positions and seems associated with the layer most exposed to the UV radiation from IRS 1; the component at 12 \kms\ is found toward the IF position and seems related to the foreground molecular gas; finally, a component at 8.5 \kms\ is only detected toward the MP2 position, most likely related to a low-UV irradiated PDR. We derived the column density of the deuterated species (together with their hydrogenated counterparts), and determined the deuterium fractions as $D_\mathrm{frac}$=[XD]/[XH]. The values of $D_\mathrm{frac}$ are around 0.01 for all the observed species, except for HCO$^+$ and N$_2$H$^+$ which have values 10 times lower. The values found in Mon~R2 are similar to those measured in the Orion Bar, and are well explained with a pseudo-time-dependent gas-phase model in which deuteration occurs mainly via ion-molecule reactions with H$_2$D$^+$, CH$_2$D$^+$ and C$_2$HD$^+$. Finally, the [H$^{13}$CN]/[HN$^{13}$C] ratio is very high ($\sim11$) for the 10 \kms\ component, which also agree with our model predictions for an age of $\sim$0.01 to a few 0.1~Myr.}
{The deuterium chemistry is a good tool for studying the low-mass and high-mass star-forming regions. However, while low-mass star-forming regions seem well characterized 
with $D_{\rm frac}$(N$_{2}$H$^{+}$) or $D_{\rm frac}$(HCO$^{+}$), a more complete chemical modeling is required to date massive star-forming regions. This is due to the higher gas temperature together with the rapid evolution of massive protostars.}

             
   \maketitle

 
%

\section{Introduction}

\subsection{Deuteration and star formation}

During the past decade, the chemistry of deuterium has become an important tool for understanding the formation of stars and planets. However, the deuteration processes are not fully understood yet, and important uncertainties remain in the interpretation of observational data. The cosmological ratio of the elemental abundances between deuterium and hydrogen (D/H) is approximately 1--2$\times10^{-5}$ (\citealt{roberts2000}; \citealt{linsky2006}; \citealt{pety2007}). However, higher abundances of deuterated molecules have been observed in many astrophysical environments, including cold dense cores (\citealt{guelin1982}), mid-planes of circumstellar disks (\citealt{van2003}; \citealt{guilloteau2006}), hot molecular cores (\citealt{hatchell1998}), and even photon-dominated regions (PDRs; see \citealt{leurini2006}). Several chemical pathways have been proposed to produce this deuterium enrichment.

Deuteration in the gas phase is driven by ion-molecule reactions in which deuterium and hydrogen atoms are exchanged. For a cold source with kinetic temperatures $T_{\rm k}$=10$-$20~K, the dominant process is H$_{3}^{+}$+HD$\rightleftharpoons$ H$_{2}$D$^{+}$+H$_{2}$. This reaction proceeds left-to-right with an exothermicity of $\sim$232~K (\citealt{gerlich2002}; \citealt{pagani2011}; 2013) and is very rapid at low temperatures, raising the abundance ratio [H$_{2}$D$^{+}]/[$H$_{3}^{+}$] to values much higher than 10$^{-5}$ (see e.g. \citealt{caselli2003}). The production of the deuterated ions of H$_{3}^{+}$ leads to high abundances of other deuterated species through secondary ion-molecule reactions. In addition to singly deuterated isotopologs, doubly and triply deuterated species such as NHD$_{2}$, ND$_{3}$, and D$_{2}$CO can be explained, at least partially, in terms of gas-phase synthesis (\citealt{gerin2006}; \citealt{roueff2005}). The deuteration via H$_2$D$^+$ becomes very inefficient when $T_{\rm k}$ is higher than 30~K. At temperatures $T_{\rm k}$=30$-$50~K, CH$_{3}^{+}$ and C$_{2}$H$_{2}^{+}$ react rapidly with HD, leading to the ions CH$_{2}$D$^{+}$ and C$_{2}$HD$^{+}$ (\citealt{herbst1987}; \citealt{millar1989}) via the reactions CH$_{3}$+HD$\rightleftharpoons$ CH$_{2}$D$^{+}$+H$_{2}$ ; C$_{2}$H$_{2}^{+}$+HD$\rightleftharpoons$ C$_{2}$HD$^{+}$+H$_{2}$. The left-to-right exothermicities, $\sim$390~K (\citealt{asvany2004}) and $\sim$550~K (\citealt{herbst1987}), respectively, are considerably higher than for the reaction involving H$_{3}^{+}$. Detailed gas-phase models including deuterated species have been constructed using large networks of gas-phase reactions with both the steady-state and pseudo-time-dependent pictures showing that significant deuterium fraction enhancements can be detected at temperatures up to $T_{\rm k}$$\sim$70~K (\citealt{roueff2005}; 2007; 2013). 

Deuterated isotopologs of methanol and formaldehyde have been detected in hot cores and corinos where $T_{\rm k}$$>$100~K (see e.g. \citealt{parise2002}; 2004; \citealt{fuente2005a}). In these cases, the deuteration is thought to occur on grain surfaces. The deuterium and hydrogen atoms on the grain surface react with complex molecules, leading to both deuterated and normal isotopologs (\citealt{tielens1983}, \citealt{stantcheva2003}; \citealt{nagaoka2005}). The deuterated compounds are released to the gas phase when the ice is evaporated, producing high abundances of the deuterated isotopologs of methanol and formaldehyde (\citealt{parise2002}; 2004; 2006).
 
Observationally, deuteration has been widely studied in cold pre-stellar regions and hot corinos, and the results are interpreted on the basis of the deuteration pathways explained above. Much less studied is the deuteration in warm ($T_{\rm k}$=30$-$70 K) cores, in which the dust temperature is too high for the deuteration via H$_2$D$^+$ to proceed efficiently and it is not high enough for the icy mantles to evaporate. \citet{parise2007} observed a sample of deuterated species toward a warm ($\sim$30~K) clump in the Orion Bar. They detected DCN, DCO$^+$ and HDCO and derived [XD]/[XH] fractions (hereafter $D_{\rm frac}$) of 0.01, 0.0006 and 0.006 for HCN, HCO$^+$ and H$_2$CO, respectively. Because of the good agreement with chemical models (\citealt{roueff2005}; 2007), they interpreted these high values of deuteration as the consequence of gas-phase chemistry driven by ion-molecule reactions with CH$_2$D$^+$ and C$_2$HD$^+$. Later, \citet{guzman2011} proved that photo-desorption from the grain mantles could also be an important formation mechanism in the PDRs for species like H$_2$CO, suggesting that the deuteration on grain surfaces could also contribute to the enhancement of the deuterium fraction in UV-irradiated regions. Species such as C$_2$D and DNC, which are not expected to form on grain surfaces, were not detected in the Orion Bar, and an extensive comparison with gas-phase models was not possible in this source. A complete observational study including more deuterated species is necessary to distinguish among the different deuteration mechanisms at work.
 

\subsection{Ultracompact {\sc Hii}~region in Mon~R2}

Ultracompact (UC) {\sc Hii}~regions represent one of the earliest phases in the formation of a massive star. They are characterized by extreme UV radiation (G$_\mathrm{0}>$10$^{5}$ in units of Habing field), small physical scales ($\le$0.1~pc), and are found embedded in dense molecular clumps with densities $\ge$$10^6$~cm$^{-3}$ (\citealt{hoare2007}). The UV radiation from the star forms a first layer of ionized hydrogen ({\sc Hii}), followed by the PDR.
PDRs are environments where the thermal balance and chemistry are driven by UV photons (6~eV$\le h\nu\le$13.6~eV). The PDRs associated with UC~{\sc Hii}~are characterized by extreme UV field intensities ($>$ 10$^4$ expressed in units of the Habing field G$_{0}$, see \citealt{habing1968}) and high densities (n$>$10$^5$~cm$^{-3}$). They are typically located farther than a few kpc from the Sun, which makes their observational study very difficult. However, fully comprehension is of paramount importance for our understanding of the interaction of the massive star formation process with the interstellar medium both in our Galaxy and in extragalactic objects.

\begin{figure}
\sidecaption
\includegraphics[width=9cm]{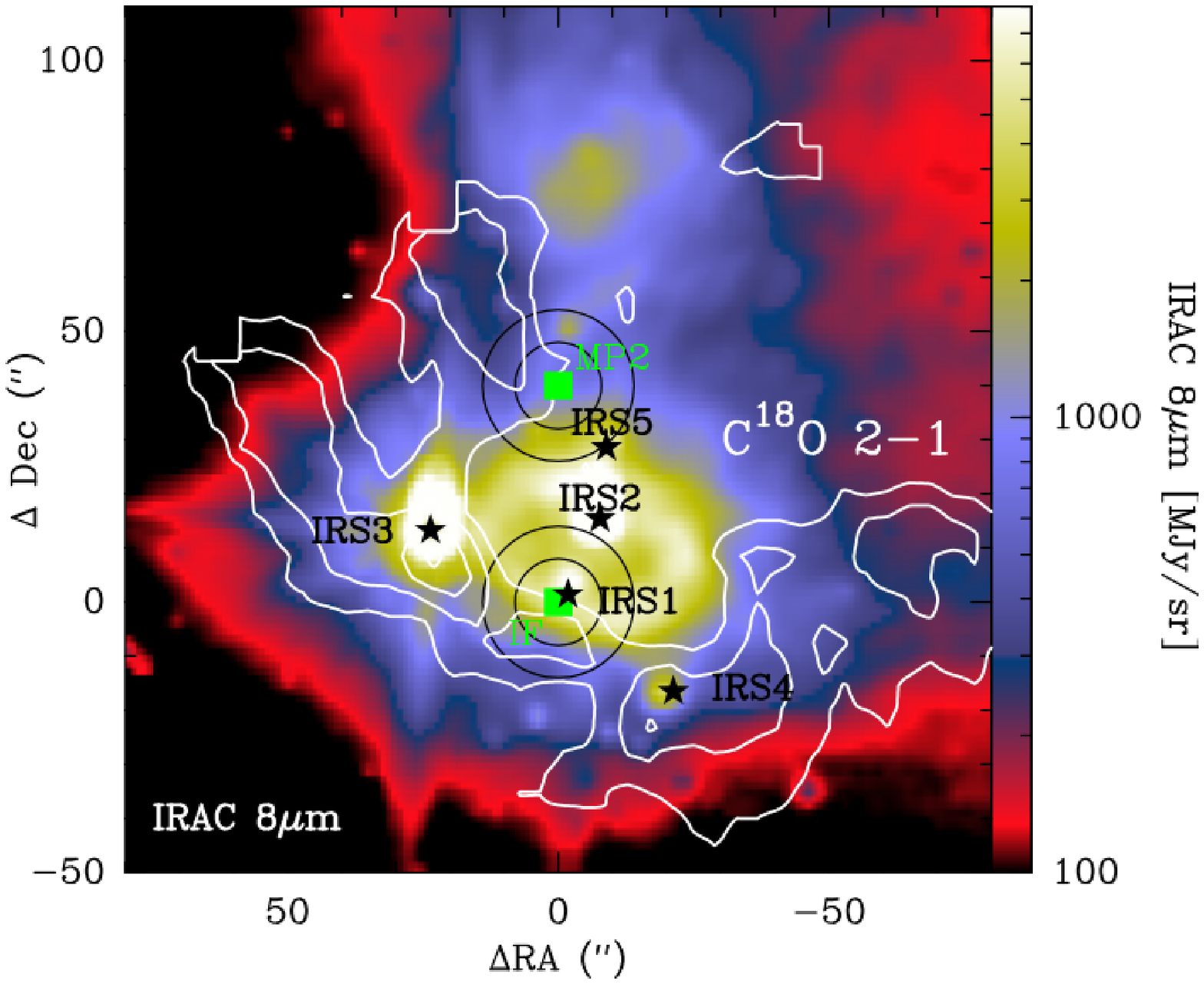}\\
\caption{In colors, the {\it Spitzer}-IRAC 8$\mu$m emission from small dust (\citealt{ginard2012}). In contours, the integrated emission between 5 and 15\kms of the C$^{18}$O 2-1 line (Pilleri \et\ 2012a). Green squares show the two positions analyzed in this paper. Black stars show the positions of the brightest infrared sources, following the nomenclature of Henning \et\ (1992). The beams at 3mm ($\sim$29$\arcsec$) and 2mm ($\sim$16$\arcsec$) toward the IF and MP2 positions are drawn.}
\label{figure1:A}
\end{figure}

Monoceros~R2 (Mon~R2, \citealt{vanden1966}) is the closest UC~{\sc Hii}~region (d=830~pc; \citealt{herbst1976}), and can be spatially resolved with the current instrumentation in the mm and IR domains. Due to its brightness and proximity, this source is an ideal case to study the physical and chemical conditions in an extreme PDR and can be used as pattern for other PDRs that are illuminated by a strong UV field.
The UC~{\sc Hii} region has a cometary shape and its continuum peaks toward the infrared source Mon~R2 IRS~1. Previous millimeter and far-infrared spectroscopic and continuum studies (\citealt{henning1992}; \citealt{giannakopoulou1997}; \citealt{tafalla1997}; \citealt{choi2000}; \citealt{rizzo2003}; 2005; \citealt{pilleri2012}; 2013) showed that the molecular emission presents an arc-like structure surrounding the {\sc Hii}~region, with the bulk of the emission to the SW (see Fig.~\ref{figure1:A}). The advent of the new IR facilities {\it Spitzer} and {\it Herschel}, together with the high spatial resolution at mm wavelengths provided by the IRAM-30m telescope have allowed us to probe the PDR in the interface between the ionized and the molecular gas (\citealt{rizzo2003}; 2005; \citealt{berne2009}; \citealt{pilleri2012}; 2013; \citealt{ginard2012}). These studies showed that the UC~{\sc Hii} region is surrounded by a dense PDR ($n>$10$^5$~cm$^{-3}$) that is well detected in the pure rotational lines of H$_2$, the mid-infrared bands of polycyclic aromatic hydrocarbons (PAHs) and in the e\-mi\-ssion of rotational lines of reactive molecular ions such as CO$^+$ and HOC$^+$. In addition, a second PDR is detected $40\arcsec$ north from IRS~1, which corresponds to a second molecular peak (hereafter, MP2). The position MP2 is well detected in the PAH emission at 8~$\mu$m and its chemical properties are similar to those of low- to mild-UV i\-rra\-diated PDRs such as the Horsehead (see Fig.~\ref{figure1:A}; \citealt{ginard2012}, \citealt{pilleri2013}).

In this paper, we present an extensive study of deuterated compounds in the PDRs around the UC~{\sc Hii} Mon~R2 that includes DCN, DNC, DCO$^+$, C$_2$D, HDCO, D$_2$CO, NH$_2$D, and N$_2$D$^+$. We focus our study on the chemistry toward the ionization front (hereafter IF) and the second PDR in MP2 position and discuss the implications of our observations. In Sects. 2 and 3 the observational setup and results are described. In Sect. 4, we calculate column densities and abundances for each species, while in Sects. 5 and 6 we discuss our main results. Finally, Sect. 7 presents our conclusions.

\section{Observations and data reduction}

The data presented in this paper are part of the 3, 2, and 1~mm spectral surveys we have carried out using the IRAM-30m telescope at Pico Veleta (Spain) toward the PDRs around the UC~{\sc Hii} region Mon~R2. The observations were performed in three observational blocks in 2012 January, September, and December. All observations were performed in dual polarization using the EMIR receivers (\citealt{carter2012}) with the fast Fourier transform spectrometer (FTS) at 200~kHz of resolution (\citealt{klein2012}). During the observations we pointed on the strong nearby quasar (0605-058) every 2 hours and checked the focus on a planet every 4 hours. Pointings and focus corrections were stable throughout the whole run. A line calibrator was observed for every tuning to check that the intensities of the lines were correct. The emission from the sky was substracted using an OFF position free of molecular emission ($+400'',-400''$). The offsets are given relative to the coordinates of the IF position: $\alpha$(J2000)$=06\mathrm{h}07\mathrm{m}46.2\mathrm{s}$, $\delta$(J2000)$=-06^{\circ}23'08.3''$. For the on-the-fly (OTF) maps the OFF position was observed every 2 minutes for 20 seconds. 
For the single pointing observations the OFF position was observed in position switching mode every 30 seconds. Throughout the paper, we use the main-beam brightness temperature ($T_{\mathrm{MB}}$) as intensity scale. To do so, we multiplied our spectra in $T_{\mathrm{A}}^{*}$ by a factor $F_{\rm{eff}}/B_{\rm{eff}}$\footnote{The values of the $F_{\rm{eff}}$ and $B_{\rm{eff}}$ where taken from the IRAM-30m webpage;\\ http://www.iram.es/IRAMES/mainWiki/Iram30mEfficiencies}. A summary of the observational parameters is shown in Table~A.1.

In the first observational block we covered the frequency range from 202.00~GHz to 265.00~GHz. In this block we used the OTF observing mode to cover a 120$''\times 120''$ region centered on the IF position. In the second observational block we performed single-pointing observations toward both the IF and MP2 positions, to cover the frequency ranges of 84.00 -- 99.78~GHz and 104.00 -- 111.96~GHz for the 3mm band; 132.80 -- 136.89~GHz, 141.80 -- 145.89~GHz and 150.60 -- 154.68~GHz for the 2mm band; 210.30 -- 218.00~GHz and 225.90 -- 233.70~GHz for the 1mm band. In the last observational block we made single-pointing observations in the IF and MP2 position (molecular peak at offset [0$''$,40$''$]), to cover the frequency ranges of 250.50 -- 258.30~GHz and 266.20 -- 274.00~GHz.  

The data were reduced using the CLASS/GILDAS package\footnote{See http://www.iram.fr/IRAMFR/GILDAS for more information about the GILDAS software.} (\citealt{pety2005}). Initial inspection of the data revealed a few single-channel spikes, platforming of individual FTS units, but otherwise clean baselines. To fix the spikes problem we flagged individual channels and filled them with white noise corresponding to the rms measured by a baseline fit. To correct for the platforming, a zero-order baseline was subtracted from each FTS sub-bands of each spectrum using a dedicated procedure provided by IRAM. In addition, a second-order baseline was needed for almost all the detected lines. All resulting spectra were smoothed to a velocity resolution of 0.65~\kms. 



\begin{figure*}
\sidecaption
\includegraphics[width=9cm]{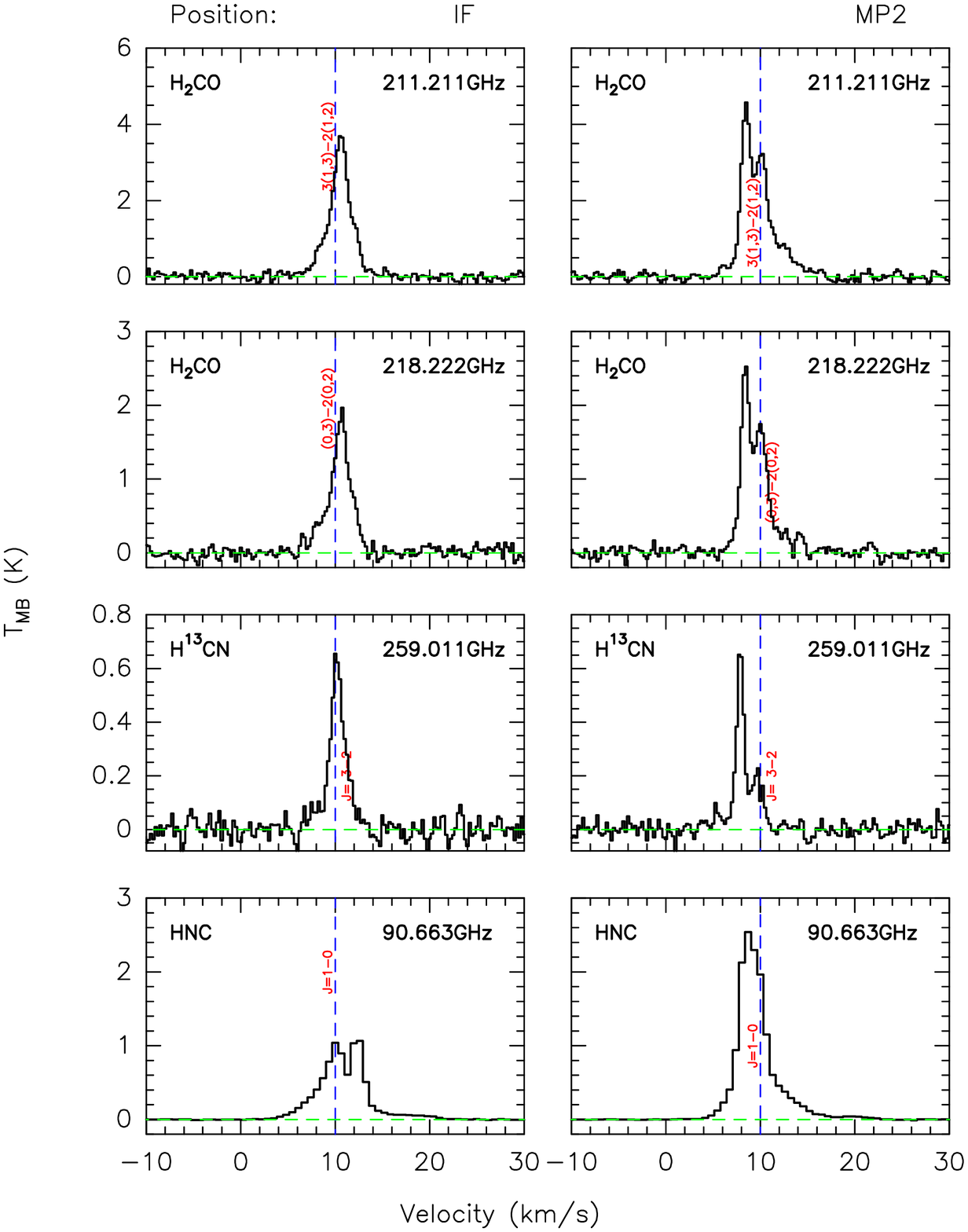}
\includegraphics[width=9cm]{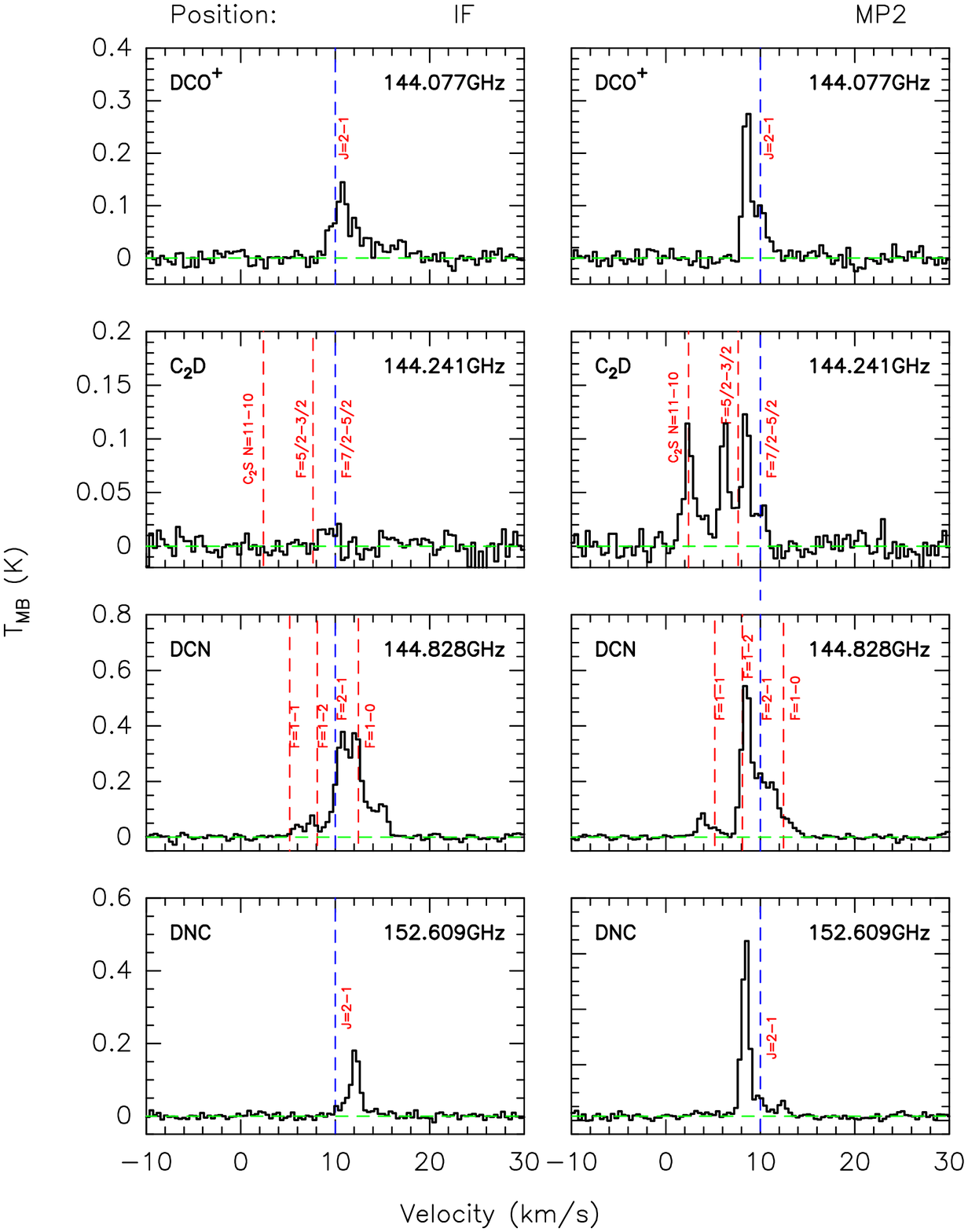}\\ 
\caption{Observed spectra toward the IF and MP2 position. The blue dashed line shows the velocity of 10~km~s$^{-1}$ relative to the rest frequency of the transition. Two different velocity components can be distinguished at each position. {\bf Left:} transitions of H$_{2}$CO (211.211 and 218.222~GHz), H$^{13}$CN at 259.011~GHz and HNC at 90.663~GHz (from top to bottom) at the IF and MP2 positions. {\bf Right:} observed spectra for the deuterated species DCO$^{+}$, C$_{2}$D, DCN and DNC transitions toward the IF and MP2 positions. The velocities are relative to the frequencies 144.077, 144.241, 144.828, and 152.609~GHz (from top to bottom).}
\label{figure1}
\end{figure*}

\begin{figure*}
\sidecaption
\includegraphics[width=17.5cm]{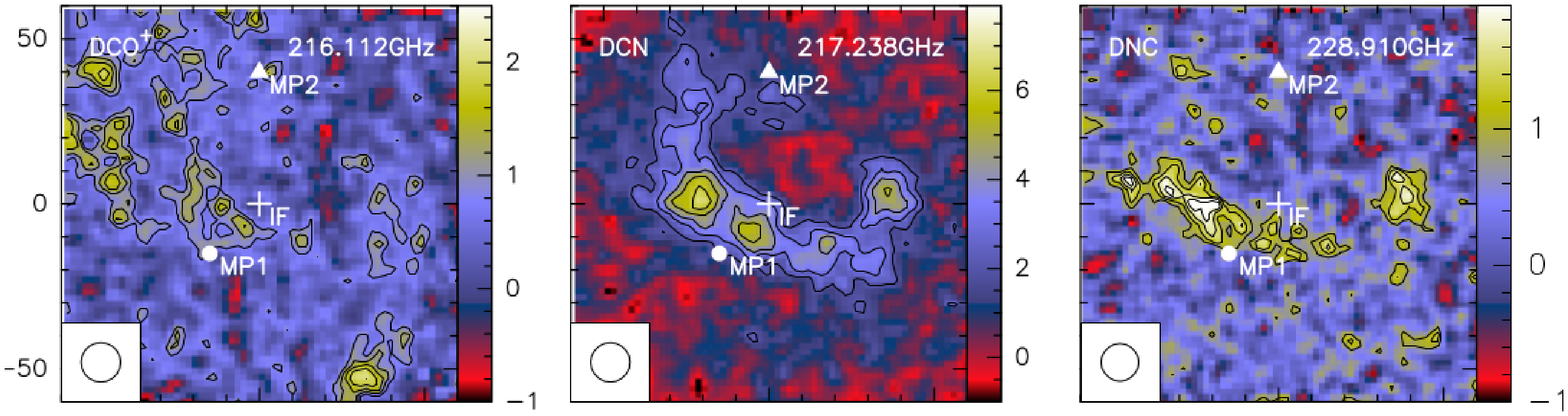}\\
\includegraphics[width=17.5cm]{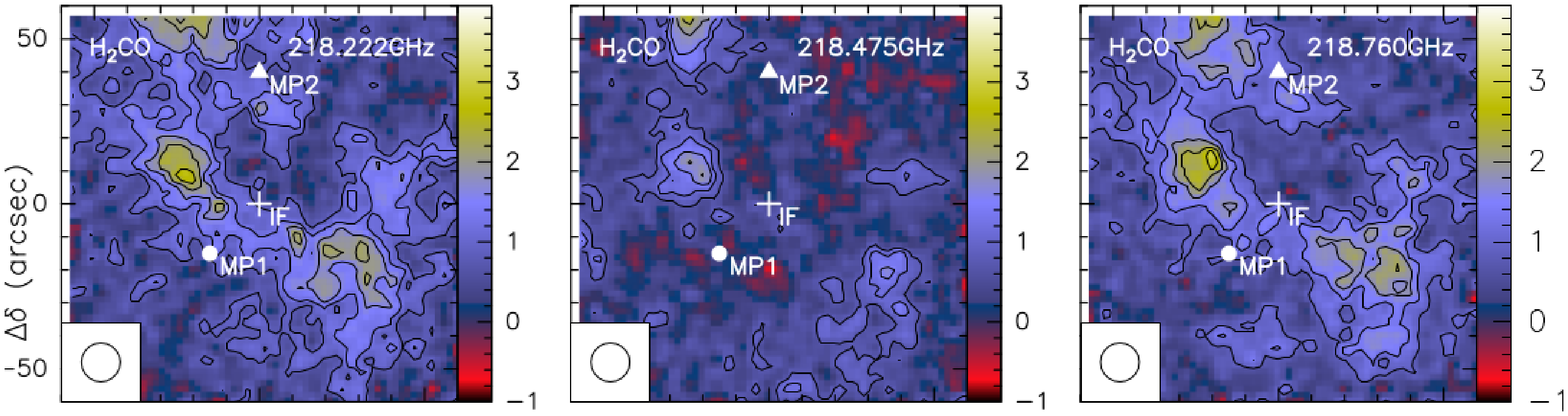}\\
\includegraphics[width=17.5cm]{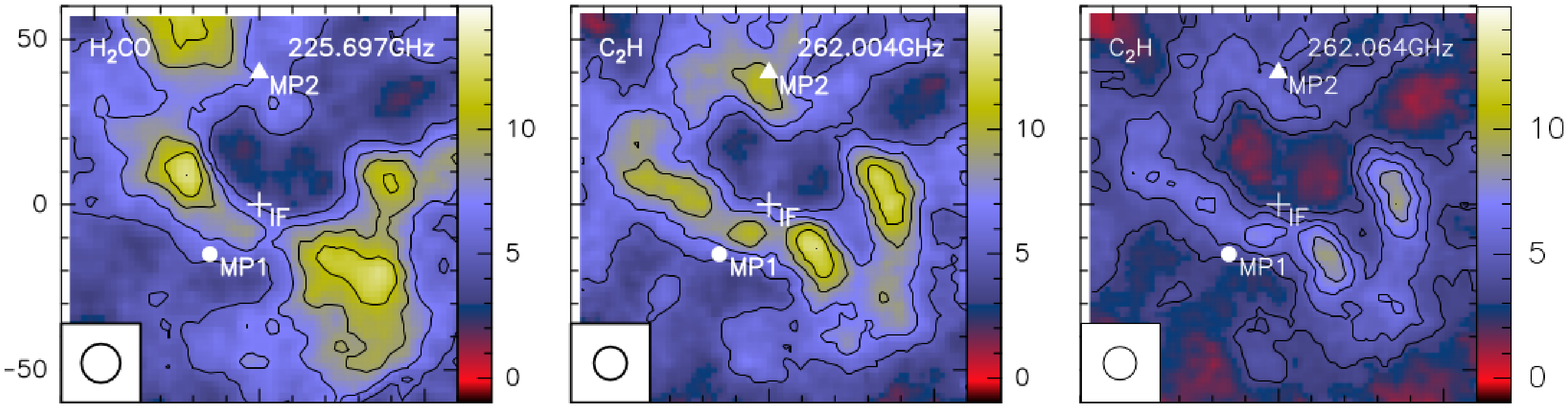}\\
\includegraphics[width=17.28cm]{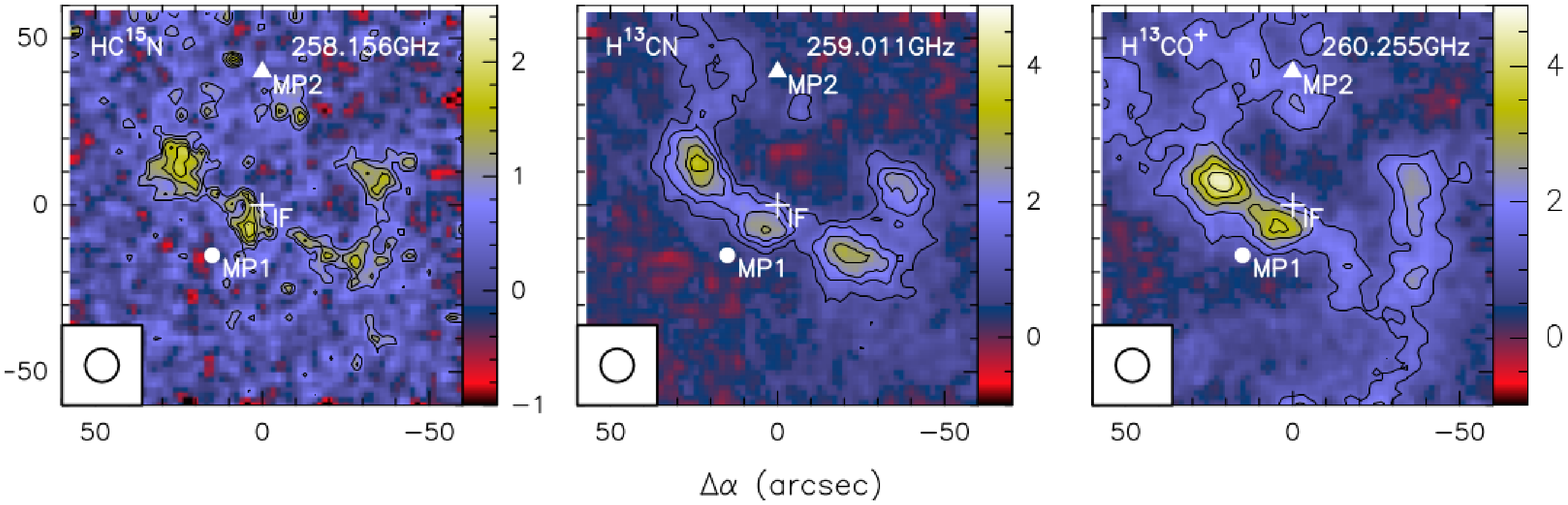}\\
\caption{2$'\times$2' OTF maps of DCO$^{+}$ (at 216.112~GHz), DCN (at 217.238~GHz), DNC (at 228.910~GHz), H$_2$CO (218.222, 218.475, 218.760 and 225.697~GHz), C$_2$H (at 262.004 and 262.064~GHz), HC$^{15}$N (at 258.156~GHz), H$^{13}$CN (at 259.011~GHz) and H$^{13}$CO$^{+}$ (at 260.255~GHz). The cross marks the IF position, the triangle marks the MP2 position, and the circle shows the MP1 position at the offset (-15",15"). The contour levels are 40\% to 100\%, in steps of 15\% of the peak intensity; $\sim2$~K kms$^{-1}$ for DNC; $\sim3$~K kms$^{-1}$ for DCO$^{+}$ and HC$^{15}$N; $\sim4$~K kms$^{-1}$ for H$_{2}$CO at $\sim218.00$~GHz; $\sim5$~K kms$^{-1}$ for H$^{13}$CN and H$^{13}$CO$^{+}$; $\sim7$~K kms$^{-1}$ for DCN; $\sim15$~K kms$^{-1}$ for H$_{2}$CO and C$_{2}$H at$\sim262.00$~GHz.}
\label{f16:maps} 
\end{figure*}

\section{Results}


\subsection{Observed spectra}

Figures~\ref{f1:CCH}--\ref{f15:HDCO} show the spectra of hydrogenated (Figs.~\ref{f1:CCH}--\ref{f8:H18COp}) and deuterated (Figs.~\ref{f12:CCD}--\ref{f15:HDCO}) species toward the IF and MP2 positions (i.e., the offsets [$+$0\arcsec,$+$0\arcsec] and [$+$0\arcsec,$+$40\arcsec], respectively). The spectra show a different velocity profile toward the two positions. Two velocity components were found at each position for almost all the molecules. At the IF position we detect a velocity component at $\sim$10~km~s$^{-1}$ and a second one at $\sim$12~km~s$^{-1}$; while at the MP2 position, the two velocity components correspond to $\sim$8~km~s$^{-1}$ and $\sim$10~km~s$^{-1}$. The component at 10 \kms\ is detected at both positions and seems associated with the layer most exposed to the UV radiation from IRS~1 (see section 3.3). The component at 12 \kms\ is found toward the IF position but it is related to the SW part of the molecular cloud. Finally, the component at 8.5 \kms\ is only detected toward the MP2 position and related to a low-UV irradiated PDR.

All the hydrogenated species are detected toward the two positions, and with a few exceptions, \eg\ H$^{13}$CN at 259.011~GHz and HC$^{15}$N at 258.156~GHz, the lines are more intense at the MP2 than at the IF position (see Fig.~2). The deuterated species DCN, DNC, DCO$^+$, C$_2$D, HDCO, and NH$_2$D are detected at both positions. We have detected N$_2$D$^+$ only toward the IF position, but the lack of detection toward the MP2 position is very likely due to the poor quality of our spectrum at this frequency. D$_2$CO is only detected toward the MP2 position.

Initial inspection of the spectra reveals some differences in the intensity line ratios between the two velocity components at each position.
For instance, toward the IF position the H$^{13}$CN (3$\rightarrow$2) line is more intense at 10 \kms. However, the intensity of the HNC (1$\rightarrow$0) line is similar for the two components (see Fig.~2). To study these differences quantitatively, we fitted a Gaussian to each velocity component using the CLASS software of the GILDAS package. Table~A.2 lists the results of the Gaussian fits for the deuterated species, and Tables~A.3 and A.4 list the Gaussian fits for the hydrogenated species. 

\begin{figure*}
\sidecaption
\includegraphics[width=19cm]{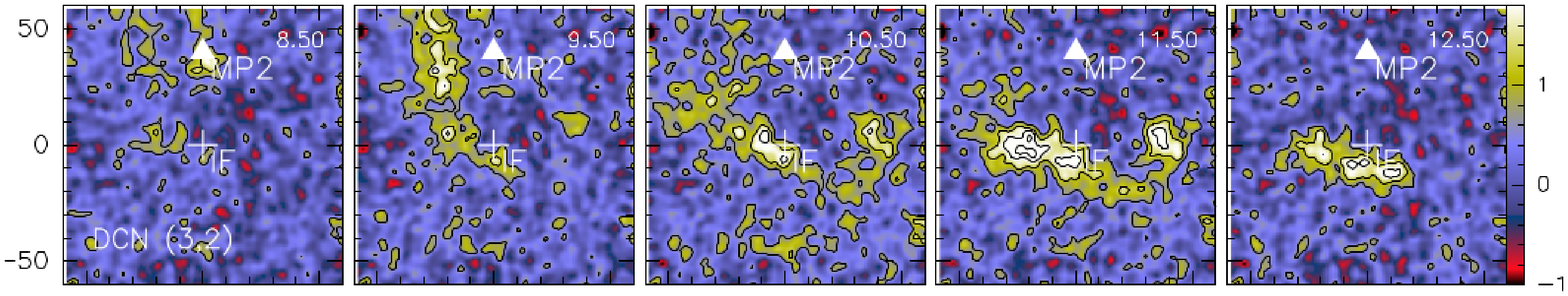} \\
\includegraphics[width=19cm]{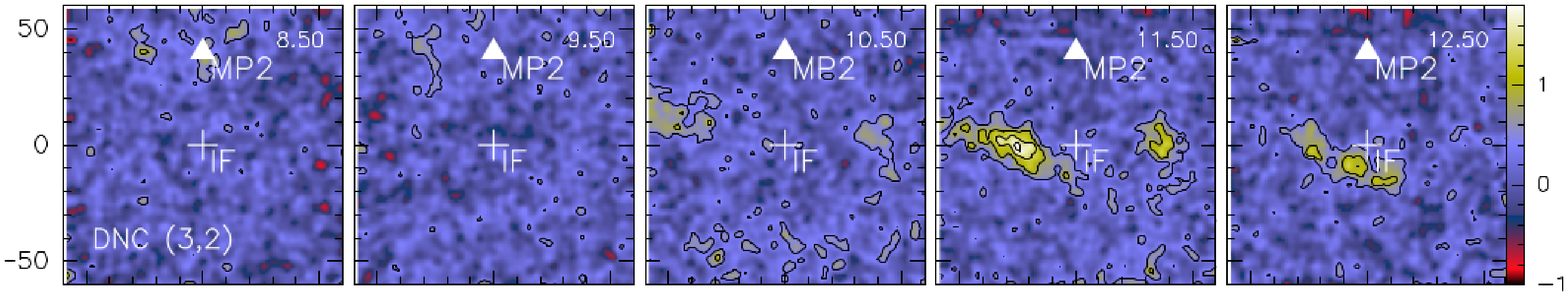} \\
\includegraphics[width=19cm]{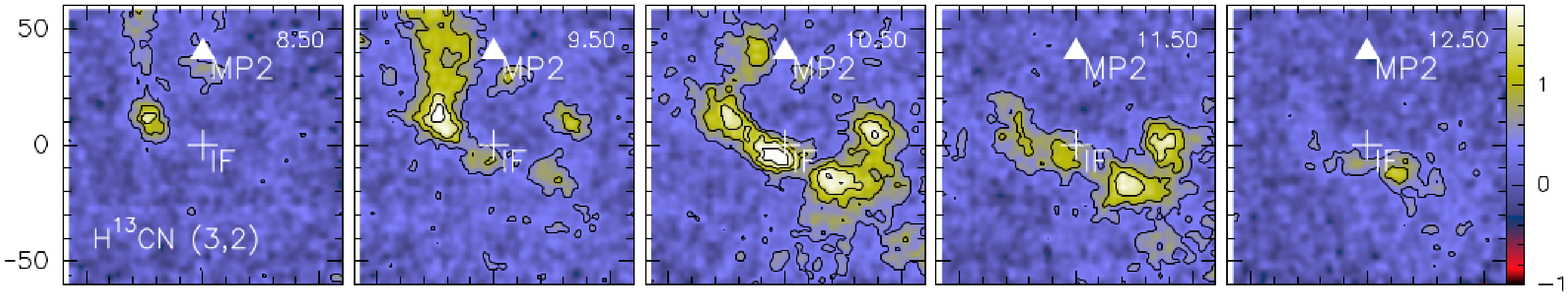} \\
\includegraphics[width=19cm]{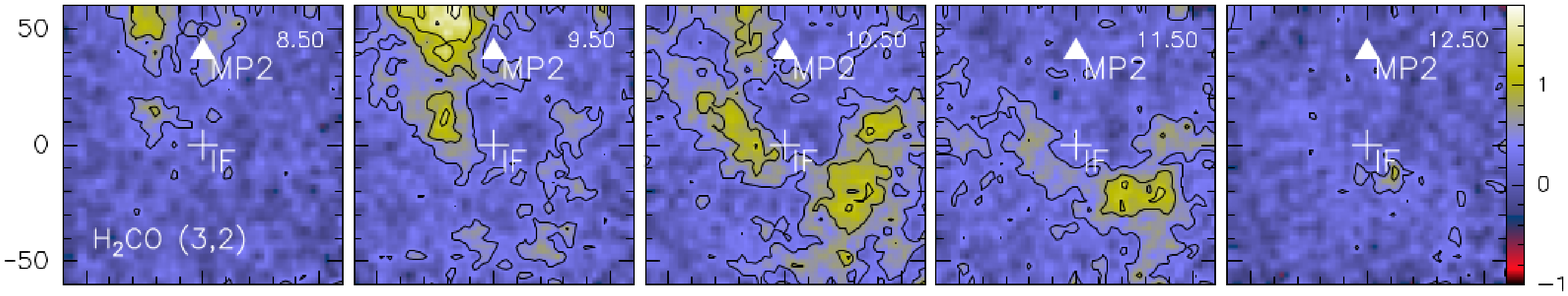}     \\
\includegraphics[width=19cm]{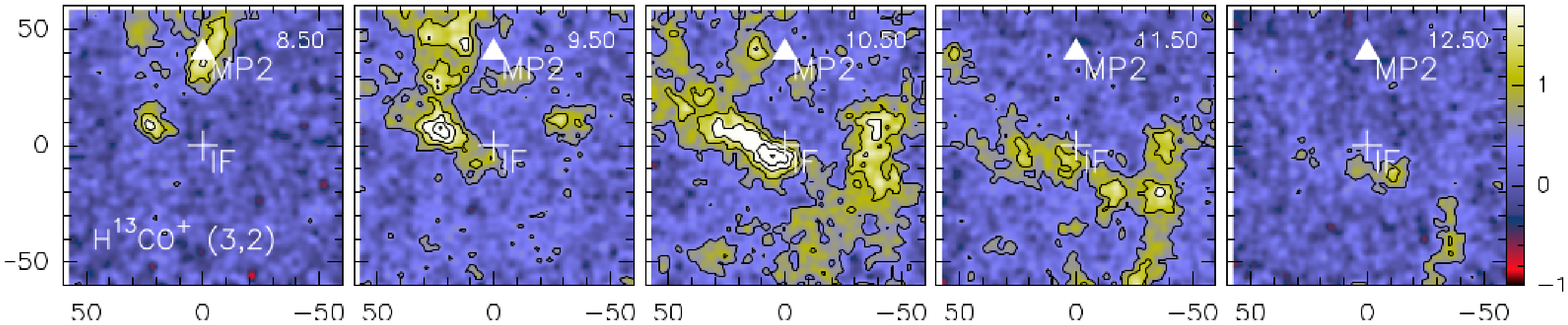}  
\caption{Channel maps for the molecules DCN at 216.112~GHz, DNC at 217.238~GHz, H$^{13}$CN at 259.011~GHz, H$_{2}$CO at 218.960~GHz and H$^{13}$CO$^{+}$ at 260.255~GHz, from top to bottom. The velocities for every channel are 8.5, 9.5, 10.5, 11.5, and 12.5 km s$^{-1}$ from left to right. The peak intensity is $\sim2$~K kms$^{-1}$ for each line.}
\label{f16:maps}
\end{figure*}

\subsection{Integrated intensity maps}

The maps carried out with the IRAM-30m telescope allow us to characterize the spatial distribution of different deuterated and hydrogenated species in Mon~R2. In Fig.~3 we show the integrated line intensity maps of different transitions of DCO$^{+}$, DCN, DNC, H$_2$CO, C$_2$H, HC$^{15}$N, H$^{13}$CN, and H$^{13}$CO$^{+}$. All the maps have an angular resolution between 9\arcsec\ and 11\arcsec, allowing a direct comparison. 
The spatial distribution of all these molecules is quite similar, consisting of an arc structure of $\sim$30\arcsec\ in radius, pointing to the SE and opened to the NW, following the cometary shape of the nebula. H$_2$CO is a very abundant species in PDRs and therefore presents a very intense and extended structure. This molecule extends up to $\sim$1\arcmin\ from the IF position and consists of three peaks at the positions ($-$25\arcsec,$+$55\arcsec), ($+$10\arcsec,$-$25\arcsec), and ($+$20\arcsec,$+$35\arcsec) surrounded by faint emission. The C$_2$H molecule is also very abundant and presents an extended structure, with an intense peak toward the MP2 position. The emission of both, the H$_2$CO and C$_2$H molecules, come from the inner clumps and the envelope of the region, but seems to be dominated by a lower density envelope. The HC$^{15}$N, H$^{13}$CN, and H$^{13}$CO$^{+}$ molecules are less abundant and their distributions are more compact, they present peaks at the offsets ($+$10\arcsec,$+$25\arcsec), ($-$10\arcsec,$-$10\arcsec), and ($-$20\arcsec,$+$20\arcsec). The emission of DCN and DNC species are compact and intense, while the DCO$^{+}$ emission is weaker and more extended.

\subsection{Spectral maps}

Figure~4 shows the intensity maps at different velocities for DCN (3$\rightarrow$2), DNC (3$\rightarrow$2), H$^{13}$CN (3$\rightarrow$2), HN$^{13}$C (3$\rightarrow$2) and H$^{13}$CO$^+$ (3$\rightarrow$2). In all the species we find the three velocity components described above. The emission at $\sim$8 \kms\ is located in the NW, the gas emitting at $\sim$10 \kms\ is surrounding the {\sc Hii} region, and the gas in the SW is at a velocity of $\sim$12 \kms\ (see Fig.~4). The transition from one component to another is abrupt and does not occur smoothly, as expected in the case of a global and coherent rotation. Moreover, two velocity components coexist at many positions, for example at the IF and MP2 positions. This suggests that the observed velocity behavior is the result of the superposition of three filaments around the expanding UC~{\sc Hii} region or, alternatively, the existence of a twisted filament, with the gas at $\sim$12 \kms\ and  at $\sim$8 \kms\, located at the front and the back of the UC~{\sc Hii} region as observed from the Sun.


\section{Analysis}


\subsection{Molecular column densities}

\begin{table*}[]
\begin{center}
\caption{Column densities and excitation temperatures derived from rotation diagrams.}
\begin{tabular}{ l c c c c  c c  c c c c c c c }
\hline\hline
&\multicolumn{5}{c}{IF position}
&
&\multicolumn{5}{c}{MP2 position}
\\ 
\cline{2-6}\cline{8-12}
&\multicolumn{2}{c}{10.5~km~s$^{-1}$} 
&\multicolumn{3}{c}{12~km~s$^{-1}$} 
&
&\multicolumn{2}{c}{8.5~km~s$^{-1}$}
&\multicolumn{3}{c}{10~km~s$^{-1}$}   
\\ \cline{2-3}\cline{5-6}\cline{8-9}\cline{11-12}
    
&$T_{\mathrm{rot}}$ 
&$N$ 
&
&$T_{\mathrm{rot}}$ 
&$N$ 
&
&$T_{\mathrm{rot}}$ 
&$N$ 
&
&$T_{\mathrm{rot}}$ 
&$N$
\\
Species  
&(K) 
&($10^{12}\mathrm{cm}^{-2}$)
&
&(K) 
&($10^{12}\mathrm{cm}^{-2}$)
&
&(K) 
&($10^{12}\mathrm{cm}~\mathrm{s}^{-1}$)
&
&(K) 
&($10^{12}\mathrm{cm}~\mathrm{s}^{-1}$)
\\
\hline\hline
C$_2$D 		 &19$^{a}$ &$3.09\pm1.09$ &        	         &\ldots&\ldots       &  	 &20$^{+12}_{-8}$	 &9.60$^{+3.9}_{-2.00}$ 	&  &19$^{a}$  	&$3.0\pm0.490$  		\\ 	 	
\\		 
DCN 		 &44$^{+10}_{-10}$ 	 & 2.09$^{+0.75}_{-0.60}$&  	&8$^{+6}_{-2}$		&0.67$^{+0.10}_{-0.14}$  & 			 &12$^{+10}_{-5}$&  1.40$^{+0.20}_{-0.40}$  & &31$^{+10}_{-10}$	  		&  0.40$^{+0.020}_{-0.15}$ 	 	 		\\     			     
\\
DNC 		 &45$^{+5}_{-5}$	 &$0.091^{+0.070}_{-0.060}$& 	&45$^{+10}_{-5}$&$0.434^{+0.050}_{-0.010}$ & 		 &12$^{+2}_{-2}$&0.46$^{+0.020}_{-0.090}$&& \ldots &\ldots \\ 
\\
DCO$^{+}$ 	 &19$^{+10}_{-10}$	 &$0.111$$^{+0.087}_{-0.011}$&  	&19$^{a}$		&0.201$\pm0.90$   & 	     &31$^{+10}_{-12}$&0.27$^{+0.040}_{-0.095}$ & &12$^{+10}_{-4}$	   		&  0.10$^{+0.034}_{-0.013}$ 		 		\\   
\\
D$_{2}$CO 	 &38$^{a}$ 	 			 &$<23.70$ 		&		&38$^{a}$    			&$<15.00$     & 	     &38$^{a}$ 			 &$2.50\pm3.760$ &    &38$^{a}$  	   		&1.97$\pm3.113$  	\\
\\
HDCO  	     &38$^{a}$ 	 		 &$0.430\pm0.126$ 	&	&38$^{a}$	    &$0.745\pm0.126$      & 			 &49$^{+16}_{-8}$ &  2.30$^{+1.46}_{-0.99}$  & &38  	   		&$0.32\pm0.055$  	\\  
\\
NH$_2$D 	 &19$^{a}$  	 	 &$0.389\pm0.128$  & 	&\ldots			&\ldots       		 	 & 		 &19$^{a}$  	 &$0.87\pm0.129$  &   &19$^{a}$   	&0.36$\pm0.129$      \\
\\
N$_2$D$^{+}$ &\ldots 		 	 &\ldots   		   &	&19$^{a}$    	&0.065$\pm0.012$ 	&  	 &19$^{a}$  	 &$<0.16$    	&	 &19       		&$<0.16$             \\      
\hline
C$_2$H  &30$^{+34}_{-10}$		&158.00$^{+50}_{-44}$    &		 &8$^{+1}_{-1}$  &455.00$^{+145}_{-98}$    &  		 &12$^{+3}_{-2}$&120.00$^{+38}_{-27}$     &	 &19$^{+9}_{-5}$		&213.00$^{+69}_{-46}$	  \\
\\
H$^{13}$CN &14$^{+5}_{-4}$		&  2.0$^{+0.380}_{-0.380}$    & 		 &6$^{+5}_{-3}$  &0.4015$^{+0.180}_{-0.180}$   & 		 & 7$^{+3}_{-2}$&  1.00$^{+0.200}_{-0.350}$	& & 8$^{+4}_{-2}$	 	&  2.50$^{+0.300}_{-0.460}$\\ 
\\
HC$^{15}$N &22$^{+3}_{-3}$		& 0.38$^{+0.060}_{-0.030}$    &   	 &19$^{a}$   	 &$0.330\pm0.042 $ &	 & 9$^{+1}_{-1}$&  0.45$^{+0.050}_{-0.030}$  &&12$^{+3}_{-2}$		&  0.17$^{+0.040}_{-0.030}$\\
\\
HN$^{13}$C  &19$^{a}$				&$0.176\pm0.017$     & 				 &19$^{a}$  	 & $0.396\pm0.057$  &  	 &12$^{+1}_{-1}$&  0.43$^{+0.016}_{-0.020}$ & & 6$^{+2}_{-1}$		&  0.12$^{+0.026}_{-0.030}$\\
\\
H$^{15}$NC 	&19$^{a}$  			&$0.057\pm0.041$      &				 &19$^{a}$    	 & $0.177\pm0.058$ & 		 &13$^{+2}_{-1}$&  0.15$^{+0.010}_{-0.040}$  &&\ldots &\ldots\\
\\
H$^{13}$CO$^{+}$	 &16$^{+1}_{-1}$		&0.96$^{+0.032}_{-0.030}$     &			 &25$^{+11}_{-7}$&  0.14$^{+0.010}_{-0.036}$   &  &12$^{+1}_{-1}$&  1.49$^{+0.382}_{-0.410}$ & & 9$^{+1}_{-1}$  		&  0.31$^{+0.002}_{-0.002}$\\
\\
HC$^{18}$O$^{+}$ 	 &16$^{+2}_{-2}$		&0.129$^{+0.016}_{-0.015}$    & 	 	 &\ldots  		 &\ldots  	&  &13$^{+1}_{-1}$&  0.12$^{+0.010}_{-0.014}$	 &&\ldots&\ldots  	  					  \\      
\\
H$_{2}$CO  		 	 &55$^{+20}_{-18}$		&67.1$^{+182.9}_{-13.72}$	     &		 &\ldots		 &\ldots      &  		 &44$^{+17}_{-9}$ & 46.15$^{+9.5}_{-7.20}$ 	& &39$^{+14}_{-8}$       & 31.99$^{+11.15}_{-8.40}$\\    
\\
H$_2$$^{13}$CO		 &38$^{a}$     			&$<3.10$  		     &					 &38$^{a}$  	 &$<2.84$  & 	 &38$^{a}$  	&$2.74\pm02.983$	      	 & &\ldots 					&\ldots 						  \\
\\
N$_{2}$H$^{+}$  &\ldots	 	 				&\ldots    		     &					 &19$^{a}$    	 &$4.28\pm0.046$  &   &19$^{a}$   	&$3.29\pm0.054$     	 &&\ldots  					&\ldots  						  \\
\hline
\end{tabular}
\scriptsize
\begin{list}{}{}
\item $^{a}$ Fixed temperature to calculate the column density.
\item 
\end{list}
\label{T5}
\end{center}
\end{table*}
To derive reliable column densities, we combined our 1mm maps with single-pointing observations obtained at 3mm and 2mm toward both the IF and MP2 positions. To do that, we convolved the 1mm maps to the angular resolution of the lower excitation transitions at 3mm or 2mm. The beam averaged column densities were calculated assuming that the emission comes from a layer with uniform physical conditions and using the rotational diagram technique. This technique is valid for optically thin emission and assumes the same excitation temperature for all transitions. Within this approximation, 

 \begin{equation}
N_\mathrm{u}=\frac{1.94\times10^{3}\,\nu^{2}\,\mathrm{W}}{\mathrm{A_{ul}}},
\end{equation}
 \begin{equation}
N_\mathrm{tot}={\frac{N_{\rm u}\,Q\,{\rm exp}\frac{E_{\rm u}}{T_{\mathrm{rot}}}}{{g_{\rm u}}}} ,
\end{equation}
where $N_\mathrm{u}$ is the column density, in cm$^{-2}$, of the upper level `u' of the corresponding transition, $\nu$ is the frequency of the transition in GHz, $\mathrm{W}$ is the area derived from the Gaussian fitting in K~kms$^{-1}$ (Table~A.1-4), $\mathrm{A_{ul}}$ is the Einstein coefficient for spontaneous emission, $N_{\mathrm{tot}}$ is the total column density of the molecule in cm$^{-2}$, $E_\mathrm{u}$ is the energy of the upper level in K, $T_\mathrm{rot}$ is the rotation temperature in K, and $Q$ is the partition function, which depends on  $T_\mathrm{rot}$. The values of $Q$ were taken from the Cologne Database for Molecular Spectroscopy (CDMS webpage\footnote{http://www.astro.uni-koeln.de/cdms}; \citealt{muller2001}; 2005). Since our approximation requires optically thin emission, we used the $^{13}$C and $^{18}$O isotopologs to calculate the column density of the hydrogentated species, assuming [$^{12}$C]/[$^{13}$C]=50 (\citealt{savage2002}; \citealt{ginard2012}) and [$^{16}$O]/[$^{18}$O]=500\footnote{The values of the [$^{12}$C]/[$^{13}$C] and [$^{16}$O]/[$^{18}$O] ratios correspond to the galactocentric distance of Mon~R2, \ie\ $\sim$9~Kpc.} (\citealt{wilson1994}; \citealt{ginard2012}). We assumed a beam filling factor of 1 for all the transitions, and the obtained values were averaged values within the beam of the lower energy transitions (at 3mm or 2mm, depending on the species). 

Table~1 shows the results of the rotational diagrams. Rotation temperatures vary between $\sim$10~K and $\sim$40~K. These differences are due to the different dipole moments of the molecules ($\mu$$\sim$1$-$5~Debye) and to the different spatial resolution of the lower energy transitions. For instance, the rotation temperature obtained from H$^{13}$CN lines corresponds to a 29$\arcsec$ beam, while the rotational temperature of DCN corresponds to a 16$\arcsec$ beam. The 29$\arcsec$ beam encompasses a larger fraction of cold gas than the 16$\arcsec$ one, and the average rotation temperature is therefore lower (see Fig. 3). This different spatial resolution is also the cause of the higher rotation temperature measured for the deuterated species of HNC compared with that of the $^{13}$C isotopolog. Moreover, the high rotation temperatures found in DCN and DNC towards IF position confirm that the high deuterium fractions are associated with dense clumps around the {\sc Hii}~region at temperatures $\sim$50 K instead of the cooler envelope. In the case of HCO$^+$, the rotation temperature is similar for the $^{13}$C isotopolog and the deuterated species. This is consistent with the spatial distribution of DCO$^+$ (see Fig. 3), where intense emission comes from the envelope. The rotation temperatures are similar for H$_2$CO and their related species. In this case, all the column densities of all the molecules are averaged over a $\sim$16$\arcsec$ beam. Similarly to DCN and DNC, the emission of this family of molecules comes mainly from the warm clumps around the {\sc Hii}~region.

For some compounds, only one transition is detected. In these cases, we calculated the column density assuming a fixed value of the rotation temperature. For these cases we considered three rotational temperatures (T$_{\rm rot}$=10~K, T$_{\rm rot}$=19~K and T$_{\rm rot}$=38~K) and calculated the column density for every case. We found that the variation of a factor 2 in the rotation temperature affects the column density by less than a factor 2. For H$_2$$^{13}$CO and its deuterated compounds, HDCO and D$_2$CO, we used T$_{\rm rot}$=38 K (because this temperature is more similar to that derived from H$_{2}$CO) and for the remaining species we assumed T$_{\rm rot}$=19~K. 

Table~2 shows the deuterium fractions ($D_{\rm frac}$(XH)= [XD]/[XH]) for each velocity component toward the IF and MP2 positions. When possible, we derived them by comparing the $^{13}$C isotopolog with the deuterated species and assuming $^{12}$C/$^{13}$C=50. Since the abundances and excitation conditions of both molecules are similar, the derived  $D_{\rm frac}$ are not strongly affected by possible line opacity effects. As commented above, we have assumed a beam filling factor of 1 for the emission of all the studied transitions. To check the validity of our assumption and the possible impact on the estimated values of $D_{\rm frac}$, we produced maps of the [DCN (3$\rightarrow$2)]/[H$^{13}$CN (3$\rightarrow$2)] and [DCO$^+$ (3$\rightarrow$2)]/[H$^{13}$CO$^+$ (3$\rightarrow$2)] line-integrated intensity ratios (see Fig. 4).
These maps have an angular resolution of $\sim$10$\arcsec$ and provide information about the spatial variations of the values of $D_{\rm frac}$ in the region. For the two family of species, HCO$^+$ and HCN, the values of the line-integrated intensity ratio varies by less than a factor of 2 within the 29$"$ beam, suggesting that the beam-filling assumption is good enough and our values of $D_{\rm frac}$ are correct within the same factor. Toward the MP2 position, the deuterium fractions are highest toward the center, suggesting that the center is cooler and this PDR represents a clump illuminated from the outside, rather than one harboring a young star.

The only exception is the [NH$_2$D]/[NH$_3$] ratio. To calculate the [NH$_2$D]/[NH$_3$] ratio, we took the NH$_3$ column density derived by \citet{montalban1990} from the integrated intensity maps of the (1,1) and (2,2) lines observed with the Effelsberg telescope. The beam of the NH$_3$ observations was $\sim$42$"$. Taking into account possible calibration differences between the two telescopes and that the pointings are not exactly the same, we consider that the uncertainty could be as large as a factor of $\sim$5.

\begin{figure*}
\centering
\includegraphics[width=6.57cm]{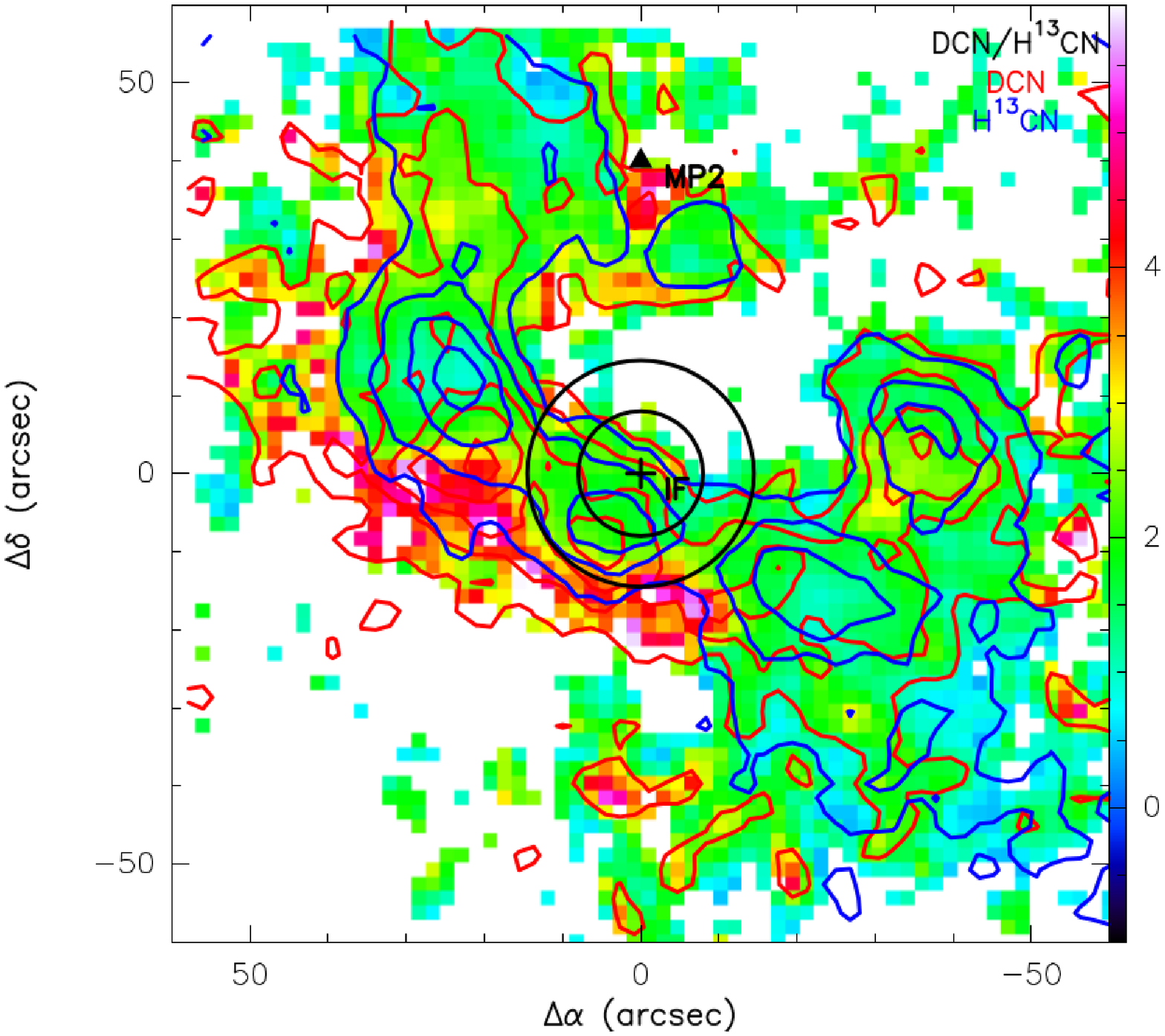} 
\includegraphics[width=5.72cm]{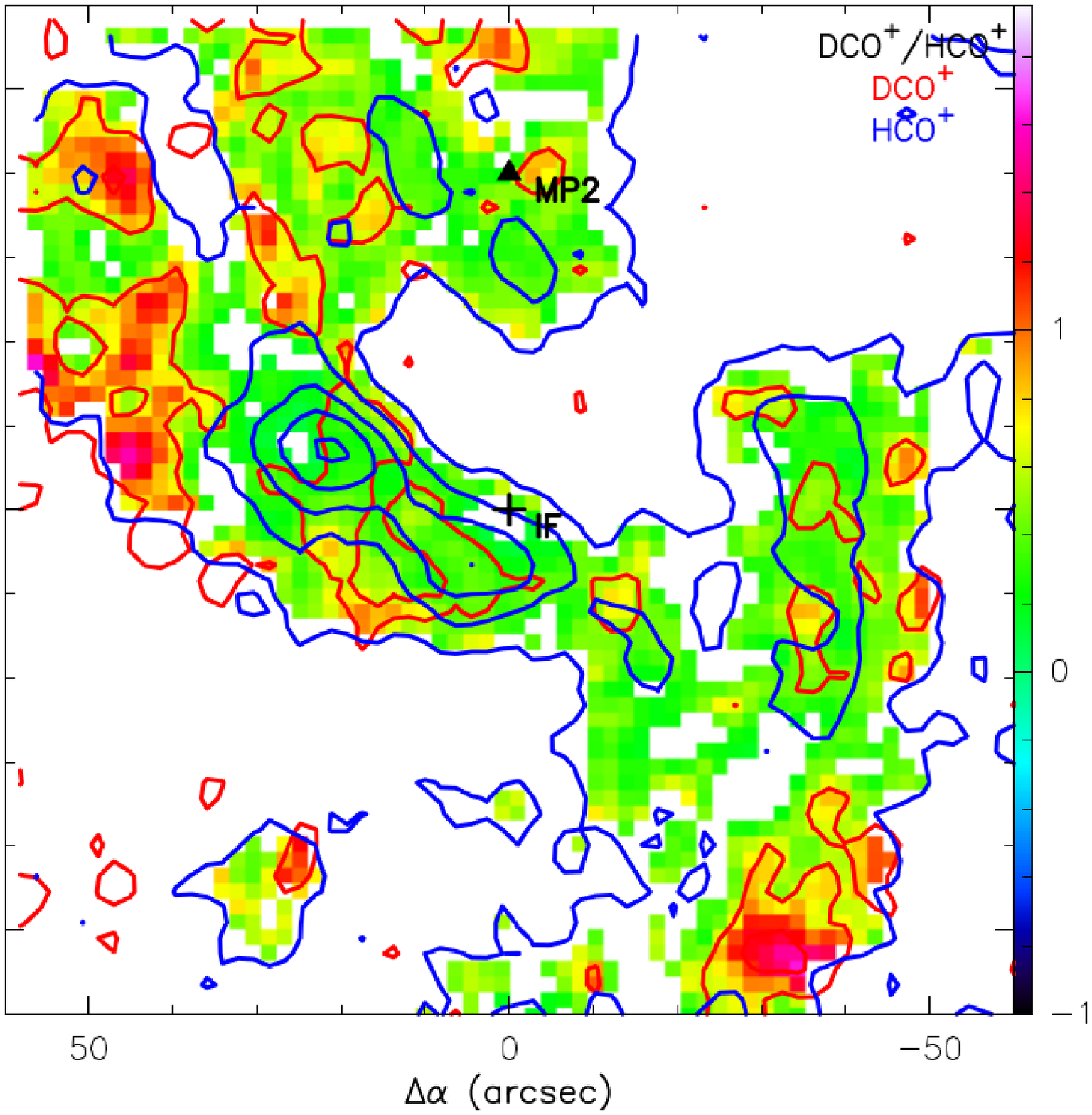}
\caption{{\it Left:} the color scale shows the [DCN]/[HCN] ratio, the red contours mark the DCN emission at 216.112~GHz, the blue contours the H$^{13}$CN emission at 259.011~GHz. {\it Right:} the color scale shows the [DCO$^{+}$]/[HCO$^{+}$] ratio, the red contours mark the DCO$^{+}$ emission at 144.07728~GHz, the blue contours the H$^{13}$CO$^{+}$ emission at 260.255~GHz. The contour levels are 5$\sigma$, 10$\sigma$, 15$\sigma$, 20$\sigma$, and 25$\sigma$. The black circles represent the 2 and 3mm beams.}
\label{ratios}
\end{figure*}


\subsection{Comparison of the IF and MP2 positions}

In this section we compare the chemistry of the different velocity components. The 10 \kms\ component presents similar abundance ratios toward the two positions, and these ratios are different from those in the 12 \kms\ and 8 \kms\ components. This component is characterized by a very high [HCN]/[HNC] abundance ratio, [HCN]/[HNC]$\sim$10. Regarding the spatial distribution, this component is boundering the {\sc Hii}~region and is the closest (in projected distance) from IRS~1. The 8 \kms\ and 12 \kms\ present [HCN]/[HNC] ratios of about $\sim$1-3, characteristic of cold or moderately warm clouds. There are no important chemical differences between the 8 \kms\ and 12 \kms\ components, suggesting that they correspond to gas at a similar kinetic temperature, probably because they are equidistant from IRS~1.

\begin{table}[]
\begin{center}
\scriptsize
\caption{Fractional abundances ratios.}
\begin{tabular}{ l c c c c c c c c c c c c c}
\hline\hline\noalign{\smallskip}
&\multicolumn{2}{c}{IF}
&
&\multicolumn{2}{c}{MP2}
\\
\cline{2-3}\cline{5-6}
Species
&10.5~km~s$^{-1}$  
&12~km~s$^{-1}$
&
&8.5~km~s$^{-1}$  
&10~km~s$^{-1}$    
\\  
\hline\hline
\noalign{\smallskip}
$\frac{\mathrm{H}^{13}\mathrm{CN}}{\mathrm{HN}^{13}\mathrm{C}}$  &$11.36$  	     &$1.013$  	         & &2.33    &20.83 \\
\\
$\frac{\mathrm{HC}^{15}\mathrm{N}}{\mathrm{H}^{15}\mathrm{NC}}$  &$6.66$  	     &$1.86$             & &$3.00$  &\ldots\\
\\
$\frac{\mathrm{H}^{13}\mathrm{CN}}{\mathrm{HC}^{15}\mathrm{N}}$	 &$5.26$  	     &$1.22$             & &2.22    &14.70 \\
\\
$\frac{\mathrm{DCN}}{\mathrm{H}^{13}\mathrm{CN}\times 50}$	 	 &2.0$\times 10^{-2}$ &3.3$\times 10^{-2}$ & &2.8$\times 10^{-2}$ &$0.32\times 10^{-2}$\\
\\
$\frac{\mathrm{DNC}}{\mathrm{HN}^{13}\mathrm{C}\times 50}$	  	 &$1.0\times10^{-2}$ &2.2$\times 10^{-2}$ & &2.1$\times 10^{-2}$ &\ldots \\ 
\\
$\frac{\mathrm{C}_{2}\mathrm{D}}{\mathrm{C}_{2}\mathrm{H}}$ 	 &1.9$\times 10^{-2}$ &\ldots        & &8$\times 10^{-2}$    &$1.4\times 10^{-2}$ \\

\\
$\frac{\mathrm{HDCO}}{\mathrm{H}_{2}\mathrm{CO}}$ 		 &$0.63\times10^{-2}$ &\ldots  	         & &$4.9\times 10^{-2}$   &$1.0\times 10^{-2}$\\

\\
$\frac{\mathrm{HDCO}}{\mathrm{H}_{2}^{13}\mathrm{CO}\times 50}$  &$>0.3\times10^{-2}$  &$>0.5\times10^{-2}$  & &$1.7\times 10^{-2}$   &\ldots\\

\\
$\frac{\mathrm{DCO}^{+}}{\mathrm{H}^{13}\mathrm{CO}^{+}\times 50}$	 &$0.23\times10^{-2}$ &2.8$\times 10^{-2}$              & &0.4$\times 10^{-2}$   &0.6$\times 10^{-2}$\\ 
\\
$\frac{\mathrm{DCO}^{+}}{\mathrm{HC^{18}O}^{+}\times 500}$	 &$0.17\times10^{-2}$ &\ldots               & &$0.45\times10^{-2}$    &\ldots \\ 

\\
$\frac{\mathrm{D_{2}CO}}{\mathrm{HDCO}}$          & $<55.89$ 		& $<20.13$	&& 1.08		&6.15	\\
\\
$\frac{\mathrm{D_{2}CO}}{\mathrm{{H}_{2}^{13}\mathrm{CO}\times 50}}$  & $<0.152$  		&$<0.010$	 	&&$0.018$	&\ldots	\\
\\
$\frac{\mathrm{N_{2}D}^{+}}{\mathrm{N_{2}H}^{+}}$ &\ldots 		&0.015		 	&			 &$<0.05$ 	&\ldots	\\
\\
$\frac{\mathrm{NH_{2}D}}{\mathrm{NH_{3}}^{*}}$        &$0.39\times10^{-2}$  &\ldots &               &$0.62\times10^{-2}$  &\ldots  \\

\hline\hline
\end{tabular}
\end{center}
\noindent
\\
\scriptsize
$^{*}$NH$_{3}$ column densities were taken from Montalban \et\ 1990.\\
\begin{list}{}{}
\item 
\item 
\end{list}
\label{T6}
\end{table} 

  
The values of the deuterium fractions are quite similar for the three velocity components. Values around 0.01 are found for all of the observed species except for DCO$^+$ and N$_{2}$D$^+$ which present a deuterium fraction 10 times lower. The largest differences are found for C$_2$D, $D_{\rm frac}$(C$_2$H) being $\sim$4 times higher in the 8 \kms\ component toward the MP2 position than in the others. The hydrogenated compound of this species is also very abundant in this component, suggesting that opacity effects could contribute to this higher value of the deuterium fraction. We would need to observe the $^{13}$C isotopolog to obtain a more accurate value of $D_{\rm frac}$(C$_2$H). The deuterated species N$_2$D$^+$ and D$_2$CO have only been detected in the more shielded components at 12.0 \kms\ (N$_2$D$^+$) and 8 \kms\ (D$_2$CO), again consistent with these velocity components being associated with colder gas, farther from IRS~1. 

\begin{figure*}
\centering
\includegraphics[width=10cm]{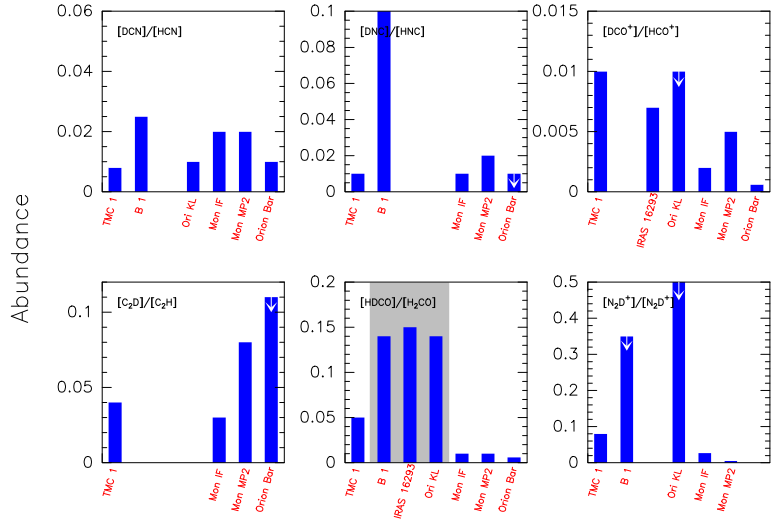}\\ 
\caption{Histogram comparing the deuterium fractions in different regions: the dark cloud TMC 1, the young protostellar object Barnard 1, the hot corino IRAS+16293-2422, the hot core Orion KL, and the PDRs in the Orion Bar and Mon~R2. References for the plotted values are shown in Table~3.}
\label{histo}
\end{figure*}

\subsection{Comparison with other sources}

In Table~3 we compare the deuterium fractions measured in Mon~R2 with those in some prototypical objects. In particular, we compare with a dark cloud (TMC1; \citealt{turner2001}), a young protostellar object (Barnard 1; \citealt{daniel2013}; \citealt{gerin2001}, \citealt{marcelino2005}), a hot corino (IRAS 16293-2422; \citealt{lis2001}; \citealt{loinard2000}; 2001; \citealt{tine2000}), and a hot core (Orion KL; \citealt{turner1990}). While the deuterium fractions of HCN and HNC are quite similar for all the sources ([DCN]/[HCN]~$\sim$~[DNC]/[HNC]~$\sim0.01$), the deuterium fractions of H$_2$CO, HCO$^+$, N$_2$H$^+$ and NH$_3$ do change. Values of the [HDCO]/[H$_2$CO] ratio measured toward hot cores and hot corinos are found to be higher ($\sim0.15$) than those in dark clouds ($\sim0.05$) and PDRs ($\sim0.01$). This indicates that the deuteration of formaldehyde proceeds more efficiently on the grain surfaces and the deuterated compounds are released in to the gas phase when the dust is heated to high temperatures ($>$100~K). The low values of $D_{\rm frac}$ found in Mon~R2 suggest that surface chemistry and subsequent ice evaporation is not the main deuteration pathway, as in the case of hot corinos (see Fig. 6).

In contrast, the [DCO$^{+}$]/[HCO$^{+}$] ratio is higher in dark clouds ($\sim$0.01) than in hot cores ($\sim$0.007) and the PDRs associated with Mon~R2 and the Orion Bar ($\sim$0.002). This result is also supported by the study carried out by \citet{pety2007} toward the Horsehead nebula. They found that the HCO$^+$ deuterium fraction is larger than 2\% in the cold dense core and lower than 0.1\% in the PDR. The deuteration of HCO$ ^+$ proceeds via ion molecule reactions in the gas phase that are quite sensitive to the gas temperature. The behavior observed in Fig. 6 can be explained as the consequence of the increasing kinetic temperature. Note that for IRAS 16293-2422, the observed DCO$^+$ more likely comes from the envelope in which the hot corino is embedded.

For the other N-bearing molecules the [N$_{2}$D$^{+}$]/[N$_2$H$^{+}$] ratio is higher in the very first evolutionary stages of the stellar formation, represented the by the dark cloud TMC1 ($\sim$0.08) and the young protostellar object Barnard 1 ($\sim$0.3), than in more evolved objects such as Mon~R2 ($\sim$0.03). This result agrees with the recent study of \citet{fontani2011}, in which the authors derived the N$_2$H$^+$ deuterium fraction for a large sample of massive young stellar objects. They showed that the N$_2$H$^+$ deuteration is lower ($\sim$0.02) in objects associated with ultracompact {\sc Hii}~regions (i.e., similar to our PDRs). 

The situation is less clear for the NH$_3$. From the ratios listed in Table~\ref{literature}, the [NH$_{2}$D]/[NH$_3$] ratio is similar in hot cores ($\sim$0.06), dark clouds ($\sim$0.02) and PDRs ($\sim$0.06), but it reaches a value of $\sim$0.33 toward Barnard 1 (see Table 3). The young protostellar object (YSO) Barnard 1 is a special object where the deuterium fractions are known to be higher than in other cold cores. Daniel \et\ (2013) interpreted the high [NH$_{2}$D]/[NH$_3$] ratio as a time evolution effect, the consequence of its peculiar evolutionary stage, with less than 500 yr after the formation of the early hidrostatic core. This interpretation was based on calcuations by Aikawa \et\ (2012).
\begin{table*}[] 
\begin{center}
\caption{DX/HX ratio comparison}
\begin{tabular}{ l c c c c c c c c c c c c c c }
\hline\hline\noalign{\smallskip}
&\multicolumn{2}{c}{Mon~R2$^{a}$}
&Ori~Bar$^{b}$ 
&TMC1$^{c}$
&Barnard~1$^{d}$
&IRAS~16293$^{e}$
&Ori~KL$^{f}$
\\
\cline{2-3} 
&IF
&MP2
&(Clump~3)
&
&
\\
\hline\hline
\noalign{\smallskip}
$\mathrm{H}^{13}\mathrm{CN}$/$\mathrm{HN}^{13}\mathrm{C}$     &$10$  	      		&2.33					&$2.5$ 		    	&$0.9$--1.5  &1.04   &\ldots         &\ldots  \\
DCN/HCN                           							  &$0.02$ 	      		&0.03					&$0.01$ 	   		&$0.008$     &0.025  &0.01 			 &\ldots  \\
DNC/HNC                             						  &$0.01$ 	       		&0.02					&$<0.01$    	    &$0.01$      &0.11   &\ldots 		 &\ldots  \\
$\mathrm{DCO}^{+}$/$\mathrm{HCO}^{+}$           	          &$0.2\times 10^{-2}$  &$0.5\times 10^{-2}$	&$0.6\times 10^{-3}$&$0.01$      &\ldots &$0.7\times 10^{-2}$&0.14 \\
$\mathrm{C}_{2}\mathrm{D}$/$\mathrm{C}_{2}\mathrm{H}$ 		  &$0.02$	       		&0.08					&$<0.11$     	    &$0.03$--0.06&\ldots &\ldots		 &\ldots  \\
$\mathrm{HDCO}$/$\mathrm{H}_{2}\mathrm{CO}$    				  &0.01		       		&0.01					&$0.6\times 10^{-2}$&$0.05$      &0.14   &0.15  		 &0.14 \\
$\mathrm{D_{2}CO}/\mathrm{HDCO}$ 							  &$<38.47$	       		&1.00					&\ldots		    	&\ldots      &0.40   &0.3 			 &0.021\\
$\mathrm{N_{2}D}^{+}/\mathrm{N_{2}H}^{+}$	 			 	  &0.015$^{*}$	       	&$<5.0\times 10^{-2}$	&\ldots		    	&0.08	     &0.35   &\ldots		 &$<0.30$\\
$\mathrm{NH_{2}D}/\mathrm{NH_{3}}$	  		 				  &$0.39\times10^{-2}$ 	&$0.6\times 10^{-2}$	&\ldots		    	&0.02	     &0.63   &\ldots		 &0.062\\
\hline\hline
\end{tabular}
\end{center}
\noindent
\\
\scriptsize
$^{*}$From the component at 12~km~s$^{-1}$.\\
$^{a}$This work.\\
$^{b}$\citet{parise2009}.\\
$^{c}$\citet{turner2001}.\\
$^{d}$\citet{daniel2013}; \citet{gerin2001}, \citealt{marcelino2005}.\\
$^{e}$\citet{lis2001}; \citet{loinard2000}; \citet{loinard2001}; \citet{tine2000}.\\
$^{f}$\citealt{turner1990}.\\
\begin{list}{}{}
\item 
\item 
\end{list}
\label{literature}
\end{table*} 
 
Finally, because of their similarities, it is natural to compare our results in more detail with those in the Orion Bar (\citealt{parise2009}). Both clouds are massive star-forming regions, and in the two cases the PDRs are located at the interface of the {\sc Hii}~region and the molecular cloud. The only difference is that the Orion Bar is a more evolved {\sc Hii}~region, and the densities and the radiation field in the PDR associated with it are a factor of $\sim$10 lower than in Mon~R2. Both regions have been observed using the IRAM-30m telescope, which avoids possible calibration problems when comparing lines observed with different telescopes, but the spatial scales are different because the distance to Mon~R2 is twice than to Orion. The [DNC]/[HNC] ratio is $\sim$0.01 in Mon~R2, similar to the [DCN]/[HCN] ratio. \citet{parise2009} did not detect DNC toward the so-called "clump 3" in the Orion Bar and derived an upper limit to the deuterium fraction of 0.01, which is still consistent with our results. The same can be said for C$_2$D which was not detected toward clump 3, with an upper limit of $<$0.11 for $D_{\rm frac}$(C$_2$H). For HDCO, we have the opposite behavior, the $D_{\rm frac}$(H$_2$CO) in the 10 \kms\ component is a factor of 2 lower than in the Orion Bar. Taking into account the uncertainties because of the beam-filling factor and the different distance of Mon~R2 and the Orion Bar from the Sun, we do not consider that this difference is significant. Summarizing, although the number of deuterated compounds detected in Mon~R2 is higher than in the Orion Bar, the values of $D_{\rm frac}$ agree, within the uncertainties, in both PDRs.

The main difference between Mon~R2 and Orion Bar is the different [HCN]/[HNC] ratio, which is a factor of $\sim$3 higher in the 10 \kms\ Mon~R2 component. It is well known that the [HCN]/[HNC] ratio depends on the gas tempeature with values $\sim$1 in cold clouds and higher in warm regions. This is consistent with the values shown in Table~3. The [H$^{13}$CN]/[HN$^{13}$C] ratio in the 10 \kms\ component in Mon~R2 is significantly higher than in the Orion Bar, in line with the different gas kinetic temperatures ($\sim$30~K in the Orion Bar and $\sim$50~K in Mon~R2). As discussed in Section 5, this difference can also be due to the different timescales and densities, because the Orion Bar is a more evolved object than Mon~R2. Within Mon~R2, the [H$^{13}$CN]/[HN$^{13}$C] ratio is higher in the 10 \kms\ component than in the PDR toward MP2 position. The lower rotation temperature of the HCN and the HNC isotopologs toward the MP2 position suggests that the gas kinetic temperature and/or density are lower at the latter.
 
\section{Chemical model}

We compare our observational results with a pseudo-time-dependent gas phase chemical model. We chose this code to be consistent with previous conclusions by Pilleri \et\ (2012) about early-time chemistry for small hydrocarbons.

One concern could be the influence of the UV radiation on the deuteration (not accounted by our pseudo-time-dependent model). The detection of the reactive ions CO$^{+}$ and HOC$^{+}$ (Rizzo \et\ 2003) and the pure H$_{2}$ rotational lines (\citealt{berne2009}) toward the IF position showed that the UV radiation is impinging into molecular gas at this
region. We used the steady-state PDR Meudon code to investigate the influence of UV photons on the deuterium fractions. The UV photons do not influence the deuterium fractions directly, but through the increased gas temperature because of the UV heating. With high UV fields, $\sim10^{5}$ Habing field, the gas temperatures are very high in the first layers of the PDR, and consequently, the deuterium fractions are very low. Since the deuteration via ion-molecule reactions is not efficient for temperatures $>$70~K, the emission of the deuterated species is more likely arising from shielded (A$_V$$\sim$8-10~mag) dense clumps where the gas kinetic temperature is $\sim$50~K. At these moderate temperatures, 50~K, the high deuterium fractions measured toward Mon~R2 cannot be explained with steady-state chemistry, and we needed to use a time-depedent code to account for them.

The chemical network used in our pesudo-time-dependent code includes recent modifications of the reaction rate coefficients involving nitrogen, as reported in Wakelam \et\ (2013) following recent experiments on neutral-neutral reactions at low temperatures by Daranlot \et\ (2011; 2013) and is based on previous studies by Roueff \et\ (2005; 2013) and Pagani \et\ (2011). The ortho/para ratio of H$_2$ (hereafter OPR) is considered as an input parameter that does not evolve with time, and influences the reverse reaction of the H$_3$$^{+}$ fractionation reaction with HD, where the presence of ortho H$_2$ reduces the endothermicity of the reaction by a significant amount of energy (170.5~K), and then the
deuteration efficiency as first pointed out by Pagani \et\ (1992). The contribution of para- and ortho-H$_2$ is also explicitly introduced in the initial step of nitrogen hydride chemistry, the N$^{+}$ + H$_2$ reaction, following the prescription by Dislaire \et\ (2012). Within this approximation, the chemical network includes 214 species and 3307 chemical reactions.

To fit the observed abundance ratios we considered a grid of chemical models with the temperature fixed at 50~K and varied the density, initial conditions, and OPR. We considered two densities: n$_{\mathrm{H}}=3\times 10^{5}$ cm$^{-3}$ and n$_{\mathrm{H}}=2\times 10^{6}$~cm$^{-3}$, which are representative of the range of densities measured in this region (Pilleri \et\ 2012, 2013). Table~4 shows the elemental abundances of the low (A and B) and high (C and D) metallicity models. In the model, the adsorption and evaporation of molecules on/from the grains are not included. When molecules such as H$_2$CO or CH$_3$OH evaporate, the elemental abundances of O and C in the gas phase increase. Given the limitation of our model, we mimicked this situation with the high metallicity case.

We also used a constant OPR because it is difficult to deal with the full ortho-para chemistry in the time-dependent code, but we considered different values of OPR: 0.3, that is the equilibrium value at T$_{\mathrm{K}}=50$~K, $1\times10^{-2}$, $1\times10^{-3}$ and $1\times10^{-4}$ for every model. The results of these models were quite similar for t$<$1 Myr. The OPR mainly affects the [NH$_2$D]/[NH$_3$] ratio. We can fit the observed values with both OPR$=1\times10^{-2}$ and 0.3.

In Fig.~7, we show the results of our pseudo-time-dependent model. The first important conclusion is that all the models give values of D$_{frac}$ and the [HCN]/[HNC] ratio very far from the observed ones in the steady state, \ie\ for ages $>$1 Myr. We find, however, values closer to our observations for ages between 10$^{4}$ to a few 10$^{5}$~yr. Similarly, Pilleri \et\ (2013) showed that the abundances of small hydrocarbons are better explained with an early time chemistry. This age, from 10$^{4}$ to a few 10$^{5}$~yr,  also agrees with the typical ages of UC {\sc Hii} regions. The data suggest that the collapse and the chemical evolution of the region occur quickly. Nevertheless, it is important to stress the fact that our models (from A to D) consider that the temperature and the density of the region are constant when in fact these parameters are variable. As a consequence, our age estimate is illustrative and mainly proves that the deuterium chemistry is out of equilibrium. The second conclusion is that the observations
cannot be fitted with the high metallicity case, \ie\ the models C and D. This is also consistent with the emission from the deuterated molecules coming from shielded clumps where the temperature is not high enough to fully evaporate the ice mantles.

The deuterium fractions are not very dependent on the assumed density within the range considered in our grid of models (see Fig.~7), but it is the [HCN]/[HNC] ratio. The variation of the [HCN]/[HNC] ratio comes from the differences in the atomic oxygen abundance. HCN does not react with O, whereas we allow the reaction; HNC$+$ O $\rightarrow$ NH $+$ CO. This reaction was introduced by Pineau des Forets \et\ (1990), but has never been measured in the laboratory, however. When O increases, the [HNC]/[HCN] ratio decreases drastically. The high [HCN]/[HNC] ratio measured in the 10.5 km~s$^{-1}$ component is higher than the values predicted by high-metallicity models. The [HCN]/[HNC] ratio is very dependent on the time and density (\eg\ models A and B in Fig.~7). Taking into account that we expect a density gradient in the region, we consider that there is a reasonable agreement with the low-metallicity case. This also suggests that the different [HCN]/[HNC] ratios measured among the three velocity components in Mon~R2 could be the consequence of small differences in density and gas kinetic temperature.

The best fit is obtained model A (see Fig.~7). As commented above, this model assumes constant density and temperature. To obtain a more realistic view, we ran two more models (E and F) consisting of two sequential phases at a different temperature and OPR: (i) in phase 1, we ran a model with T$=$15~K and OPR$=1\times 10^{-4}$ to simulate the cold collapse phase, (ii) then in phase 2, we ran a second model with T$=$50~K and OPR$=1\times10^{-2}$, using as input the abundances of the previous models for the ages 10$^{5}$~yr (model E) and 10$^{6}$~yr (model F). Table~4 summarizes the parameters of all the models. Model E is quite similar to model A, while model F is very different from the others. This model provides a good fit for the [C$_2$D]/[C$_2$H], [DNC]/[HNC], [DCN]/[HCN], and [NH$_2$D]/[NH$_3$] ratios, but it fails for the other observed ratios. The main disagreement of model F are the low absolute abundances of the molecular species, in particular, the C$_2$H abundance is two orders of magnitude lower than those obtained for the other models. This indicates that the collapse phase prior to the formation of the UC {\sc Hii} region should be fast enough to avoid the gas to achieve the steady state (see also Pilleri \et\ 2012). In model E, we assumed a constant density of $2\times 10^{6}$~cm$^{-3}$ during phase I. The chemical equilibrium is reached faster at higher densities. In a real time-dependent model in which the density is low at the beginning and increases with time, the collapse time could be longer than 0.1 Myr.

Figure~A.13 shows the abundances fits of HCN, HCO$^{+}$, and C$_{2}$H relative to H$_{2}$ considering a gas kinetic temperature of T$_{k}=50$~K and OPR=1$\times$10$^{-2}$ (from model A to model D). The red line represents the [HCN]/[H$_{2}$] ratio, the purple line shows the [HCO$^{+}$]/[H$_{2}$] ratio, and the blue line shows the [C$_{2}$H]/[H$_{2}$] ratio. The dashed lines shows the ratios of observational results toward MonR2 for the 10.5 \kms\ component at the IF position.

As commented above, the best fit is obtained for the model A and the worst one is the model D (see Fig.~A.13). The observational abundances relative to H$_{2}$ were calculated considering N(C$^{18}$O)$=7.3\times 10^{15}$~cm$^{-2}$ and X(C$^{18}$O)$=1.7\times 10^{-7}$ for the IF position (Ginard \et\ 2012).

\begin{figure*}
\centering
\includegraphics[width=5.5cm]{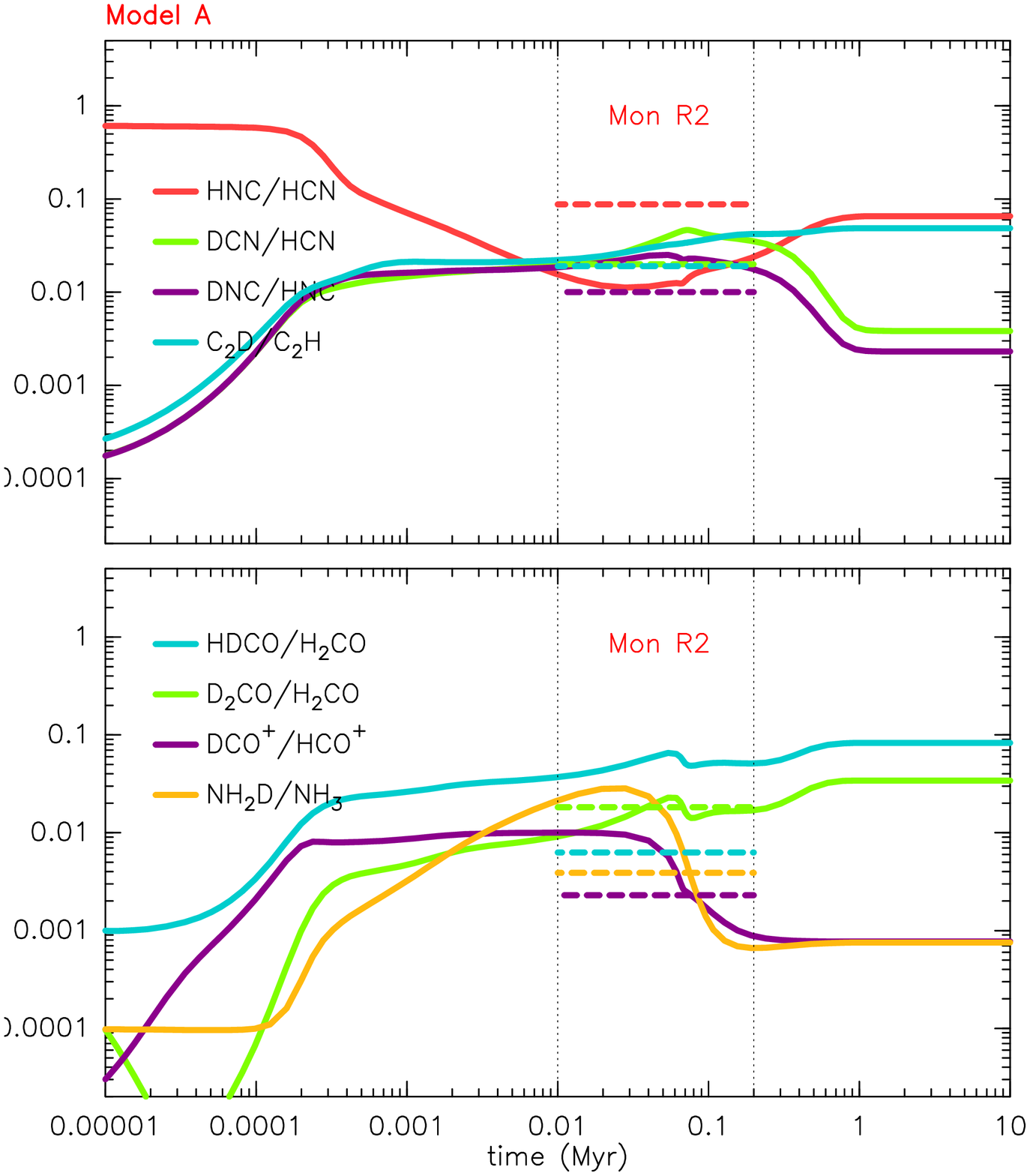}  
\includegraphics[width=5.25cm]{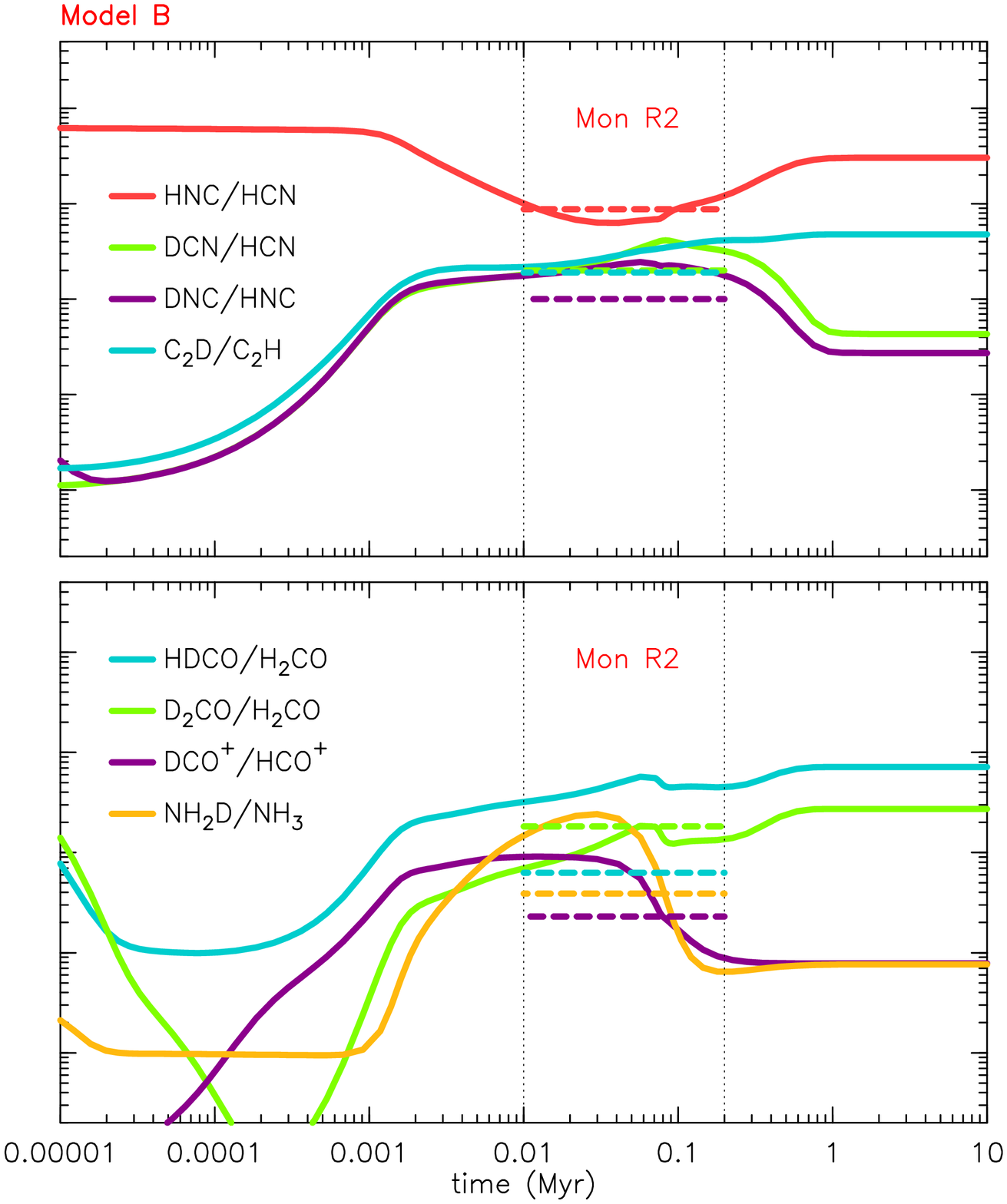} \\
\includegraphics[width=5.5cm]{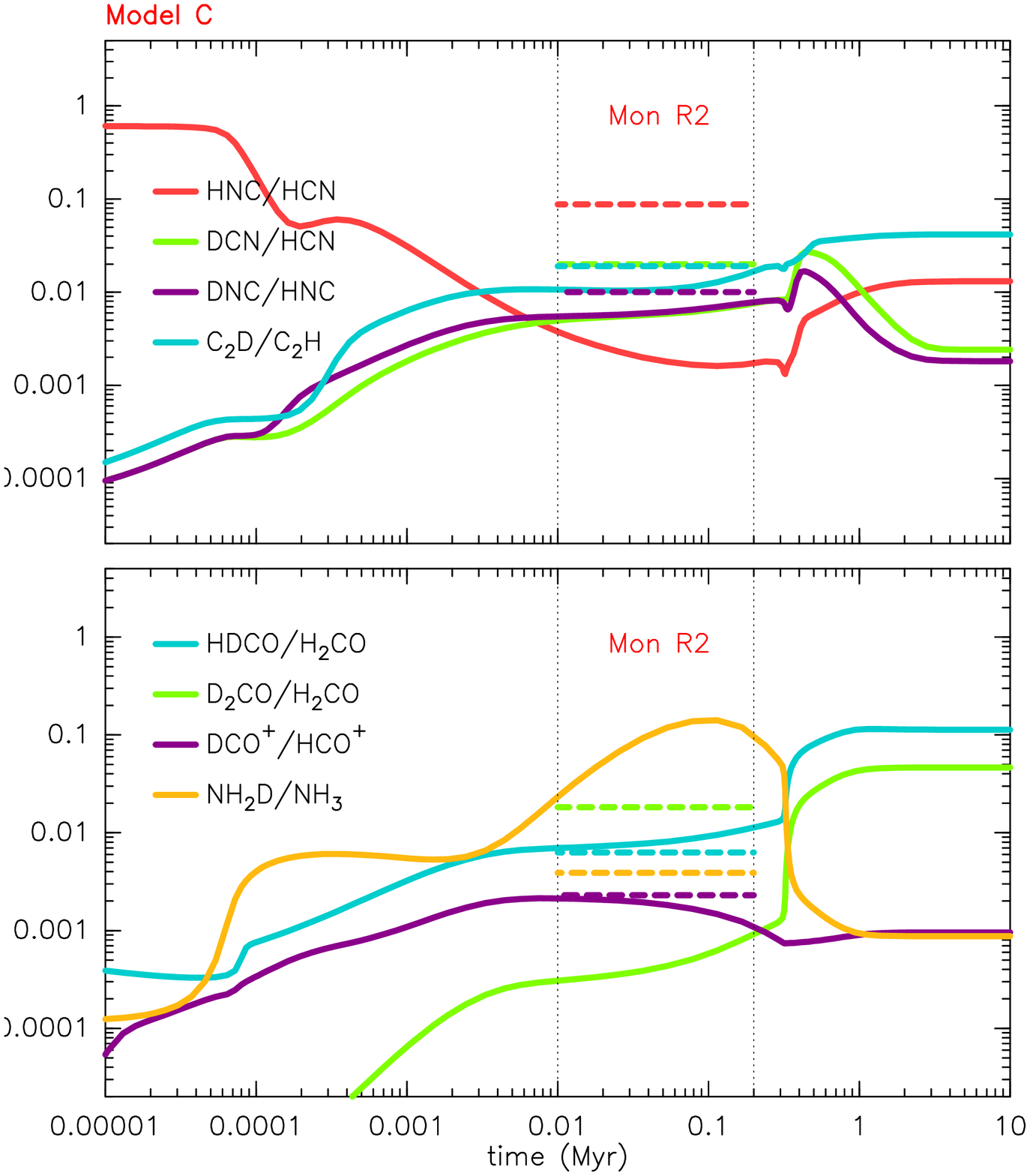}  
\includegraphics[width=5.25cm]{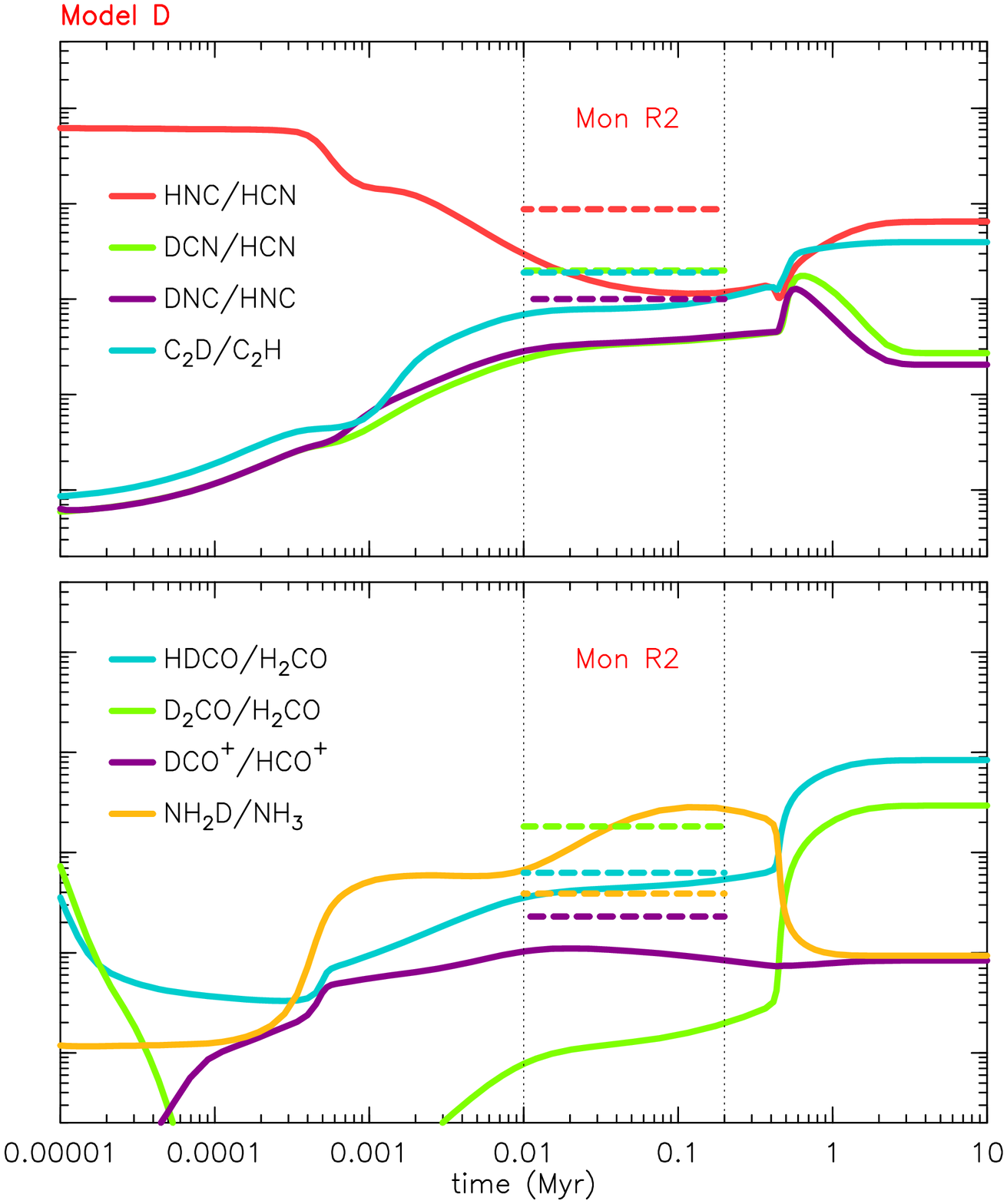} \\
\includegraphics[width=5.5cm]{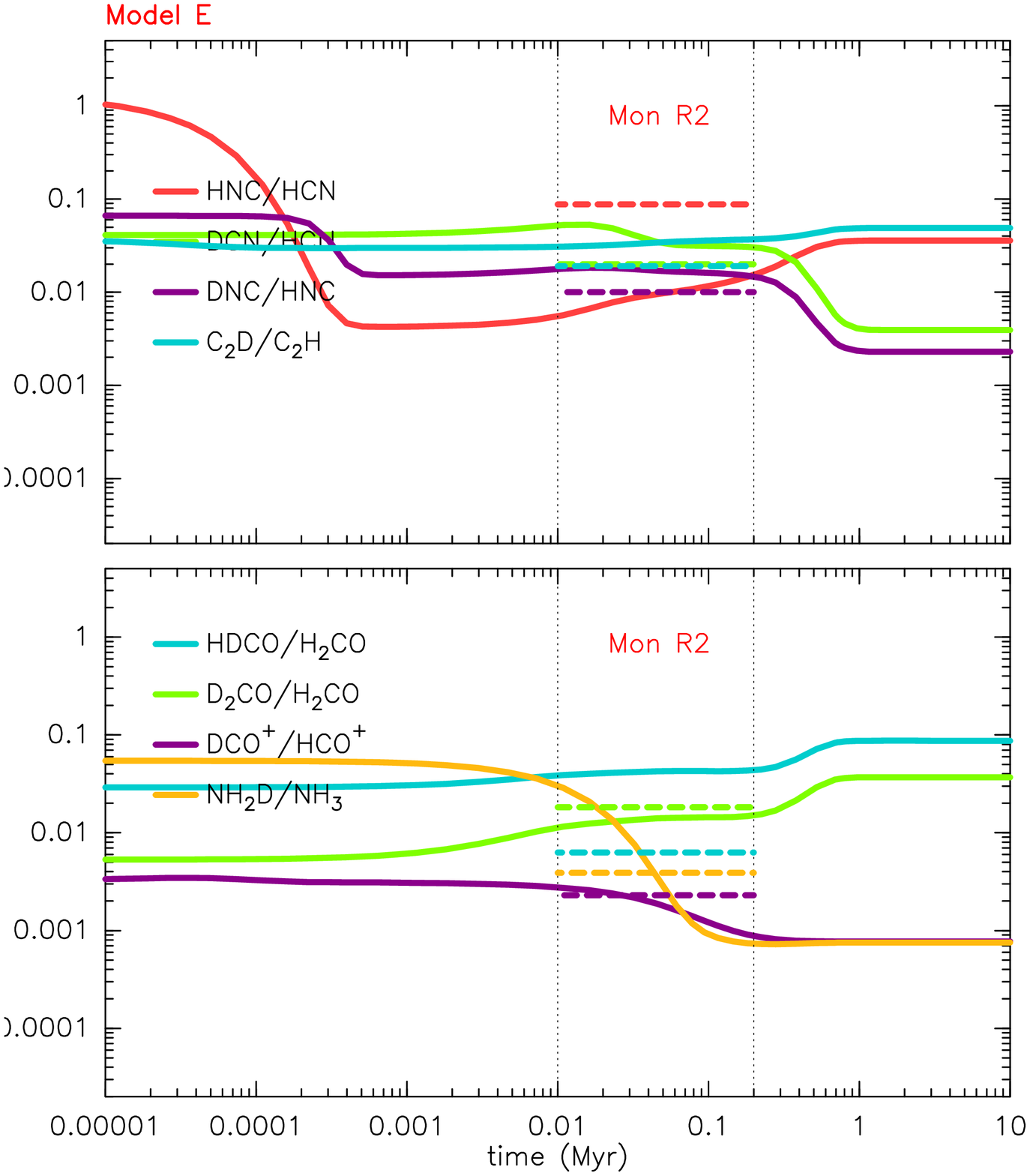}  
\includegraphics[width=5.25cm]{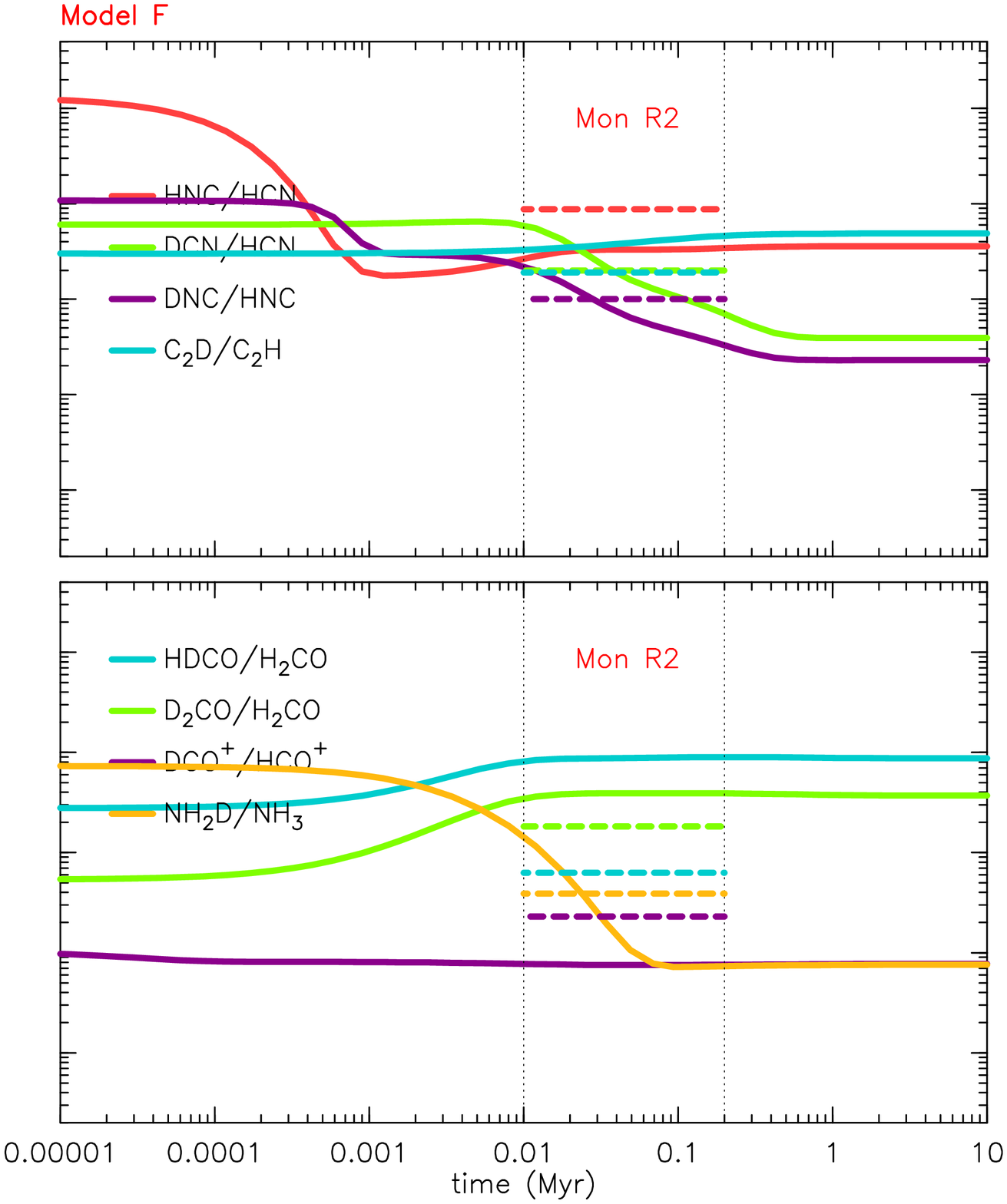}     
\caption{Chemical models considering a gas kinetic temperature of T$_{k}=50$~K and OPR=1$\times$10$^{-2}$ (from model A to model D). Models E and F consider a temperature T$_{k}=50$~K, but their initial abundances are obtained using another model (with T$_{k}=15$~K and OPR=1$\times$10$^{-4}$) at a time of $10^5$~yr (for model E) and $10^6$~yr (for model F). At the top part of the figures of every model, the red line represents the [HNC]/[HCN] ratio, the green represents the [DCN]/[HCN] ratio, the purple line shows the [DNC]/[HNC] ratio, and the blue line shows the [C$_{2}$D]/[C$_{2}$H] ratio. In the bottom part of the figures of every model, the blue line represents the [HDCO]/[H$_{2}$CO] ratio, the green line shows the [D$_{2}$CO]/[H$_{2}$CO] ratio, the purple line shows the [DCO$^{+}$]/[HCO$^{+}$] ratio, and the yellow one shows the [NH$_{2}$D]/[NH$_{3}$] ratio. The dashed lines shows the ratios of observational results toward MonR2 for the 10.5 \kms\ component at the IF position.}
\label{models}
\end{figure*}
 
\begin{table*}[]
\begin{center}
\scriptsize
\caption{Model parameters.}
\begin{tabular}{ l c c  c c  c c  c c cc cc }
\hline\hline\noalign{\smallskip}
&
&Model A 
&
&Model B
&
&
&Model C
&
&Model D\\
\cline{3-3}\cline{5-5}\cline{8-8}\cline{10-10}
Parameter
\\  
\hline\hline\noalign{\smallskip}
T$_{k}$	  &Temperature       		  &50~K						  &&50~K			             &&&50~K					   &&50~K			               \\
$n_{H}$   &H density         		  &$2\times 10^{6}$ cm$^{-3}$ &&$3\times 10^{5}$ cm$^{-3}$   &&&$2\times 10^{6}$ cm$^{-3}$ &&$3\times 10^{5}$ cm$^{-3}$   \\
He/H	  &Helium abundance 		  &0.1						  &&0.1					    	 &&&0.1						   &&0.1  				           \\
O/H		  &Oxygen abundance  		  &$1.8\times 10^{-4}$		  &&$1.8\times 10^{-4}$	    	 &&&$3.3\times 10^{-4}$		   &&$3.3\times 10^{-4}$		   \\
C/H	      &Carbon abundance  		  &$7.3\times 10^{-5}$		  &&$7.3\times 10^{-5}$	    	 &&&$1.3\times 10^{-4}$		   &&$1.3\times 10^{-4}$		   \\
N/H 	  &Nitrogen abundance		  &$2.1\times 10^{-5}$		  &&$2.1\times 10^{-5}$	    	 &&&$7.5\times 10^{-5}$		   &&$7.5\times 10^{-5}$		   \\
S/H 	  &Sulfur abundance  		  &$8\times 10^{-8}$ 		  &&$8\times 10^{-8}$	         &&&$1.8\times 10^{-5}$ 	   &&$1.8\times 10^{-5}$ 		   \\
Fe/H 	  &Iron abundance    		  &$2\times 10^{-8}$ 		  &&$2\times 10^{-8}$ 	         &&&$2\times 10^{-8}$ 		   &&$2\times 10^{-8}$ 		   \\
D/H 	  &Deuterium fraction		  &$1.5\times 10^{-5}$ 		  &&$1.5\times 10^{-5}$          &&&$1.5\times 10^{-5}$ 	   &&$1.5\times 10^{-5}$ 		   \\
ortho/para ratio &OPR	   		      &$1\times 10^{-2}$		  &&$1\times 10^{-2}$            &&&$1\times 10^{-2}$		   &&$1\times 10^{-2}$      	   \\
$\zeta$   &Cosmic ray ionization rate &$5\times 10^{-17}$~s$^{-1}$&&$5\times 10^{-17}$~s$^{-1}$  &&&$5\times 10^{-17}$~s$^{-1}$&&$5\times 10^{-17}$~s$^{-1}$   \\
\hline\hline
\noalign{\smallskip}
&
&
&Model E 
&
&&&
&Model F \\
\cline{3-5}\cline{8-10}
Parameter
&
&
Phase 1
&
&
Phase 2$^{a}$
&
&
&
Phase 1
&
&
Phase 2$^{b}$
\\
\hline\hline
\noalign{\smallskip}
T$_{k}$		  &Temperature                 &15~K						&&50~K							&&&15~K							&&50~K							\\
$n_{H}$       &H density                   &$2\times 10^{6}$ cm$^{-3}$	&&$2\times 10^{6}$ cm$^{-3}$ 	&&&$2\times 10^{6}$ cm$^{-3}$ 	&&$2\times 10^{6}$ cm$^{-3}$	\\
He/H		  &Helium abundance            &0.1							&&\ldots 			    		&&&0.1							&&\ldots						\\
O/H			  &Oxygen abundance            &$1.8\times 10^{-4}$			&&\ldots 						&&&$1.8\times 10^{-4}$			&&\ldots				  	    \\
C/H			  &Carbon abundance            &$7.3\times 10^{-5}$		    &&\ldots						&&&$7.3\times 10^{-5}$			&&\ldots						\\
N/H 		  &Nitrogen abundance          &$7.5\times 10^{-5}$			&&\ldots						&&&$7.5\times 10^{-5}$			&&\ldots						\\
S/H 		  &Sulfur abundance            &$2.1\times 10^{-5}$  		&&\ldots 		 				&&&$2.1\times 10^{-5}$ 			&&\ldots						\\
Fe/H 		  &Iron abundance          	   &$8\times 10^{-8}$			&&\ldots 						&&&$8\times 10^{-8}$ 			&&\ldots			    		\\
D/H 		  &Deuterium fraction          &$2\times 10^{-8}$ 			&&\ldots						&&&$2\times 10^{-8}$ 			&&\ldots						\\
ortho/para ratio &OPR			  		   &$1\times 10^{-4}$ 			&&$1\times 10^{-2}$ 			&&&$1\times 10^{-4}$ 			&&$1\times 10^{-2}$				\\
$\zeta$       &Cosmic ray ionization rate  &$5\times 10^{-17}$~s$^{-1}$	&&$5\times 10^{-17}$~s$^{-1}$	&&&$5\times 10^{-17}$~s$^{-1}$	&&$5\times 10^{-17}$~s$^{-1}$	\\
\hline\hline
\end{tabular}
\end{center}
\noindent
\\
\scriptsize
$^{a}$ Initial abundances of phase 2 were taken from phase 1 at a time of $10^5$~yr.\\
$^{b}$ Initial abundances of phase 2 were taken from phase 1 at a time of $10^6$~yr.\\
\label{modelo-tabla}
\end{table*}

\section{Discussion: can $D_{\rm frac}$(N$_2$H$^+$) be used as an evolutionary indicator in massive star-forming regions?}

Deuteration has been used extensively as an evolutionary indicator in low-mass star-forming cores. $D_{\rm frac}$(N$_2$H$^+$) is found to be $>$0.1 in starless cores close to the onset of gravitational collapse (\citealt{crapsi2005}). After the formation of the star, $D_{\rm frac}$(N$_2$H$^+$) decreases as the core evolves (\citealt{emprechtinger2009}; \citealt{caselli2008}). The reason is that $D_{\rm frac}$ and N(H$_2$D$^+$) are very dependent on the gas temperature in the range 10~K$-$30~K in which the CO frozen out on the grain surfaces is released to the gas phase.

It is not clear, however, that $D_{\rm frac}$(N$_2$H$^+$) can be used as evolutionary indicator in the high-mass regime. Fontani \et\ (2011) showed that the $D_{\rm frac}$(N$_2$H$^+$) in massive starless cores is $\sim$0.2, i.e., as high as in low-mass pre-stellar cores, but it drops to values of $\sim$0.04 for high-mass protostars and UC~{\sc Hii} regions. Less clear results were obtained by \citet{miettinen2011} toward a sample of IRDCs. They found that CO was not depleted in the observed sources and that $D_{\rm frac}$(N$_2$H$^+$) was lower than in low-mass starless cores (see \citealt{tan2013}, for a recent review).

Our results for Mon~R2 suggest that the deuterium fractions of molecules in massive protostars are dominated by ion-molecule reactions and are strongly time dependent. When the gas temperature increases over 20~K and the CO on the icy mantles is released to the gas phase, $D_{\rm frac}$(N$_2$H$^+$) decreases rapidly and reaches the steady-state value in less than 0.1~Myr. Even in the starless phase, if the core is turbulent and the temperature is higher than 20~K, $D_{\rm frac}$(N$_2$H$^+$) can reach values lower than those typical in the low-mass starless cores. The same is true for $D_{\rm frac}$(HCO$^+$). Other molecules such as HCN and HNC need about 1 Myr to reach the steady the state and could be useful to age massive protostellar cores and UC~{\sc Hii} regions. $D_{\rm frac}$(H$_2$CO) in only useful to prove the hot core phase. Once all the icy mantles are evaporated, $D_{\rm frac}$(H$_2$CO) reaches the steady-state value in a short time. 

\section{Conclusions}

We have determined the values of $D_{\rm frac}$ for a large sample of molecules (C$_2$H, HCN, HNC, HCO$^{+}$, H$_2$CO, N$_{2}$H$^{+}$, and NH$_3$) toward the PDRs IF and MP2 position, around the UC~{\sc Hii} region Mon~R2. Our results can be summarized as follows:

\begin{itemize}
\item Our observations show that Mon~R2 presents a complex morphological and kinematical structure. We detected two velocity components toward the IF and MP2 positions. The component at 10 km$\mathrm{s}^{-1}$ is detected at both positions and seems associated with the layer most exposed to the UV radiation from IRS~1. The component at 12 km$\mathrm{s}^{-1}$ is only detected toward the IF position and is related to the bulk of the molecular cloud with its maximum in the SW. The component at 8.5 km$\mathrm{s}^{-1}$ is only detected toward the MP2 position and is related to the second low-UV PDR described by \citet{ginard2012} and \citet{pilleri2013}. 
\item There are no important differences between the values of $D_{\rm frac}$ toward the two PDRs. This is interpreted in the scenario of the deuterated compounds coming from dense and warm clumps with gas kinetic temperatures of 50~K, instead from the most exposed PDR layers. 
\item Values of $D_{\rm frac}$ of $\sim$0.01 are found for HNC, HCN, C$_2$H and H$_2$CO, and $<$0.001 for HCO$^+$, N$_2$H$^+$, and NH$_3$. These values are consistent with the predictions of the gas-phase model at an early time, $\sim$0.1 Myr. This time is consistent with the ages estimated for UC~{\sc Hii} regions on the basis of statistical studies.
\item The deuterium chemistry is a good chemical clock in both the low-mass and high-mass regime. However, the values of $D_{\rm frac}$(N$_2$H$^+$) and  $D_{\rm frac}$(HCO$^+$) cannot provide a good estimate of the evolutionary stage of massive protostar regions because this abundance ratio reaches the steady-state value on a short scale time after all the CO is released from the grains into the gas phase. We need to use the values of $D_{\rm frac}$ for different molecules (\eg\ HCN, HNC and C$_{2}$H) with longer chemical scales time to provide accurate age estimates.\\
\\
\end{itemize}

\begin{acknowledgements}

We acknowledge \'A. S\'anchez-Monge for useful comments and suggestions. We also thank J. A. Toal\'a for a critical reading of the manuscript. We thank the Spanish MINECO for funding support from grants CSD2009-00038, AYA2009-07304, and AYA2012-32032.
 
\end{acknowledgements}


 
\begin{appendix}
\section{tables and figures}

\begin{figure*}
\centering
\includegraphics[width=15cm]{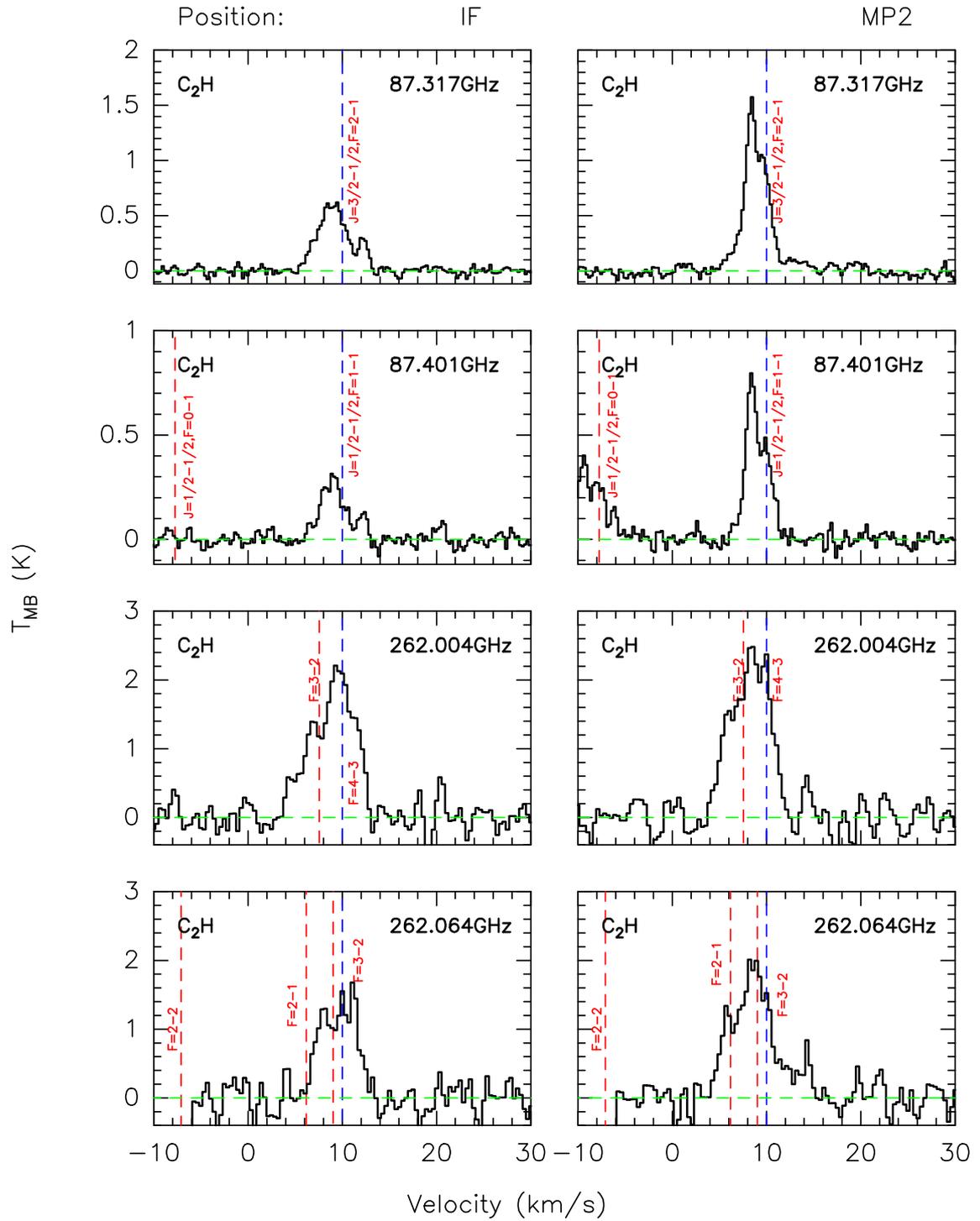}
\caption{C$_2$H spectra at the IF and MP2 positions. The red dashed lines mark different C$_2$H transitions. The blue dashed line shows the velocity of 10km~s$^{-1}$, relative to the frequencies 87.317, 87.401, 262.004, and 262.064~GHz, from top to bottom.}
\label{f1:CCH}
\end{figure*}

\begin{figure*}
\centering
\includegraphics[width=15cm]{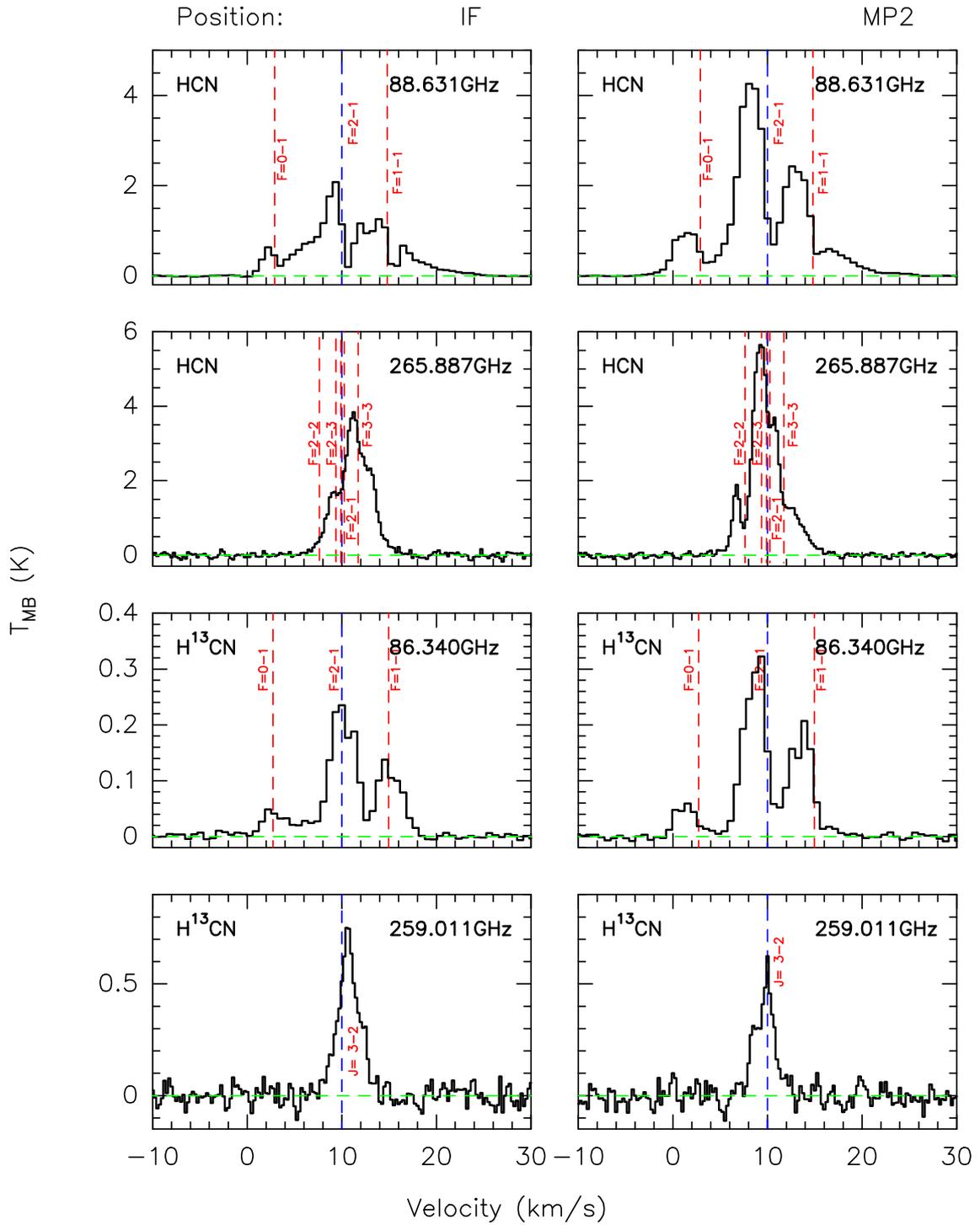}
\caption{Panels 1 and 2: HCN spectra at the IF and MP2 positions. The red dashed lines mark different HCN transitions.The blue dashed line shows the velocity of 10km~s$^{-1}$, relative to the rest frequency of the transitions, 88.631 and 265.88~GHz. Panels 3 and 4: H$^{13}$CN spectra at the IF and MP2 positions, the red dashed lines mark different H$^{13}$CN transitions. The blue dashed line shows the velocity of 10km~s$^{-1}$, relative to the frequencies 86.340 (panel 3) and 259.054~GHz (panel 4). }
\label{f2:HCN}
\end{figure*}

\begin{figure*}
\centering
\includegraphics[width=15cm]{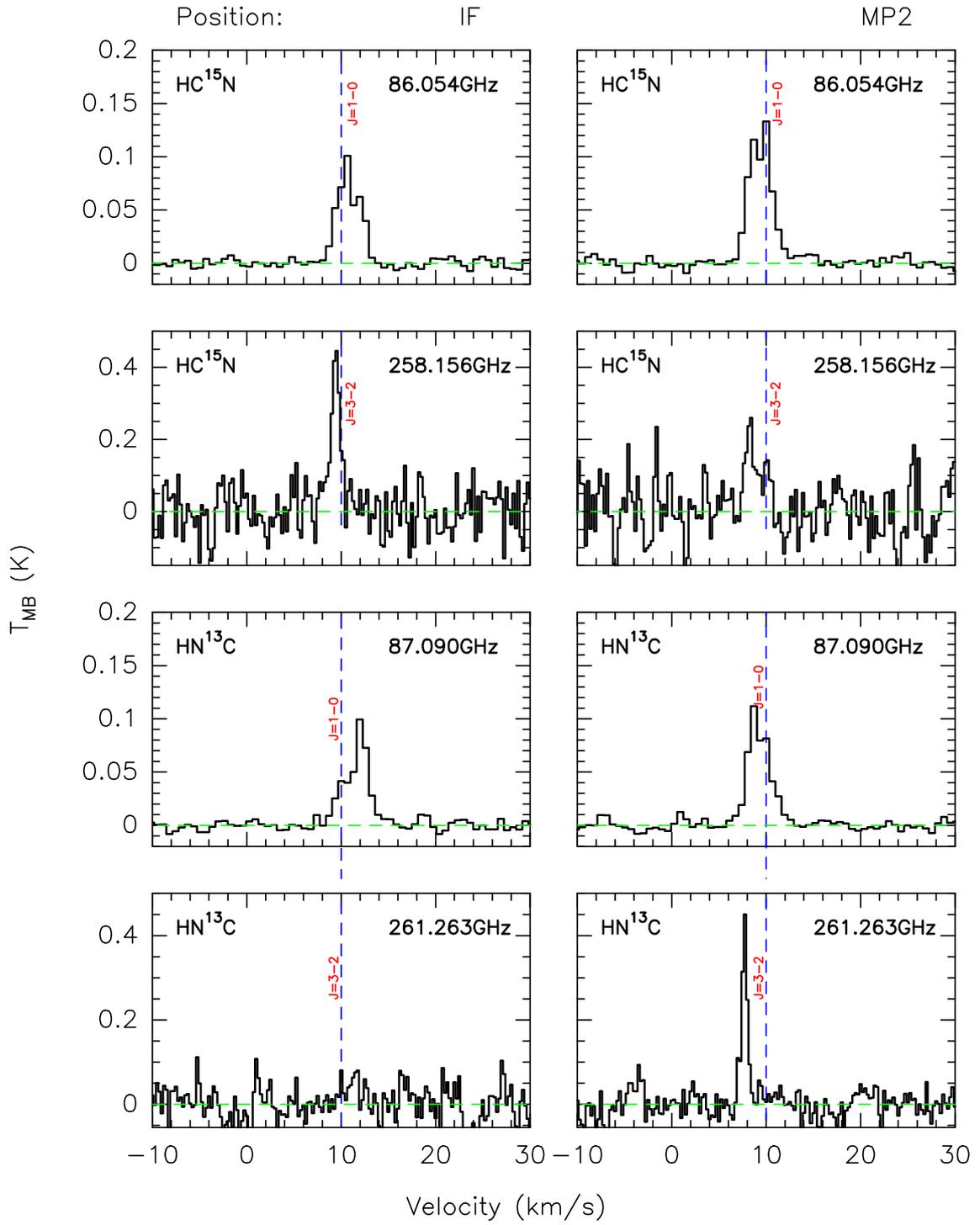}
\caption{Panels 1 and 2: HC$^{15}$N spectra at the IF and MP2 positions. The blue dashed line shows the velocity of 10km~s$^{-1}$ relative to 86.054, and 258.156~GHz, for the panel 1 and 2, respectively. Panels 3 and 4: HN$^{13}$C spectra at the IF and MP2 positions at 87.090, and 261.263~GHz.}
\label{f3:H13-15-CN}
\end{figure*}

\begin{figure*}
\centering
\includegraphics[width=15cm]{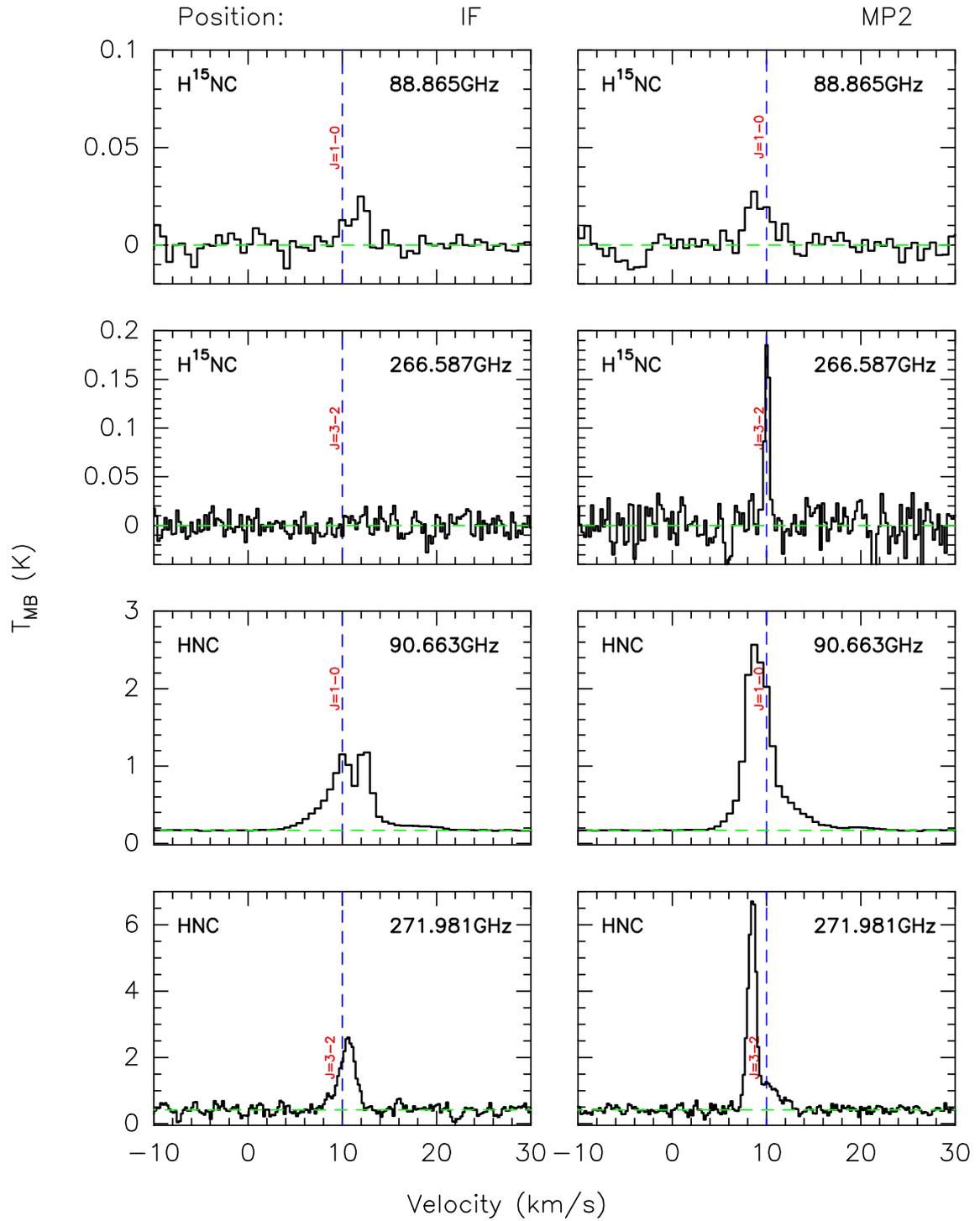}
\caption{Panels 1 and 2: H$^{15}$NC spectra at the IF and MP2 positions. Panels 3 and 4: HNC spectra \textbf{at} the IF and MP2 poitions. The blue dashed line shows the velocity of 10km~s$^{-1}$ relative to 88.865, 266.587, 90.663, and 271.981~GHz, respectively.}
\label{f4:HN15C}
\end{figure*}

\begin{figure*}
\centering
\includegraphics[width=15cm]{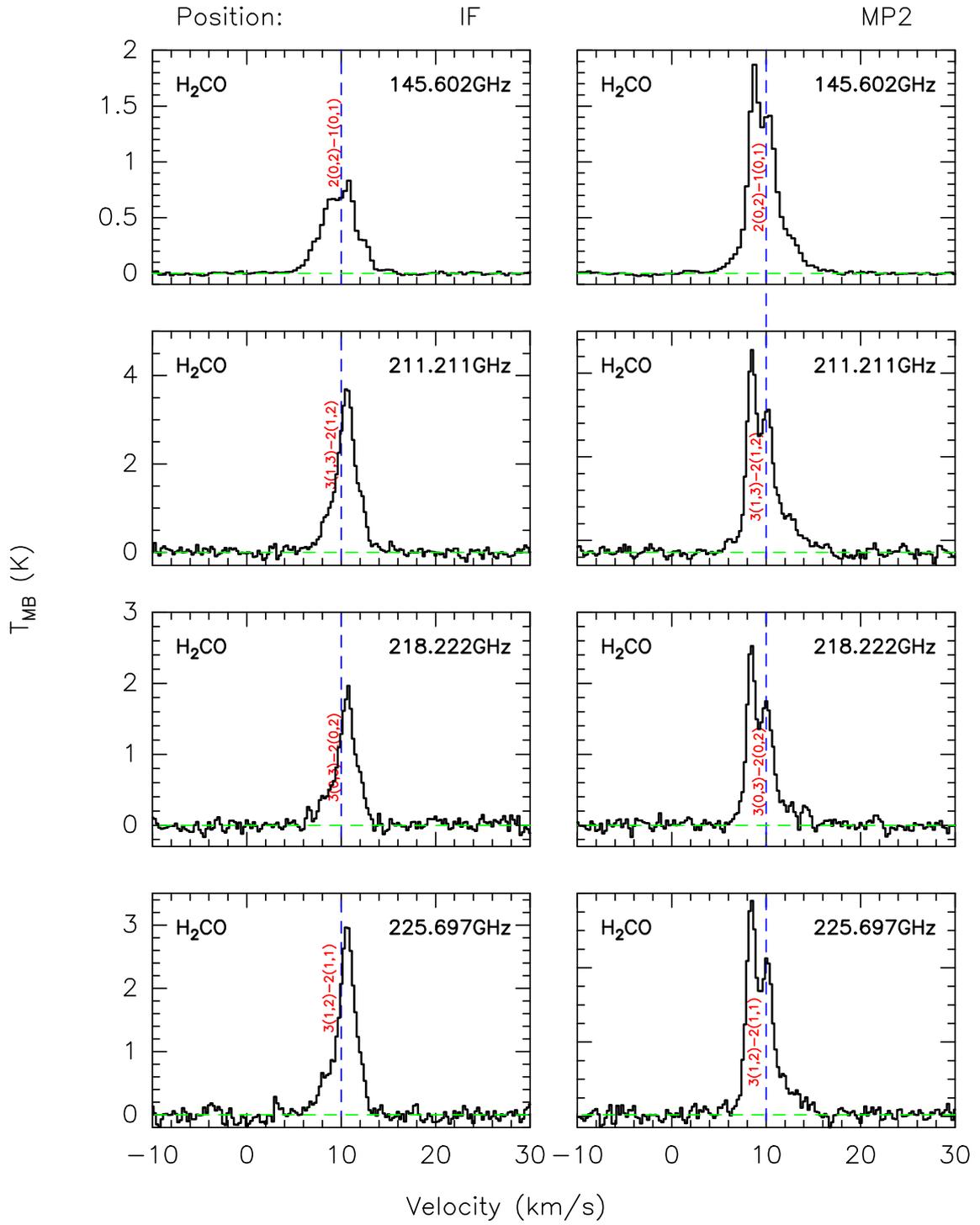}
\caption{H$_2$CO spectra at the two positions IF and MP2. The blue dashed line shows the velocity of 10km~s$^{-1}$, relative to the frequencies 145.602, 211.211, 218.222, and 218.697~GHz, from top to bottom.}
\label{f5:H2CO}
\end{figure*}

\begin{figure*}
\centering
\includegraphics[width=15cm]{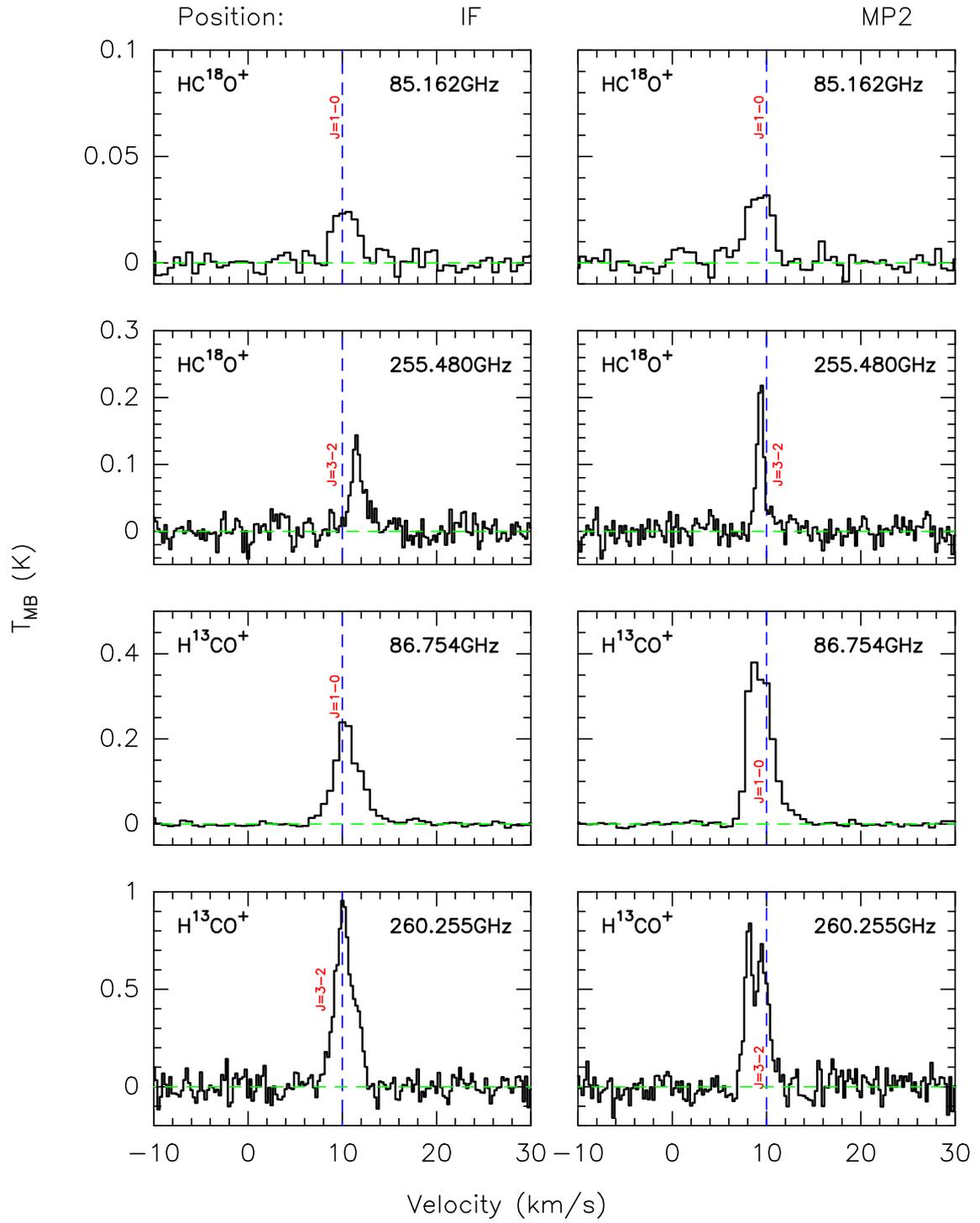}
\caption{Panels 1 and 2: HC$^{18}$O$^{+}$ spectra at the two positions IF and MP2. Panels 3 and 4: H$^{13}$CO$^{+}$ spectra at the two positions IF and MP2. The blue dashed line shows the velocity of 10km~s$^{-1}$, relative to the frequencies 85.162, 255.480, 86.745, and 260.255~GHz, from top to bottom.}
\label{f6:H2CO2}
\end{figure*}

\begin{figure*}
\centering
\includegraphics[width=15cm]{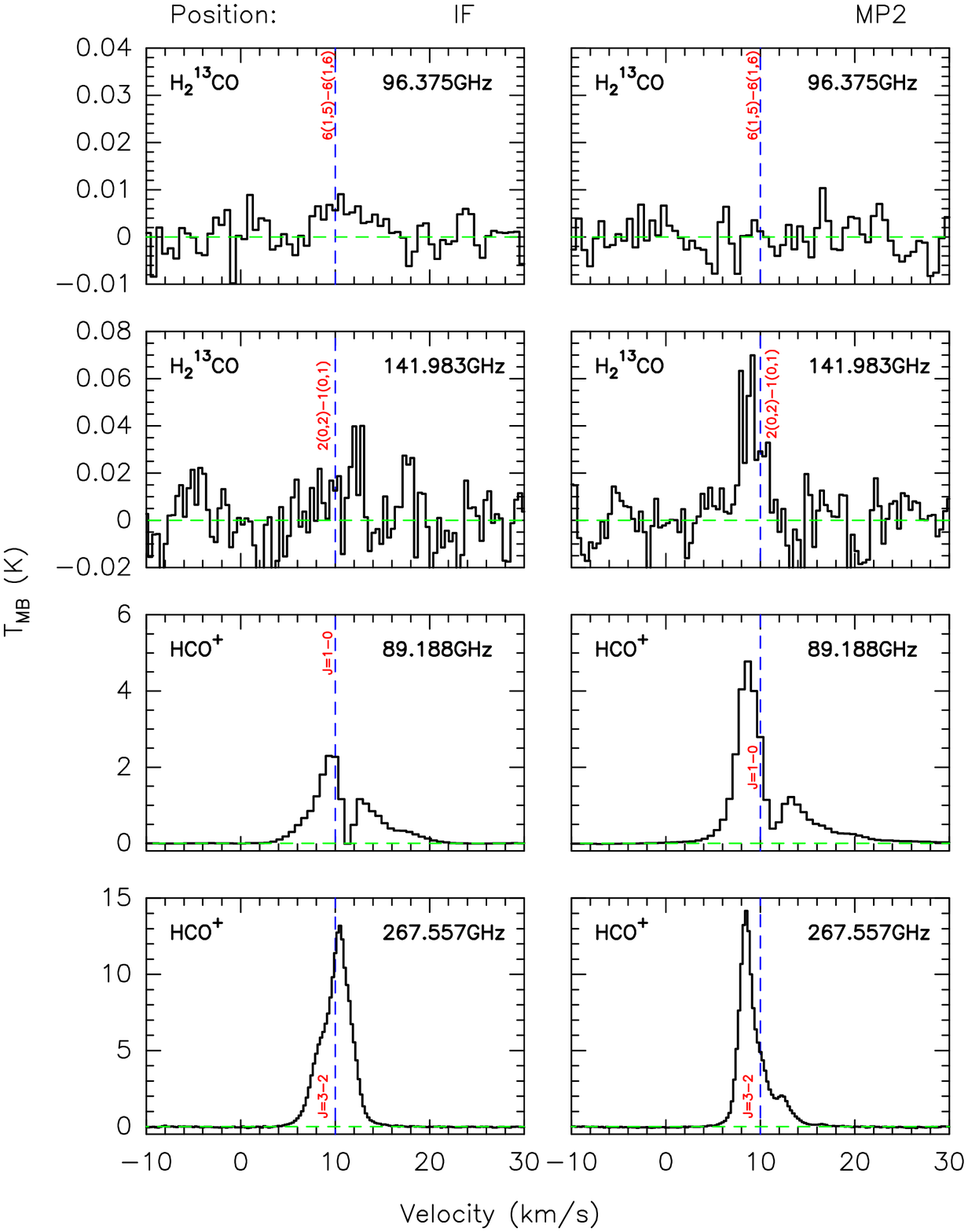}
\caption{In the panel 1 and 2 are the H$^{13}_2$CO spectra at the two positions IF and MP2. The blue dashed line shows the velocity of 10km~s$^{-1}$, relative to the frequency 96.375~GHz, in the panel 1, and 141.983~GHz, in panel 2. Panel 3 and 4 show the HCO$^{+}$ spectra toward the two positions IF and MP2. The blue dashed line shows the velocity of 10km~s$^{-1}$, relative to the frequency 89.188~GHz, in panel 3, and 267.557~GHz, in panel 4.}
\label{f7:H132CO}
\end{figure*}

\begin{figure*} 
\centering
\includegraphics[width=15cm]{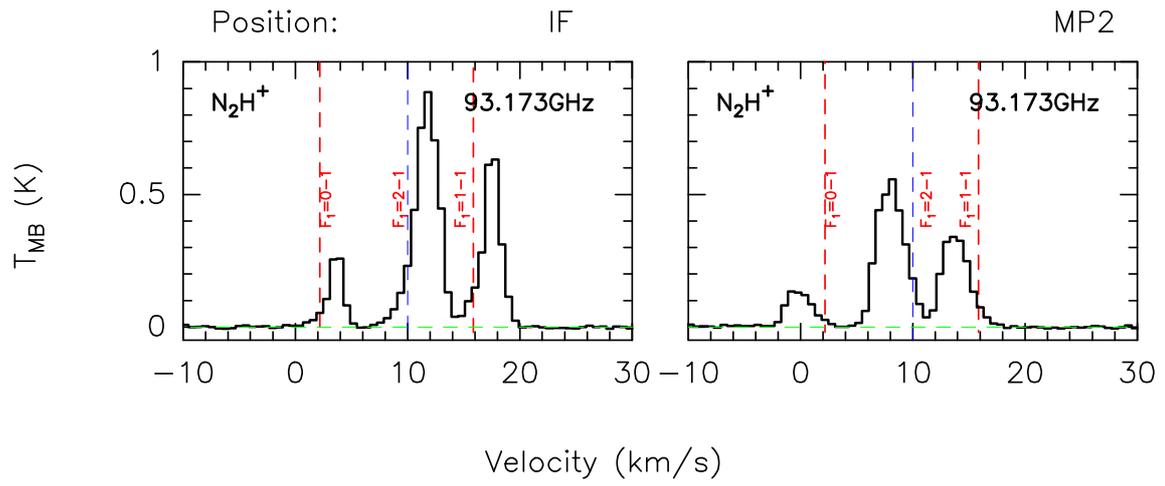}
\caption{N$_2$H$^{+}$ spectra at 93.173~GHz at the two positions IF and MP2. The red dashed lines mark different transitions of this molecule. The blue dashed line shows the velocity of 10km~s$^{-1}$ relative to the frequency of 93.137 GHz.}
\label{f8:H18COp}
\end{figure*}

\begin{figure*}
\centering
\includegraphics[width=15cm]{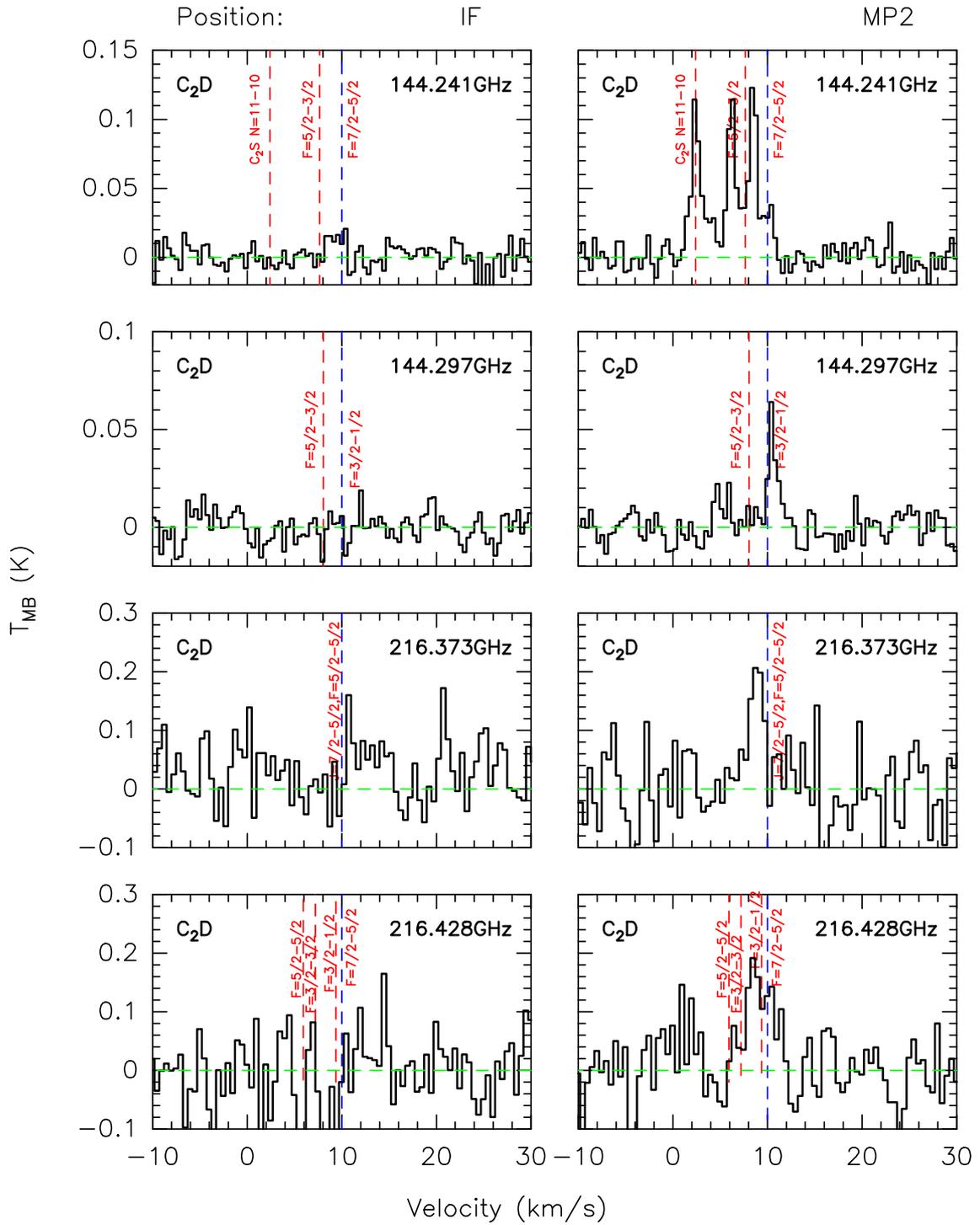}
\caption{C$_2$D spectra at the two positions IF and MP2. The red dashed lines mark different C$_2$D transitions. The blue dashed line shows the velocity of 10km~s$^{-1}$, relative to the frequencies 144.241, 144.297, 216.373, and 216.428~GHz, from top to bottom.}
\label{f12:CCD}
\end{figure*}

\begin{figure*}
\centering
\includegraphics[width=15cm]{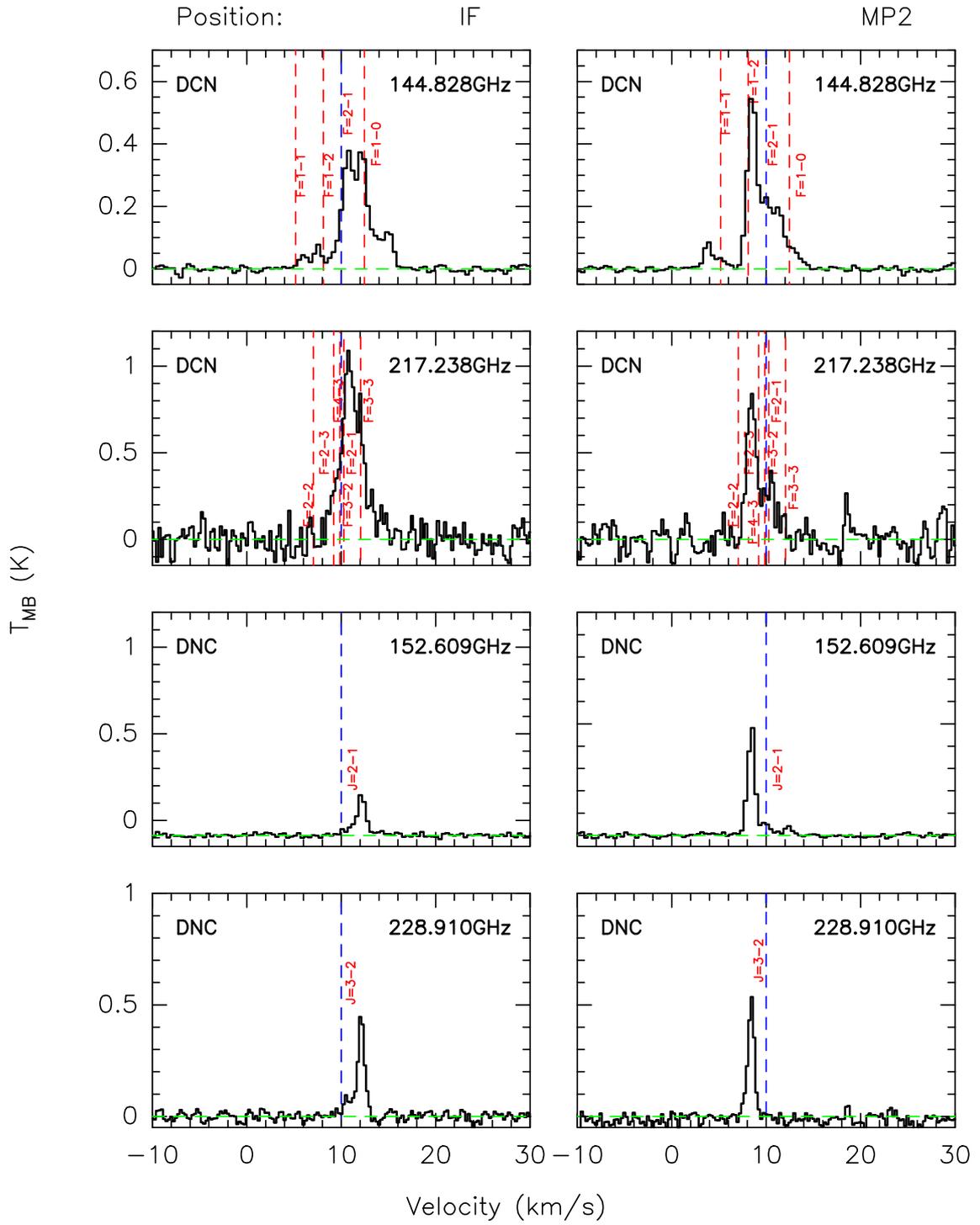}
\caption{DCN and DNC spectra at the two positions IF and MP2. The red dashed lines mark different transitions. The blue dashed line shows the velocity of 10km~s$^{-1}$, relative to the frequencies 144.828, 217.238 (for DCN), 152.609, and 228.91~GHz (for DNC), from top to bottom.}
\label{f13:DCN-DNC}
\end{figure*}

\begin{figure*}
\centering
\includegraphics[width=15cm]{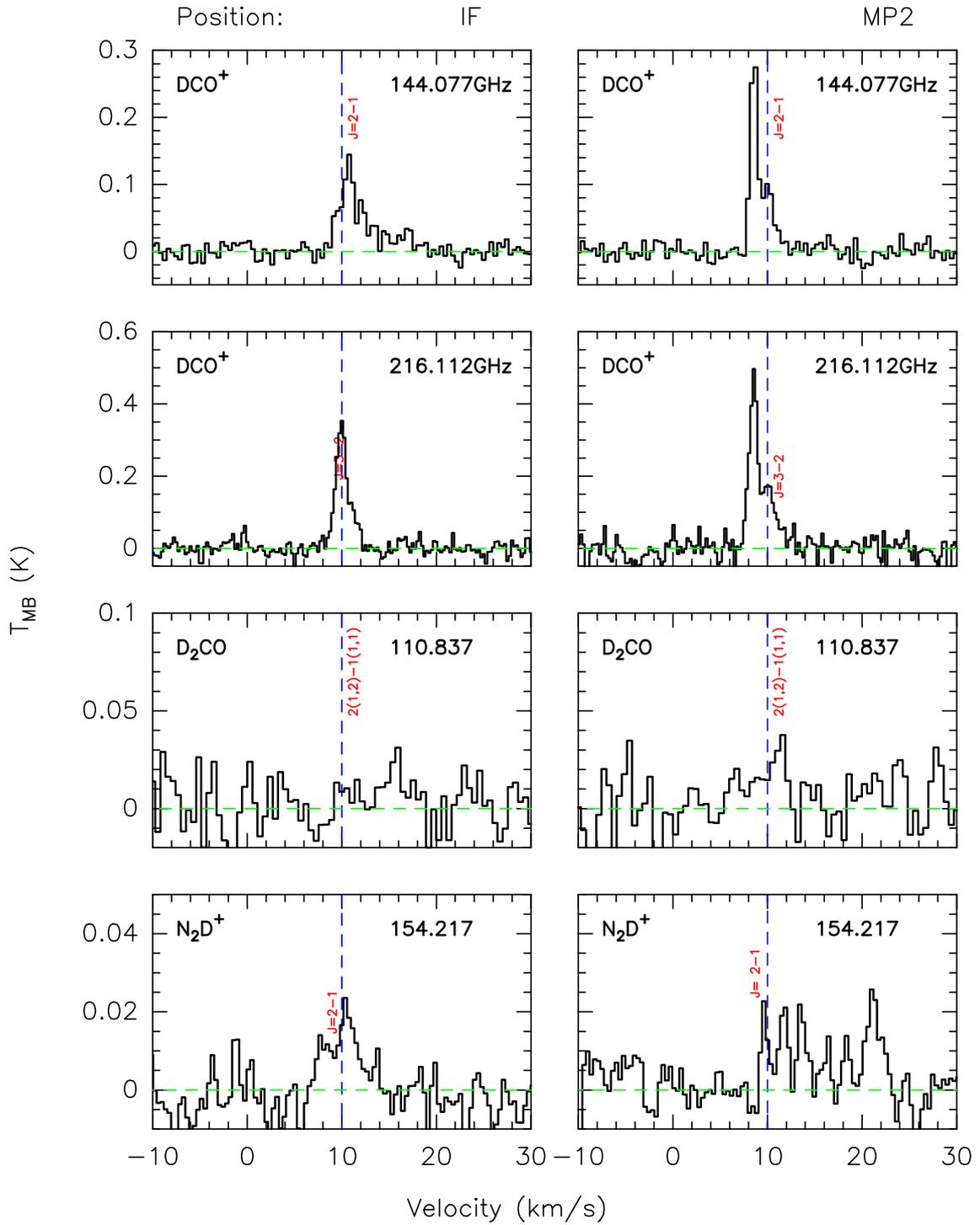}
\caption{Panels 1 and 2: DCO$^{+}$ spectra at the two positions IF and MP2. The blue dashed line shows the velocity of 10km~s$^{-1}$, relative to the frequencies 144.077 (in the top panel) and 216.112~GHz (in the bottom panel). Panel 3: D$_{2}$CO spectra at the two positions IF and MP2. The blue dashed line shows the velocity of 10km~s$^{-1}$, relative to the frequency 110.837~ GHz. Panel 4: N$_{2}$D$^{+}$ spectra at the two positions IF and MP2. The blue dashed line shows the velocity of 10km~s$^{-1}$, relative to the frequency 154.217~ GHz.}
\label{f14:DCOp}
\end{figure*}

\begin{figure*}
\centering
\includegraphics[width=15cm]{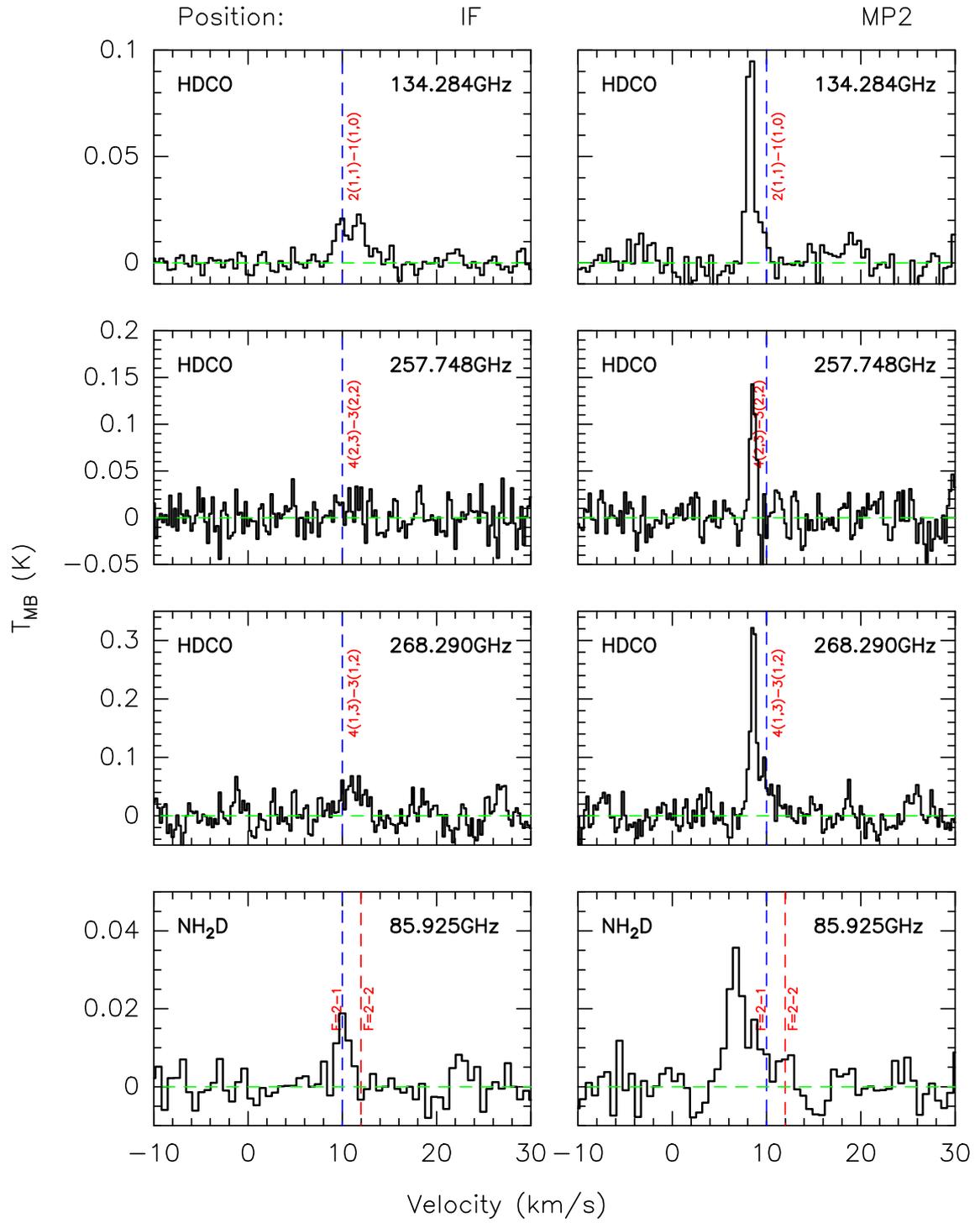}
\caption{Panels 1 to 3: HDCO spectra at the two positions IF and MP2. The blue dashed line shows the velocity of 10km~s$^{-1}$, relative to the frequencies 134.284, 257.748, and 268.290~GHz. Panel 4: NH$2$D spectra at IF and MP2 poitions, the blue dashed line shows the velocity of 10km~s$^{-1}$, relative to 85.925~GHz.}
\label{f15:HDCO}
\end{figure*}

In this section we show all the spectra of the hydrogenated (Figs.~\ref{f1:CCH}--\ref{f8:H18COp}) and deuterated (Figs.~\ref{f12:CCD}--\ref{f15:HDCO}) species toward the IF and MP2 positions. The tables of the parameters (Table A.1), the tables of the Gaussian fits (Tables A.2 --A.4), and the molecular abundances of HCN, HCO$^{+}$, and C$_{2}$ relative to H$_{2}$ predicted by models A to D (Figure A.13). 
 
\begin{table*}[]
\begin{center}
\scriptsize
\caption{Observational parameters}
\begin{tabular}{ l c c c c c c c c c c}      
Species
&Freq    
&HPBW 
&B$\mathrm{_{eff}}$
&\multicolumn{2}{c}{Position}
&OM$\mathrm{^{a}}$
&Res$\mathrm{^{b}}$
&rms
&rms$\mathrm{^{c}}$
\\
&[MHz]
&[$"$]
&
&
&
&
&[kms$^{-1}$]
&[mK]
&[mK]
\\
\hline\hline
\noalign{\smallskip}
C$_2$D 	  &144241.93  &16  	 & 0.74         & IF & MP2   		          &SP  &0.40 & 6  & 4  	\\
 		  &144243.05  &16  	 & 0.74         & IF & MP2    		          &SP  &0.40 & 6  & 4 	\\
 	 	  &144296.72  &16  	 & 0.74         & IF & MP2     		          &SP  &0.40 & 8  & 6  	\\
 	 	  &144297.66  &16  	 & 0.74         & IF & MP2     		          &SP  &0.40 & 8  & 6  	\\ 
 		  &216373.32  &11	 & 0.63         & \multicolumn{2}{c}{MAP}     &OTF &0.50 &56  &47	\\
  		  &216428.76  &11	 & 0.63         & \multicolumn{2}{c}{MAP}     &OTF &0.50 &56  &44	\\ 
  		  &216430.34  &11	 & 0.63         & \multicolumn{2}{c}{MAP}     &OTF &0.50 &56  &45	\\
  		  &216431.26  &11	 & 0.63         & \multicolumn{2}{c}{MAP}     &OTF &0.50 &56  &45	\\
DCN 	  &144826.82  &16 	 & 0.74         & IF & MP2     		          &SP  &0.40 & 8  & 7   	\\
  	 	  &144828.00  &16	 & 0.74         & IF & MP2     		          &SP  &0.40 & 8  & 7  	\\
  	  	  &144830.33  &16	 & 0.74         & IF & MP2    		          &SP  &0.40 & 8  & 7  	\\
  	  	  &217238.30  &11 	 & 0.63         & \multicolumn{2}{c}{MAP}     &OTF &0.25 &64  &39	\\
DNC 	  &152609.74  &16    & 0.74         & IF & MP2                    &SP  &0.38 & 7  & 5  	\\
  	 	  &228910.48  &11 	 & 0.60         & \multicolumn{2}{c}{MAP}     &OTF &0.25 &18  &14	\\ 
DCO$^{+}$ &144077.28  &16    & 0.74         & IF & MP2    		          &SP  &0.40 &10  & 7  	\\
	 	  &216112.58  &11  	 & 0.63         & \multicolumn{2}{c}{MAP}     &OTF &0.25 &65  &43	\\
D$_{2}$CO &110837.80  &29    & 0.78         & IF & MP2    		          &SP  &0.54 &14  &12  	\\
HDCO 	  &134284.83  &16 	 & 0.78         & IF & MP2     		          &SP  &0.43 & 4  & 3	\\  
  	 	  &256585.53  &9 	 & 0.53         & \multicolumn{2}{c}{MAP}     &OTF &0.25 &16  &10	\\
  	  	  &257748.70  &9 	 & 0.53         & \multicolumn{2}{c}{MAP}     &OTF &0.25 &17  & 8	\\
  	  	  &268290.56  &9 	 & 0.53         & IF & MP2    		          &SP  &0.25 &23  &18      \\ 
NH$_2$D   & 85926.27  &29 	 & 0.81         & IF & MP2     		          &SP  &0.65 & 4  & 4	\\
N$_2$D$^{+}$ &154217.00  &16 & 0.74         & IF & MP2     		          &SP  &0.38 & 6  & 5	\\
 			 &231321.82  &9  & 0.58         & IF & MP2     		          &SP  &0.25 & 30 &20	\\

\hline
C$_2$H 	    	  &87316.92    	&29 	  & 0.81  & IF & MP2  		        &SP	 &0.25  &37  &28\\
 	   	  &87328.62 	&29  	  & 0.81  & IF & MP2 		        &SP	 &0.25  &26  &16\\
           	  &87401.98 	&29 	  & 0.81  & IF & MP2 		        &SP	 &0.25  &33  &26\\
     	   	  &87407.35 	&29 	  & 0.81  & IF & MP2   		        &SP	 &0.25  &25  &15\\
           	  &262004.26  	&9	  & 0.53  &\multicolumn{2}{c}{MAP}      &OTF     &0.25  &99  &65\\
           	  &262006.40 	&9	  & 0.53  &\multicolumn{2}{c}{MAP}      &OTF     &0.25  &99  &65\\
   	    	  &262064.84 	&9	  & 0.53  &\multicolumn{2}{c}{MAP}      &OTF     &0.25  &123 &95\\
  	    	  &262067.33   	&9	  & 0.53  &\multicolumn{2}{c}{MAP}      &OTF     &0.25  &123 &95\\
 	   	  &262208.43    &9	  & 0.53  &\multicolumn{2}{c}{MAP}      &OTF     &0.25  &130 &58\\
H$^{13}$CN  	  &86338.76	&29	  & 0.81  & IF & MP2   		        &SP	 &0.65  & 5  & 5\\
  	   	  &86340.18	&29	  & 0.81  & IF & MP2 		        &SP	 &0.65  & 5  & 5\\
  	   	  &86342.27	&29	  & 0.81  & IF & MP2  		        &SP	 &0.65  & 5  & 5\\
  	   	  &259011.82 	&9	  & 0.53  &\multicolumn{2}{c}{MAP}      &OTF     &0.25  &42  &25\\
HC$^{15}$N	  &86054.96	&29   	  & 0.81  & IF & MP2  		        &SP	 &0.65  & 6  & 6\\
 	    	  &258156.99	&9	  & 0.53  &\multicolumn{2}{c}{MAP}      &OTF     &0.25  &67  &39\\ 
HCN	 	  &88631.60	&29  	  & 0.81  & IF & MP2   		        &SP	 &0.65  & 9  & 9\\ 
	  	  &265887.04	&9	  & 0.53  & IF & MP2  		        &SP	 &0.25  &62  &35\\
HNC   	    	  &90663.56  	&29 	  & 0.81  & IF & MP2  		        &SP	 &0.65  & 5  & 5\\ 
  	    	  &271981.14   	&9 	  & 0.53  & IF & MP2  		        &SP	 &0.21  &148 &105\\
HN$^{13}$C  	  &87090.82 	&29 	  & 0.81  & IF & MP2  		        &SP	 &0.65  & 5  & 5\\
  	    	  &261263.48 	&9 	  & 0.53  &\multicolumn{2}{c}{MAP}      &OTF     &0.25  &40  &26\\
H$^{15}$NC 	  &88865.71 	&29 	  & 0.81  & IF & MP2  		        &SP	 &0.65  & 5  & 5\\ 
 	   	  &266587.80  	& 9 	  & 0.53  & IF & MP2  		        &SP	 &0.25  &10  &6 \\ 
HCO$^{+}$  	  &89188.52   	&29 	  & 0.81  & IF & MP2  		        &SP	 &0.65  &6   &6 \\ 
 	    	  &267557.52  	& 9 	  & 0.53  & IF & MP2  		        &SP	 &0.25  &34  &29\\ 
H$^{13}$CO$^{+}$  & 86754.28 	& 9 	  & 0.81  & IF & MP2  		        &SP	 &0.65  & 5  & 5\\   
 		  &260255.33  	& 9	  & 0.53  &\multicolumn{2}{c}{MAP}      &OTF     &0.25  &55  &33\\   
 HC$^{18}$O$^{+}$  & 85162.23	&29 	  & 0.81  & IF & MP2   		        &SP	 &0.65  & 4  & 4\\   
                  &255480.21  	& 9 	  & 0.63  &\multicolumn{2}{c}{MAP}      &OTF     &0.25  &14  & 8\\  
H$_2$CO  	  &145602.94   	&16 	  & 0.74  & IF & MP2  		        &SP	 &0.40  &10  & 7\\  
 	          &211211.46 	&11 	  & 0.63  &\multicolumn{2}{c}{MAP}      &OTF     &0.25  &77  &46\\ 
	    	  &218222.19 	&11 	  & 0.63  &\multicolumn{2}{c}{MAP}      &OTF     &0.25  &73  &46\\   
	    	  &218475.63	&11 	  & 0.63  &\multicolumn{2}{c}{MAP}      &OTF     &0.25  &68  &45\\  
 	    	  &218760.06	&11	  & 0.63  &\multicolumn{2}{c}{MAP}      &OTF     &0.25  &76  &47\\  
  		  &225697.77 	&11	  & 0.63  &\multicolumn{2}{c}{MAP}      &OTF     &0.25  &95  &63\\  
H$_2$$^{13}$CO	  & 96375.75 	&29 	  & 0.81  & IF & MP2  		        &SP	 &0.60  & 4  & 4\\  
  	    	  &141983.74	&16 	  & 0.63  & IF & MP2  		        &SP	 &0.40  &14  & 10\\  
		  &212811.18 	&11	  & 0.63  &\multicolumn{2}{c}{MAP}      &OTF     &0.25  &107 &65\\ 
		  &213293.56  	&11	  & 0.63  &\multicolumn{2}{c}{MAP}      &OTF     &0.25  &121 &74\\ 
		  &219908.52	&11	  & 0.63  &\multicolumn{2}{c}{MAP}      &OTF     &0.25  &108 &66\\
N$_{2}$H$^{+}$ 	  &93171.88     &29 	  & 0.81  & IF & MP2  		        &SP	 &0.60  & 4   & 4\\
   	    	  &93173.70  	&29	  & 0.81  & IF & MP2  		        &SP	 &0.60  & 4   & 4\\ 
 	    	  &93176.13 	&29	  & 0.81  & IF & MP2  		        &SP	 &0.60  & 4  & 4\\ 
\hline
\hline
\end{tabular}
\end{center}
\noindent
\\
\scriptsize
 $^{a}$Observing Modes: on the fly (OTF), single pointing (SP).\\ 
 $^{b}$Original resolution.\\ 
 $^{c}$ rms of the spectra for the smoothed resolution of 0.65 km\,s$^{-1}$. 
\begin{list}{}{}
\item 
\item 
\end{list}
\label{tableA1}
\end{table*}



\begin{table*}[]
\begin{center}
\scriptsize
\caption{Gaussian fit parameters of the deuterated molecules \textbf{at} the IF and MP2 positions.}
\begin{tabular}{ l c c c c c c c}
\hline\hline\noalign{\smallskip}
IF 
\\       
Species
&Trans 
&Freq   
&HPBW 
&Area 
&V
&Width
&$T_{\mathrm{MB}}$
\\

& 
&(MHz)
&($"$)
&(K~km~s$^{-1}$)
&(km~s$^{-1}$)
&(km~s$^{-1}$)
&(K)
\\
\hline\hline
\noalign{\smallskip}
C$_2$D 	 	 &N$=2$--1, J$=5/2$--$3/2$, F$=5/2$--$3/2$ &144243.05  &16       &$0.045\pm0.008^{a}$    &$9.518\pm0.050$   &7.3--11   		  &\\
	  		 &N$=2$--1, J$=3/2$--$1/2$, F$=5/2$--$3/2$ &144296.72  &16       &$<0.024^{b}$		 	 &rms$^{c}$=8       &W$^{d}$=1  	  &N$_{c}$$^{h}$=2.5\\  
 	 		 &N$=3$--2, J$=7/2$--$5/2$, F$=7/2$--$5/2$ &216373.32  &16$^{*}$ &$<0.147^{b}$ 			 &rms$^{c}$=56      &W$^{d}$=1  	  &N$_{c}$$^{g}$=4  \\   
	  		 &N$=3$--2, J$=5/2$--$3/2$, F$=3/2$--$3/2$ &216430.34  &16$^{*}$ &$<0.225^{b}$  		 &rms$^{c}$=56      &W$^{d}$=1        &N$_{c}$$^{g}$=4  \\  
DCN 	  	 &J$=2$--1, F$=1$--0	        	 	   &144826.82  &16		 &$0.072\pm0.009$  		 &$11.256\pm0.053$  &$0.784\pm0.092$  &$0.087$  \\ 
	  		 &J$=2$--1, F$=1$--0			 		   &	       &		 &$0.179\pm0.012$ 		 &$12.470\pm0.043$  &$1.342\pm0.110$  &$0.125$  \\
	 		 &J$=2$--1, F$=2$--1    				   &144828.00  &16		 &$0.692\pm0.021$  		 &$10.818\pm0.024$  &$1.758\pm0.064$  &$0.367$\\
	 		 &J$=2$--1, F$=2$--1			 		   &	       &		 &$0.435\pm0.010$  		 &$12.434\pm0.015$  &$1.344\pm0.038$  &$0.304$\\ 
 	 		 &J$=2$--1, F$=1$--1 			  		   &144830.33  &16		 &$0.054\pm0.021$  		 &$10.806\pm0.200$  &$1.133\pm0.530$  &$0.045$  \\
	 		 &J$=2$--1, F$=1$--1			  		   &	       &		 &$0.078\pm0.022$ 		 &$12.389\pm0.108$  &$0.986\pm0.361$  &$0.074$  \\
  	 		 &J$=3$--2, F$=2$--1					   &217238.30  &16$^{*}$ &$1.730\pm0.134$ 		 &$10.876\pm0.048$  &$1.677\pm0.170$  &$0.969$  \\ 
	  		 &J$=3$--2, F$=2$--1			 		   &	       &		 &$0.382\pm0.062$ 	  	 &$12.215\pm0.069$  &$0.874\pm0.197$  &$0.411$  \\
DNC 	 	 &J$=2$--1 								   &152609.74  &16   	 &$0.029\pm0.012$  		 &$10.596\pm0.148$  &$1.145\pm0.241$  &0.024\\	
	  		 &J$=2$--1  							   &  	       &	     &$0.206\pm0.010$ 	 	 &$12.136\pm0.024$  &$1.122\pm0.056$  &0.173\\	  
  	 		 &J$=3$--2								   &228910.48  &$16^{*}$ &$0.086\pm0.013$ 	 	 &$10.634\pm0.073$  &$0.963\pm0.169$  &0.085 \\ 
	 		 &J$=3$--2 								   &  	       &		 &$0.413\pm0.016$ 		 &$12.106\pm0.016$  &$0.889\pm0.044$  &0.437 \\ 
DCO$^{+}$ 	 &J$=2$--1								   &144077.28  &16  	 &$0.149\pm0.015$  	 	 &$10.623\pm0.066$  &$1.528\pm0.225$  &0.092 \\  	
			 &J$=2$--1 								   & 	       &		 &$0.269\pm0.029$  		 &$12.343\pm0.76 $  &$5.829\pm0.859$  &0.044 \\   	
		     &J$=3$--2								   &216112.58  &$16^{*}$ &$0.392\pm0.052$      	 &$10.745\pm0.074$  &$1.164\pm0.173$  &0.316 \\
D$_{2}$CO    &2(1,2)--1(1,1)  						   &110837.83  &29       &$<0.530^{b}$  		 &rms$^{c}$=143     &W$^{d}$=1.5  	  &N$_{c}$$^{e}$=4.6 \\
	         &2(1,2)--1(1,1) 						   &	       &		 &$<0.335^{b}$ 			 &rms$^{c}$=143     &W$^{d}$=1.5      &N$_{c}$$^{e}$=1.9 \\	
HDCO 	     &$2(1,1)$--$1(1,0)$					   &134284.83  &16 		 &$0.029\pm0.003$  	 	 &$9.859\pm0.088$   &$1.115\pm0.206$  &0.0207 \\
	 		 &$2(1,1)$--$1(1,0)$				   	   &	       &		 &$0.035\pm0.003$  		 &$11.852\pm0.081$  &$1.493\pm0.195$  &0.0223 \\
			 &$4(0,4)$--$3(0,3)$					   &256585.53  &9 	   	 &$<0.148^{b}$			 &rms$^{c}$=16      &W$^{d}$=1  	  &N$_{c}$$^{g}$=4  \\       
			 &$4(0,4)$--$3(0,3)$				 	   &	       &		 &$<0.375^{b}$	 		 &rms$^{c}$=16      &W$^{d}$=1 		  &N$_{c}$$^{g}$=4   \\     
  			 &$4(1,3)$--$3(1,2)$  					   &268292.56  &9 	   	 &$0.146\pm0.026$		 &$11.076\pm0.245$  &$2.681\pm.527$	  &$0.0513$   \\ 
NH$_2$D      &1(1,1)0s-1(0,1)0a, F=2-1				   &85926.27   &29 		 &$0.030\pm0.005$ 	 	 &$10.080\pm0.194$  &$1.491\pm0.287$  &0.0191\\		
N$_2$D$^{+}$ &J$=2$--1 								   &154217.00  &16 		 &$0.017\pm0.009$   	 & $7.881\pm0.296$  &$1.306\pm0.517$  &0.0121 \\      
	    	 &J$=2$--1  						 	   &	       &		 &$0.054\pm0.013$   	 &$10.568\pm0.220$  &$2.673\pm0.900$  &0.0191 \\  	
	     	 &J$=3$--2 								   &231321.82  & 9  	 &$<0.059^{b}$  		 &rms$^{c}$=29  	&W$^{d}$=1  	  &N$_{c}$$^{f}$=6 \\
\hline\hline\noalign{\smallskip}
MP2  
\\    
Species 
&Trans 
&Freq   
&HPBW 
&Area 
&V
&Width
&$T_{\mathrm{MB}}$
\\
& 
&(MHz)
&($"$)
&(K~km~s$^{-1}$)
&(km~s$^{-1}$)
&(km~s$^{-1}$)
&(K)
 \\ 
\hline\hline
\noalign{\smallskip}
C$_2$D 	    &N$=2$--1, J$=5/2$--$3/2$, F$=5/2$--$3/2$ &144243.05  &16 		  &$0.132 \pm0.023$	    &$8.557\pm0.044$	 &$1.116\pm0.118$  &0.112 \\
			&N$=2$--1, J$=5/2$--$3/2$, F$=5/2$--$3/2$ &144241.93  &16 		  &$0.147 \pm0.012$     &$8.488\pm0.042$     &$1.198\pm0.140$  &0.118 \\
 			&N$=2$--1, J$=5/2$--$3/2$, F$=5/2$--$3/2$ &           &   		  &$0.040 \pm0.009$     &$10.201\pm0.118$    &$1.051\pm0.240$  &0.036\\
			&N$=2$--1, J$=3/2$--$1/2$, F$=5/2$--$3/2$ &144297.66  &16  		  &$0.048 \pm0.005$     &$8.517\pm0.038$     &$1.027\pm0.181$  &0.057\\ 
			&N$=3$--2, J$=7/2$--$5/2$, F$=7/2$--$5/2$ &216373.32  &$16^{*}$   &$0.392 \pm0.073$ 	&$8.838\pm0.147$     &$1.699\pm0.398$  &0.217\\
			&N$=3$--2, J$=5/2$--$3/2$, F$=3/2$--$1/2$ &216428.76  &$16^{*}$	  &$0.575 \pm0.090$  	&$9.108\pm0.260$     &$3.287\pm0.487$  &0.164\\ 
DCN 	    &J$=2$--1, F$=1$--0 					  &144826.82  &16 	  	  &$0.261 \pm0.019$  	&$8.591\pm0.404$ 	 &$1.471\pm0.063 $ &0.183 \\
  			&J$=2$--1, F$=1$--0 					  &           &   	   	  &$0.187 \pm0.026$ 	&$10.069\pm0.189$	 &$2.713\pm0.202 $ &0.065 \\
  	 		&J$=2$--1, F$=2$--1 					  &144828.00  &16	   	  &$0.743 \pm0.008$  	&$8.537\pm0.005$ 	 &$1.238\pm0.017$  &0.564  \\
  	  		&J$=2$--1, F$=2$--1 					  &           &  	   	  &$0.176 \pm0.006$  	&$9.965\pm0.014$ 	 &$0.905\pm0.028$  &0.182  \\
  	 		&J$=2$--1, F$=1$--1 					  &144830.33  &16	  	  &$0.092 \pm0.010$  	&$8.664 \pm0.049$	 &$0.985 \pm0.084$ &0.087 \\
  	  		&J$=2$--1, F$=1$--1						  &           &  	  	  &$0.044 \pm0.012$  	&$1.100 \pm0.140$ 	 &$1.267 \pm0.236$ &0.033 \\
  	  		&J$=3$--2, F$=2$--1						  &217238.30  &$16^{*}$   &$0.974 \pm0.064$  	&$8.466\pm0.037$  	 &$1.151\pm0.089$  &0.794\\
  	  		&J$=3$--2, F$=2$--1 					  &           &  	      &$0.401 \pm0.069$  	&$10.390\pm0.117$ 	 &$1.418\pm0.293$  &0.266\\
DNC 	    &J$=2$--1  	      						  &152609.74  &16         &$0.484 \pm0.009$  	&$8.442\pm0.008$  	 &$0.936\pm0.033$  &0.486\\
	 		&J$=2$--1 	    						  & 		  &	  		  &$0.099 \pm0.012$  	&$10.016\pm0.127$ 	 &$1.716\pm0.367$  &0.055\\ 
	  		&J$=2$--1 	    						  & 		  &	   		  &$0.043 \pm0.007$  	&$12.352\pm0.071$ 	 &$0.990\pm0.200$  &0.041\\ 
  	 		&J$=3$--2 	    						  &228910.48  &$16^{*}$   &$0.451 \pm0.017$  	&$8.436\pm0.015$  	 &$0.806\pm0.037$  &0.526\\ 
DCO$^{+}$   &J$=2$--1	    						  &144077.28  &16         &$0.291 \pm0.016$  	&$8.575\pm0.023$     &$0.899\pm0.061$  &0.304 \\
			&J$=2$--1	    						  &	  	 	  &           &$0.140 \pm0.012$ 	&$10.102\pm0.056$    &$1.305\pm0.140$  &0.100 \\
 			&J$=3$--2 	    						  &216112.58  &$16^{*}$   &$0.467 \pm0.051$  	&$8.360\pm0.053$  	 &$0.976\pm0.120$  &0.453\\
	 		&J$=3$--2 	     						  &	  		  &	 		  &$0.135 \pm0.046$  	&$10.047\pm0.125$ 	 &$0.756\pm0.280$  &0.168\\
D$_{2}$CO   &2(1,2)--1(1,1)  						  &110837.83  &29         &$0.056 \pm0.042$ 	&$9.290\pm0.950$   	 &$3.193\pm2.280$  &0.016\\   
			&		   								  &	          &	  		  &$0.044 \pm0.035$ 	&$11.430\pm0.334$  	 &$1.250\pm0.684$  &0.033\\  	  
HDCO 	 	&$2(1,1)$--$1(1,0)$ 					  &134284.83  &16 	   	  &$0.103 \pm0.009$ 	&$8.300\pm0.020$  	 &$0.918\pm0.052$  &0.106\\  
  			&$2(1,1)$--$1(1,0)$ 					  & 		  & 	   	  &$0.015 \pm0.005$  	&$9.615\pm0.097$ 	 &$0.759\pm0.177$  &0.192\\  
  	 		&$4(0,4)$--$3(0,3)$ 					  &256585.53  &9 	      &$0.402 \pm0.011$  	&$8.362\pm0.077$  	 &$0.741\pm0.026$  &0.510\\
  	 		&$4(2,3)$--$3(2,2)$ 					  &257748.70  &9 	      &$0.105 \pm0.010$  	&$8.558\pm0.034$ 	 &$0.676\pm0.071$  &0.146\\
  	  		&$4(1,3)$--$3(1,2)$ 					  &268290.56  &9 	      &$0.450 \pm0.050$  	&$7.855\pm0.049$ 	 &$0.874\pm0.053$  &0.484\\ 
NH$_2$D     &1(1,1)0s-1(0,1)0a, F=2-1 				  &85926.27   &29 	      &$0.067 \pm0.005$  	&$8.704\pm0.083$ 	 &$1.812\pm0.179$  &0.035\\
 	 		&1(1,1)0s-1(0,1)0a, F=2-1 				  & 		  & 	      &$0.027 \pm0.005$  	&$11.108\pm0.145$	 &$1.200\pm0.562$  &0.022\\
N$_2$D$^{+}$&J$=2$--1 								  &154217.00  &16         &$<0.011^{b}$         &rms$^{c}$=50     	 &W$^{d}$=1 	   &N$_{c}$$^{f}$=2.6 \\ 
	    	&J$=2$--1 							 	  &	   		  &	          &$<0.011^{b}$         &rms$^{c}$=50     	 &W$^{d}$=1  	   &N$_{c}$$^{f}$=2.6 \\ 
	  	    &J$=3$--2 								  &231321.82  & 9 		  &$<0.977^{b}$         &rms$^{c}$=50     	 &W$^{d}$=1 	   &N$_{c}$$^{g}$=6 \\
\hline
\end{tabular}
\end{center}
\noindent
\scriptsize
 $^{a}$ Area calculated between two velocities.\\
 $^{b}$ Upper limit of the area, assuming a Gaussian profile.\\
 $^{c}$ Noise in mK.\\
 $^{d}$ Assumed width from the detected isotopologs.\\
 $^{e}$ Number of channels assuming a spectral resolution of D$_{v}$=0.54.\\
 $^{f}$ Number of channels assuming a spectral resolution of D$_{v}$=0.38.\\
 $^{g}$ Number of channels assuming a spectral resolution of D$_{v}$=0.25.\\
 $^{g}$ Number of channels assuming a spectral resolution of D$_{v}$=0.40.\\
 $^{*}$ Spectrum obtained by convolving the map with a Gaussian to obtain the same spatial resolution of the lower-frequency transition.
\end{table*} 
\begin{list}{}{}
\item 
\item  
\end{list}
\label{T2}

\begin{table*}[]
\begin{center}
\scriptsize
\caption{Gaussian fit parameters of the hydrogenated molecules \textbf{at} the IF position.}
\begin{tabular}{ l c c c c c c c}
\hline\hline\noalign{\smallskip}
IF 
\\        
Species 
&Trans 
&Freq   
&HPBW 
&Area 
&V
&Width
&$T_{\mathrm{MB}}$
\\
& 
&(MHz)
&($"$)
&(K~km~s$^{-1}$)
&(km~s$^{-1}$)
&(km~s$^{-1}$)
&(K)
\\
\hline\hline
\noalign{\smallskip}

H$^{13}$CN   	 &J=1--0, F=1--1		&86338.76  		&29  		&$0.301\pm0.019$ 	 &$10.468\pm0.061$		&$2.213\pm0.131$ &0.128\\
	   		 	 &J=1--0, F=1--1   	    & 	 			&  			&$0.113\pm0.018$ 	 &$12.327\pm0.072$		&$1.577\pm0.142$ &0.068\\
 	    		 &J=1--0, F=2--1  	 	&86340.18 		&29 		&$0.640\pm0.033$ 	 &$10.702\pm0.058$		&$2.490\pm0.165$ &0.241 \\
	    		 &J=1--0, F=2--1 		& 	 			&  			&$0.108\pm0.009$ 	 &$12.450\pm0.070$		&$0.752\pm0.362$ &0.128 \\
	     		 &J=1--0, F=0--1  	 	&86342.27  		&29 		&$0.127\pm0.033$  	 &$10.804\pm0.311$		&$2.611\pm0.667$ &0.046 \\
	     		 &J=1--0, F=0--1  	 	& 	 			&  			&$0.069\pm0.031$  	 &$14.075\pm0.625$		&$2.592\pm0.141$ &0.027 \\
	     		 &J=3--2  				&259011.82		&$29^{*}$   &$1.750\pm0.075$   	 &$10.640\pm0.043$		&$2.431\pm0.141$ &0.677 \\
	     		 &J=3--2  				& 	 			&  			&$0.106\pm0.046$  	 &$12.463\pm0.068$		&$0.578\pm0.217$ &0.173 \\
HC$^{15}$N  	 &J=1--0  				&86054.96		&29   		&$0.145\pm0.005$     &$10.108\pm0.033$ 		&$1.669\pm0.069$ &0.082 \\
    	    	 &J=1--0 				& 	  			&			&$0.123\pm0.008$  	 &$11.781\pm0.054$ 		&$1.753\pm0.132$ &0.061 \\
	     		 &J=3--2 				&258156.99		&$29^{*}$   &$0.520\pm0.054$  	 &$9.412\pm0.054$ 		&$1.135\pm0.153 $ &0.431\\
 HNC  	    	 &J=1--0 				& 90663.56 		&29 		&$3.141\pm0.200$  	 &$ 9.931\pm0.050^{a}$ 	&$7.785\pm0.050 $ &\\
   	    		 &J=1--0				& 	  			&			&$2.391\pm0.038$  	 &$12.350\pm0.013^{a}$ 	&$2.010\pm0.063 $ &\\
     	   	 	 &J=3--2 				&271981.14  	&9			&$0.305\pm0.489$     &$8.446 \pm0.600$   	&$0.615\pm0.890 $ &0.466 \\	
   	    		 &J=3--2 				& 	  			&			&$4.495\pm0.144$     &$10.641\pm0.029$ 		&$1.877\pm0.071 $ &2.250\\
HN$^{13}$C  	 &J=1--0  				&87090.82       &29 		&$0.074\pm0.011$     &$10.039\pm0.116$  	&$1.665\pm0.273$  &0.042\\
	    		 &J=1--0 				& 	    	    &   		&$0.166\pm0.012$     &$12.135\pm0.048$    	&$1.603\pm0.128$  &0.097\\ 
	    		 &J=3--2  				&261263.48 		&9 			&$<0.093^{b}$        &rms$^{c}$=46  		&W$^{d}$=2 		  &N$_{c}$$^{f}$=6.4 \\
 	     		 &J=3--2  				&	 			&9 			&$<0.100^{b}$        &rms$^{c}$=46  		&W$^{d}$=2  	  &N$_{c}$$^{f}$=7.5   \\  
H$^{15}$NC       &J=1--0   				& 88865.71 		&29 		&$0.0194\pm0.007 $   &$10.900 \pm0.211$   	&$1.232\pm0.224$  &0.015\\ 
  	   			 &J=1--0  				& 				& 			&$0.0602\pm0.010 $   &$11.900 \pm0.155$   	&$2.058\pm0.385$  &0.028\\
	 	         &J=3--2   				&266587.80 		& 9 		&$<0.017^{b}$        &rms$^{c}$=94 			&W$^{d}$=1  	  &N$_{c}$$^{f}$=5.5  \\ 
 	     		 &J=3--2  				& 	 			&  			&$<0.017^{b}$ 		 &rms$^{c}$=94 			&W$^{d}$=1  	  &N$_{c}$$^{f}$=5  \\
HCO$^{+}$        &J=1--0				&89188.52 		&29 		&$7.457 \pm0.042$    &$9.244 \pm0.050^{a}$  &3--11     		  & \\ 
	     		 &J=1--0  				& 				&  			&$4.466 \pm0.010$    &$12.894\pm0.050^{a}$  &11--23  		  & \\
 	   			 &J=3--2				&267557.52 		& 9 		&$6.643 \pm0.042$    &$7.868 \pm0.006$   	&$1.819\pm0.014$  &3.432 \\ 
   	    		 &J=3--2  				& 				&      		&$36.612 \pm0.041$   &$10.473\pm0.002$   	&$2.784\pm0.001$  &12.35 \\ 
H$^{13}$CO$^{+}$ &J=1--0				& 86754.28	 	&29 		&$0.769\pm0.014$     &$10.451\pm0.028$ 		&$3.230\pm0.072$  &0.224\\ 
  		 		 &J=1--0				& 				&			&$0.078\pm0.005$     &$12.529\pm0.045$ 		&$1.261\pm0.105$  &0.058\\  
		  		 &J=3--2				&260255.33 		&$29^{*}$ 	&$1.793\pm0.083$     &$10.058\pm0.041$ 		&$1.908\pm0.105$  &0.883\\   
		   		 &J=3--2    	  	    &	       		& 			&$0.299\pm0.066$     &$11.780\pm0.083$ 		&$0.987\pm0.160$  &0.285\\
HC$^{18}$O$^{+}$ &J=1--0 				& 85162.15 		&29 	    &$0.081\pm0.007$   	 &$10.313\pm0.150$  	&$2.920\pm0.300$  &0.026\\   
 		  		 &J=3--2 			 	&255480.21 		& 9 	    &$0.197\pm0.015$     &$11.556\pm0.054$ 		&$1.545\pm0.160$  &0.121\\  
H$_2$CO          &2$_{0,2}$--1$_{0,1}$ 	&145602.94   	&16 	    &$0.788\pm0.024$     &$ 8.000\pm0.024$ 		&$2.485\pm0.054$  &0.298 \\
		  		 &2$_{0,2}$--1$_{0,1}$ 	& 				&	    	&$2.620\pm0.060$     &$10.389\pm0.035$ 		&$3.291\pm0.073$  &0.758 \\
         	   	 &3$_{1,3}$--2$_{1,2}$  &211211.46 		&$16^{*}$   &$9.260\pm0.107$     &$10.567\pm0.014$ 		&$2.631\pm0.040$  &3.307 \\
 	    	     &3$_{0,3}$--2$_{0,2}$  &218222.19 		&$16^{*}$   &$0.660\pm0.110$     &$8.045 \pm0.134$ 		&$1.914\pm0.350$  &0.324 \\  
  	    	     &3$_{0,3}$--2$_{0,2}$  & 	  			&	     	&$4.140\pm0.111$     &$10.672\pm0.027$ 		&$2.254\pm0.069$  &1.725 \\    
	    	     &3$_{2,2}$--2$_{2,1}$  &218475.63  	&$16^{*}$   &$1.173\pm0.079$     &$10.621\pm0.077$ 		&$2.512\pm0.222$  &0.439 \\ 
 	     	     &3$_{2,1}$--2$_{2,0}$  &218760.06		&$16^{*}$   &$0.241\pm0.375$     &$10.043\pm0.140$ 		&$0.956\pm0.667$  &0.237 \\
 	    	     &3$_{2,1}$--2$_{2,0}$  & 	  			&		    &$0.659\pm0.426$     &$11.188\pm0.700$ 		&$2.121\pm1.019$  &0.292 \\ 
	     	     &3$_{1,2}$--2$_{1,1}$  &225697.77		&$16^{*}$   &$0.974\pm0.114$     &$ 8.141\pm0.084$ 		&$1.639\pm0.212$  &0.558 \\
 	    	     &3$_{1,2}$--2$_{1,1}$  & 	  			&	   		&$6.243\pm0.127$     &$10.664\pm0.019$ 		&$2.080\pm0.052$  &2.819 \\
H$_2$$^{13}$CO	 &6$_{1,5}$--6$_{1,6}$  & 96375.75  	&29 		&$<0.012^{b}$ 		 &rms$^{c}$=3.4 		&W$^{d}$=2  	  &N$_{c}$$^{e}$=3.0\\  
  	    	     &6$_{1,5}$--6$_{1,6}$  & 	  			&	  		&$<0.011^{b}$		 &rms$^{c}$=3.4  		&W$^{d}$=2 		  &N$_{c}$$^{e}$=2.7 \\  
		   		 &2$_{0,2}$--1$_{0,1}$  &141983.74 		&16 		&$<0.004^{b}$ 		 &rms$^{c}$=10   		&W$^{d}$=2 		  &N$_{c}$$^{e}$=3.0 \\  
  	    	   	 &2$_{0,2}$--1$_{0,1}$  & 	  			&	 		&$<0.005^{b}$ 		 &rms$^{c}$=10   		&W$^{d}$=2 		  &N$_{c}$$^{e}$=2.7 \\ 
		  		 &3$_{0,3}$--2$_{0,2}$  &212811.18 		&$16^{*}$ 	&$<0.276^{b}$ 		 &rms$^{c}$=107 		&W$^{d}$=2  	  &N$_{c}$$^{e}$=10.5 \\  
		 		 &3$_{2,1}$--2$_{2,0}$  &213293.56 		&$16^{*}$	&$<0.305^{b}$ 		 &rms$^{c}$=118  		&W$^{d}$=2 		  &N$_{c}$$^{e}$=10.5 \\  
		  		 &3$_{1,2}$--2$_{1,1}$  &219908.52 		&$16^{*}$	&$<0.214^{b}$  		 &rms$^{c}$=97   		&W$^{d}$=2  	  &N$_{c}$$^{e}$=7.7 \\  
  	    	     &3$_{1,2}$--2$_{1,1}$  & 	  			&	 		&$<0.234^{b}$  		 &rms$^{c}$=97   		&W$^{d}$=2 		  &N$_{c}$$^{e}$=9.0 \\ 
N$_{2}$H$^{+}$ 	 &J=1--0, F=1--1  		&93171.88       &29 	    &$1.280\pm0.007$     &$11.662\pm0.005$ 		&$1.843\pm0.013$  &0.652   \\
  	             &J=1--0, F=2--1  		&93173.70       &29 	    &$2.273\pm0.010$     &$11.743\pm0.005$ 		&$2.460\pm0.012$  &0.868   \\ 
  	             &J=1--0, F=0--1		&93176.13       &29 	    &$0.452\pm0.010$     &$11.470\pm0.015$ 		&$1.497\pm0.040$  &0.283   \\  	

\hline
\end{tabular}
\end{center}
\noindent
\\
\scriptsize
 $^{a}$Area calculated between two velocities.\\
 $^{b}$Upper limit of the area, assuming a Gaussian profile.\\
 $^{c}$Noise in mK.\\
 $^{d}$ Assumed width from the detected isotopologs.\\
 $^{e}$Number of channels assuming a spectral resolution of D$_{v}$=0.6.\\
 $^{f}$Number of channels assuming a spectral resolution of D$_{v}$=0.25.\\
 $^{*}$Spectrum obtained by convolving the map with a Gaussian to obtain the same spatial resolution of the lower-frequency transition.
\begin{list}{}{}
\item 
\item 
\end{list}
\label{T3}
\end{table*}
\begin{table*}[]
\begin{center}
\scriptsize
\caption{Gaussian fit parameters of the hydrogenated molecules \textbf{at} the MP2 position.}
\begin{tabular}{ l c c c c c c c}
\hline\hline\noalign{\smallskip}
MP2 
\\     
Species
&Trans 
&Freq   
&HPBW 
&Area 
&V
&Width
&$T_{\mathrm{MB}}$
\\
& 
&(MHz)
&($"$)
&(K~km~s$^{-1}$)
&(km~s$^{-1}$)
&(km~s$^{-1}$)
&(K)
\\
\hline\hline
\noalign{\smallskip}
H$^{13}$CN 				&J=1--0, F=1--1 		& 86338.76	&29	 		&$0.165 \pm0.012$ 		 &$8.144\pm0.059$   &$1.469\pm0.130$     &0.106\\
  						&J=1--0, F=1--1 		& 	 		&	 		&$0.446 \pm0.024$ 		 &$9.919\pm0.055$   &$1.984\pm0.130$     &0.211\\
	 					&J=1--0, F=2--1 		& 86340.18	&29	  	 	&$0.265 \pm0.011$ 	     &$8.285\pm0.032$   &$1.500\pm0.075$     &0.166\\
  						&J=1--0, F=2--1 		& 	 		&	  		&$0.677 \pm0.020$   	 &$9.976\pm0.024$   &$1.866\pm0.068$     &0.341\\
	 					&J=1--0, F=0--1 		& 86342.27	&29	 		&$0.130 \pm0.050$   	 &$9.219\pm0.258$   &$2.433 \pm0.270$    &0.050\\
  						&J=1--0, F=0--1 		& 	 		&	  		&$0.071 \pm0.034$   	 &$10.563\pm0.900$  &$3.374\pm0.800$   	 &0.020\\
						&J=3--2  				&259011.82 	&$29^{*}$ 	&$0.242 \pm0.028 $ 		 &$8.455 \pm0.051 $ &$0.848\pm0.111 $	 &0.268\\
 						&J=3--2 				& 	 		&	 		&$0.947 \pm0.057$   	 &$10.008\pm0.042 $ &$1.679\pm0.135$  	 &0.530 \\
HC$^{15}$N				&J=1--0 				& 86054.96	&29  	 	&$0.251 \pm0.011$  	 	 &$8.810\pm0.042$   &$1.987\pm0.108$  	 &0.119 \\
 						&J=1--0 				& 	 		&	 		&$0.105 \pm0.007$ 	 	 &$10.236\pm0.023$  &$0.973\pm0.124$  	 &0.101 \\
	 					&J=3--2					&258156.99	&$29^{*}$ 	&$0.217 \pm0.047$ 	 	 &$8.348 \pm0.099 $ &$0.893\pm0.245$  	 &0.229 \\ 
   						&J=3--2 				& 	 		&	 		&$0.164 \pm0.054$  	 	 &$10.143\pm0.279 $ &$1.693\pm0.673$  	 &0.091 \\ 
HNC  					&J=1--0 				& 90663.56 	&29 	 	&$8.390 \pm0.136$ 		 &$ 9.060\pm0.024$  &$3.131\pm0.063$  	 &2.518\\ 
   						&J=1--0  				& 	 		&	  		&$1.180 \pm0.036$ 		 &$ 13.00\pm0.051$  &$3.626\pm0.140$  	 &0.306\\
						&J=3--2					&271981.14 	&9 	  		&$7.590 \pm0.080$   	 &$8.543\pm0.005$   &$1.053\pm0.013$  	 &6.770 \\
   						&J=3--2	 				& 	 		&	  		&$1.585 \pm0.118$   	 &$10.618\pm0.074$  &$1.866\pm0.183$  	 &0.798 \\
HN$^{13}$C 				&J=1--0  				&87090.82 	&29 		&$0.196 \pm0.003$  		 &$8.589 \pm0.082$  &$1.835\pm0.111$ 	 &0.101\\
  						&J=1--0 				& 	 		&	  		&$0.104 \pm0.005$   	 &$10.400\pm0.089$  &$1.749\pm0.320$ 	 &0.056\\
	 					&J=3--2 				&261263.48 	&9 	  		&$0.320 \pm0.020$  		 &$8.349\pm0.020$   &$0.704 \pm0.057 $   &0.427\\
  						&J=3--2  				& 	 		&	  		&$0.027 \pm0.013$ 		 &$9.989\pm0.060$   &$0.360 \pm0.057 $   &0.071\\
H$^{15}$NC 				&J=1--0 				&88865.71 	&29 	  	&$0.072 \pm0.010$  		 &$9.055\pm0.152$   &$2.643 \pm0.344 $   &0.025\\ 
						&J=3--2  				&266587.80 	& 9 	  	&$0.125 \pm0.011$ 		 &$8.409\pm0.025$   &$0.588\pm0.060 $    &0.199   \\ 
HCO$^{+}$ 				&J=1--0 				&89188.52  	&29 	  	&$14.101\pm0.0107$ 		 &$8.722\pm0.010$   &$2.819\pm0.020 $    &4.698 \\ 
 						&J=1--0   				&  			&   	  	&$4.324 \pm0.084$ 		 &$13.900\pm0.036$  &$3.768\pm0.096 $    &1.078 \\
 	  					&J=3--2 				&267557.52 	& 9 	  	&$28.888 \pm0.051$ 		 &$8.620 \pm0.003$  &$2.172\pm0.005 $    &12.495  \\ 
 	 					&J=3--2   				&  			&   	  	&$7.340 \pm0.066$ 		 &$11.687\pm0.013$  &$3.235\pm0.035 $    &2.131  \\
H$^{13}$CO$^{+}$		&J=1--0 				& 86754.28 	&29 	 	&$1.198 \pm0.009$ 		 &$ 9.220\pm0.011$  &$2.865\pm0.025$ 	 &0.393 \\   
 						&J=1--0 				& 			&	 		&$0.307\pm0.008$ 		 &$10.583\pm0.021$  &$1.730 \pm0.052$ 	 &0.167 \\ 
						&J=3--2 				&260255.33  &$29^{*}$	&$1.982 \pm0.029$ 		 &$8.450\pm0.031$   &$0.926\pm0.016$  	 &2.011 \\ 
 						&J=3--2  				& 			&	  		&$0.299\pm0.035^{a}$ 	 &$10.102\pm0.078$ 	&9.5--12 \\
HC$^{18}$O$^{+}$		&J=1--0					& 85162.23	&29 	  	&$0.111\pm0.008$    	 &$9.211\pm0.107$   &$3.055\pm0.244$  	&0.034\\   
						&J=3--2  				&255480.21  & 9 	 	&$0.202\pm0.011$   	 	 &$9.348\pm0.023$   &$0.861\pm0.060$  	&0.220\\  
H$_2$CO         		&2$_{0,2}$--$_{0,1}$    &145602.94  &16 	 	&$3.214\pm0.096$    	 &$ 8.823 \pm0.022$ &$1.685\pm0.064$ 	&1.792 \\  	
	    				&2$_{0,2}$--$_{0,1}$	& 			&	 		&$2.269\pm0.098$    	 &$10.673 \pm0.038$ &$1.844\pm0.110$ 	&1.156 \\
						&3$_{1,3}$--2$_{1,2}$   &211211.46 	&$16^{*}$ 	&$6.374\pm0.120$   	 	 &$8.512 \pm0.011$  &$1.330\pm0.032$ 	&4.501  \\ 
	    				&3$_{1,3}$--2$_{1,2}$	& 			&	 		&$8.630\pm0.226$   	 	 &$10.276\pm0.030$  &$8.300\pm0.000$	&2.951  \\  
	    				&3$_{0,3}$--2$_{0,2}$   &218222.19 	&$16^{*}$	&$3.565\pm0.060$     	 &$8.499 \pm0.011$  &$1.374\pm0.029$ 	&2.437 \\   
	    				&3$_{0,3}$--2$_{0,2}$   & 			&	  		&$2.490\pm0.081$   	 	 &$10.228\pm0.023$  &$1.404\pm0.054$	&1.666  \\ 	    		
	    				&3$_{2,2}$--2$_{2,1}$   &218475.63	&$16^{*}$ 	&$0.972\pm0.098$   		 &$8.588 \pm0.079$  &$1.760\pm0.232$ 	&0.511  \\  
 	    				&3$_{2,2}$--2$_{2,1}$ 	& 			&	  		&$0.541\pm0.064$   		 &$10.331\pm0.076$  &$1.214\pm0.143$ 	&0.413  \\ 
 	     				&3$_{2,1}$--2$_{2,0}$	&218760.06	&$16^{*}$ 	&$0.676\pm0.077$   		 &$8.387 \pm0.066$  &$1.332\pm0.204$ 	&0.476  \\  
 	    				&3$_{2,1}$--2$_{2,0}$   & 			&	 		&$0.510\pm0.055$    	 &$10.102\pm0.063$  &$1.168\pm0.146$ 	&0.410  \\ 
	    				&3$_{1,2}$--2$_{1,1}$ 	&225697.77 	&$16^{*}$ 	&$5.173 \pm0.091$   	 &$8.508 \pm0.012$  &$1.461\pm0.032$ 	&3.326  \\  
  	    				&3$_{1,2}$--2$_{1,1}$   & 			&	 		&$3.498 \pm0.109$    	 &$10.285\pm0.022$  &$1.435\pm0.059$ 	&2.291  \\
H$_2$$^{13}$CO			&6$_{1,5}$--6$_{1,6}$  	& 96375.75  &29   		&$<0.002^{b}$ 			 &rms$^{c}$=5.5     &W$^{d}$=5   		&N$_{c}$$^{e}$=5.3\\ 	 
  	    				&6$_{1,5}$--6$_{1,6}$	& 			&     		&$<0.002^{b}$ 			 &rms$^{c}$=5.5     &W$^{d}$=5   		&N$_{c}$$^{e}$=3.8\\
						&2$_{0,2}$--1$_{0,1}$  	&141983.74	&16  		&$ 0.147\pm0.080^{a}$  	 &$9.150$     	    &7--12  \\
						&3$_{0,3}$--2$_{0,2}$	&212811.184 &$16^{*}$ 	&$<0.235^{b}$ 			 &rms$^{c}$=128     &W$^{d}$=5 			&N$_{c}$$^{e}$=5.4 \\    
						&3$_{0,3}$--2$_{0,2}$	& 	    	&	      	&$<0.589^{b}$			 &rms$^{c}$=128     &W$^{d}$=5  		&N$_{c}$$^{e}$=33.2\\ 	 
	 					&3$_{2,1}$--2$_{2,0}$ 	&213293.560 &$16^{*}$ 	&$<0.233^{b}$ 			 &rms$^{c}$=127     &W$^{d}$=5 			&N$_{c}$$^{e}$=5.47\\ 	 
						&3$_{2,1}$--2$_{2,0}$	& 	   		&	      	&$<0.584^{b}$ 			 &rms$^{c}$=127     &W$^{d}$=5  		&N$_{c}$$^{e}$=33.2\\ 	  
  		       			 &3$_{1,2}$--2$_{1,1}$  &219908.    &$16^{*}$ 	&$<0.189^{b}$			 &rms$^{c}$=101     &W$^{d}$=5  		&N$_{c}$$^{e}$=5.5\\     
  	    	       		 &3$_{1,2}$--2$_{1,1}$  & 	    	&	     	&$<190.000^{b}$  		 &rms$^{c}$=101     &W$^{d}$=5 			&N$_{c}$$^{e}$=5.6\\
N$_{2}$H$^{+}$ 			&J=1--0, F=1--1  		&93171.88  	&29 	 	&$1.083 \pm0.009$ 		 &$7.898\pm0.011$   &$2.849\pm0.027$ 	&0.357 \\
   	    				&J=1--0, F=2--1       	&93173.70  	&	 		&$1.746 \pm0.008$ 		 &$8.047\pm0.007$   &$2.952\pm0.016$ 	&0.556 \\ 
 	    				&J=1--0, F=0--1       	&93176.13 	&	  		&$0.400 \pm0.009$		 &$7.669\pm0.029$   &$2.680\pm0.066$ 	&0.140 \\ 
 
\hline
\end{tabular}
\end{center}
\noindent
\\
\scriptsize
 $^{a}$ Area calculated between two velocities.\\
 $^{b}$ Upper limit of the area, assuming a Gaussian profile.\\
 $^{c}$ Noise in mK.\\
 $^{d}$ Assumed width from the detected isotopologs.\\
 $^{e}$ Number of channels assuming a spectral resolution of D$_{v}$=0.6.\\
 $^{f}$ Number of channels assuming a spectral resolution of D$_{v}$=0.25.\\
 $^{*}$ Spectrum obtained by convolving the map with a Gaussian to obtain the same spatial resolution of the lower-frequency transition.
\begin{list}{}{}
\item 
\item 
\end{list}
\label{T4}
\end{table*}

\clearpage


\clearpage

\clearpage
 


\begin{figure*}
\centering
\includegraphics[width=6.86cm]{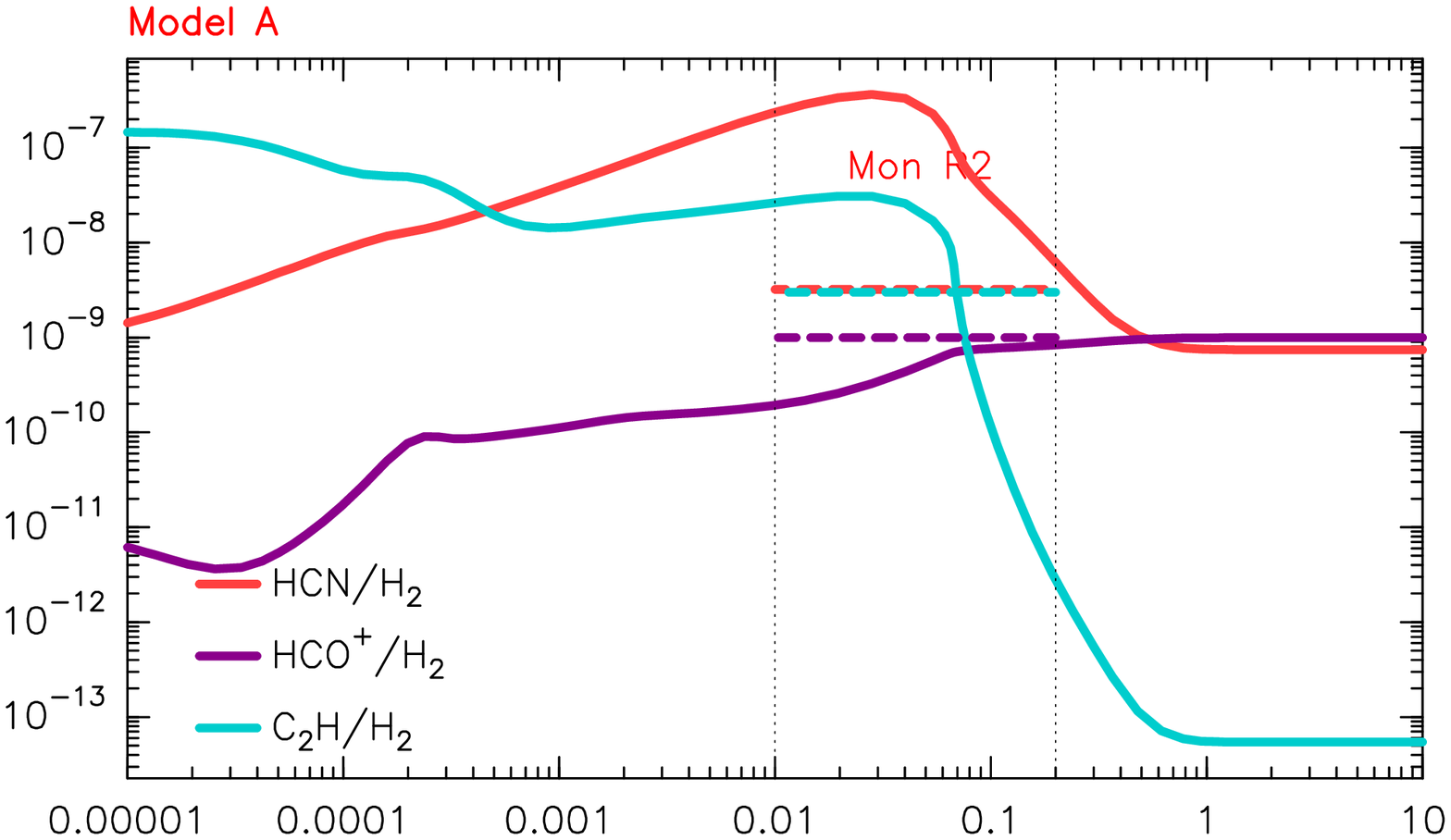}  
\includegraphics[width=6.86cm]{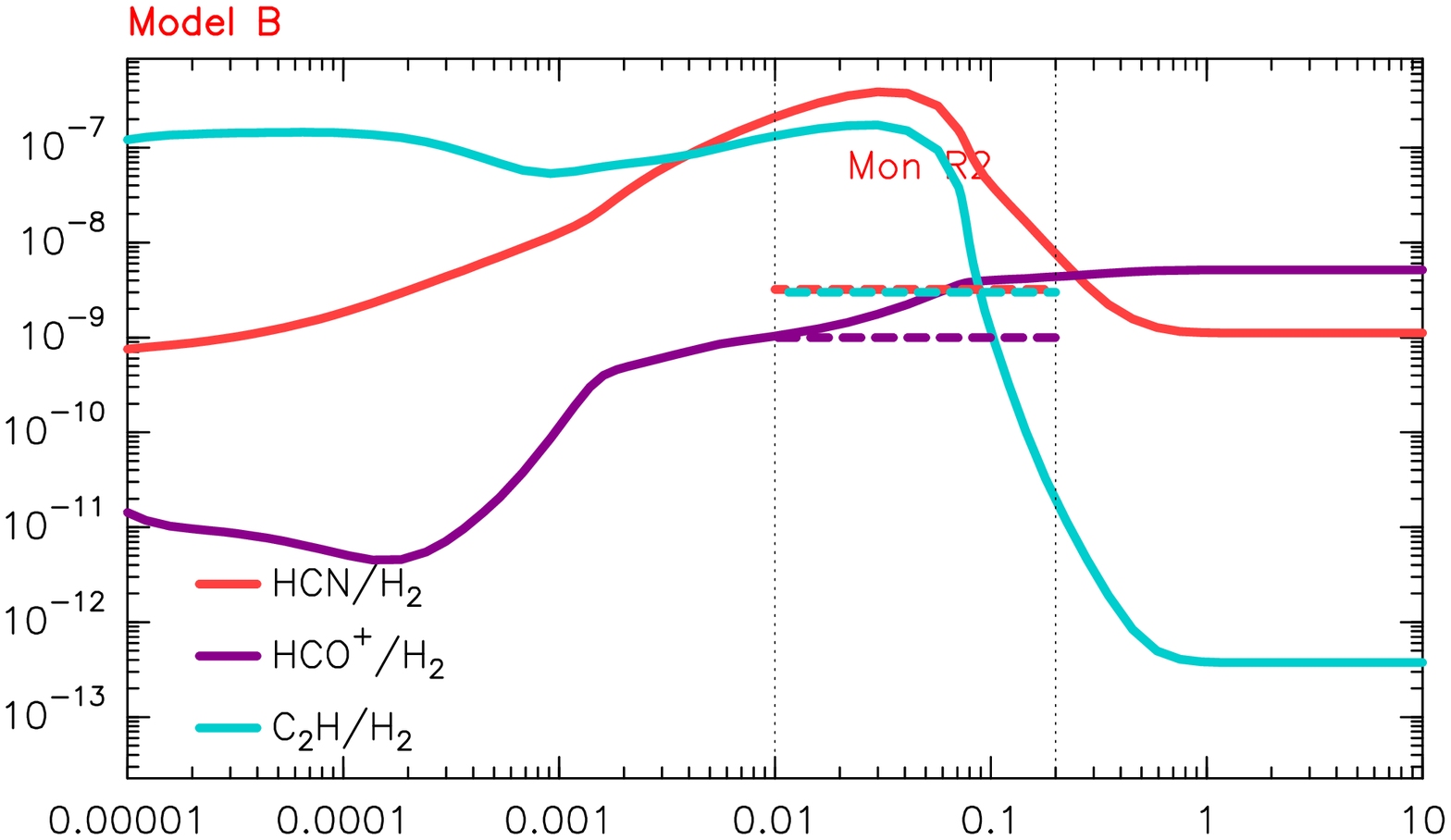} \\
\includegraphics[width=6.86cm]{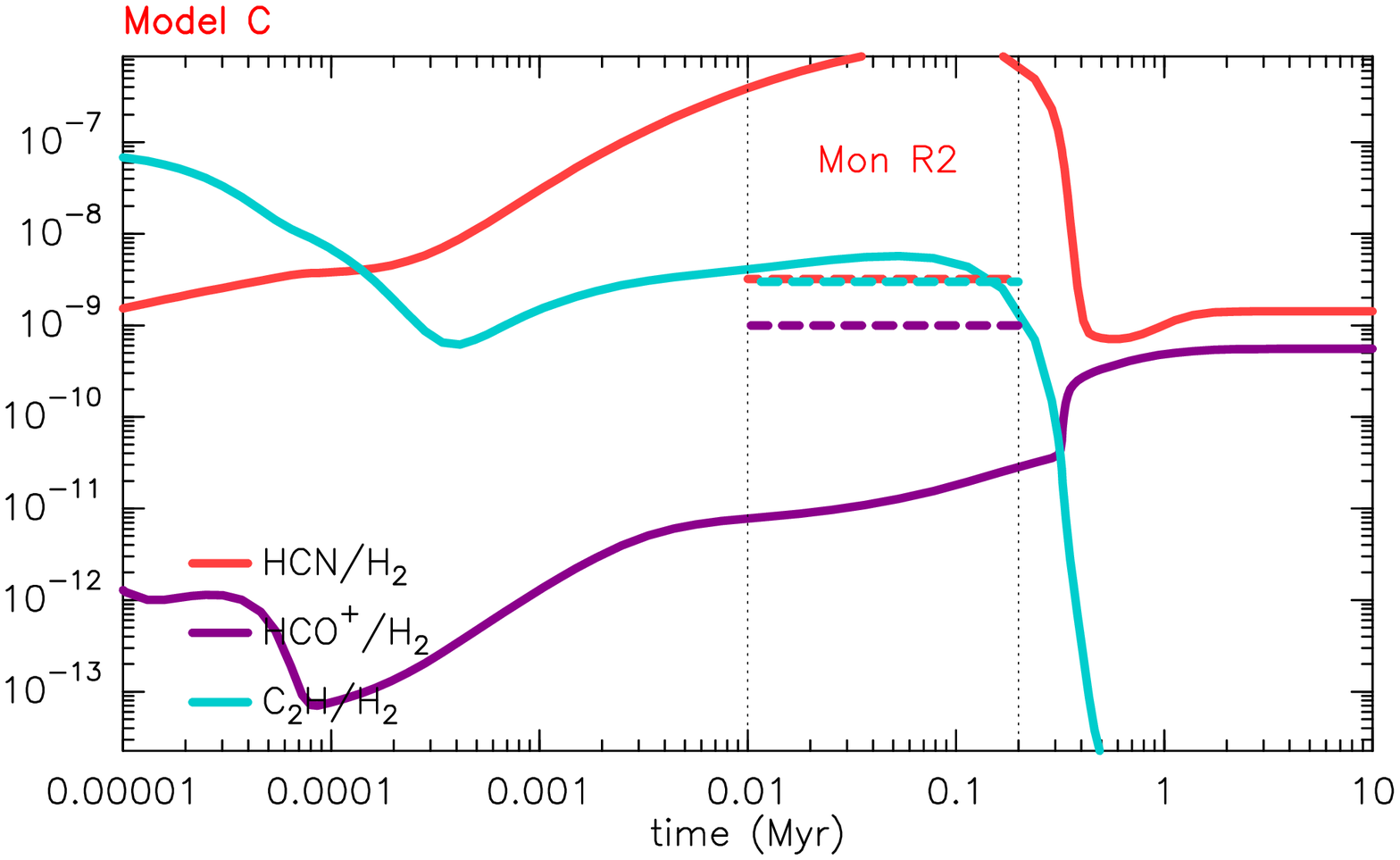}  
\includegraphics[width=6.86cm]{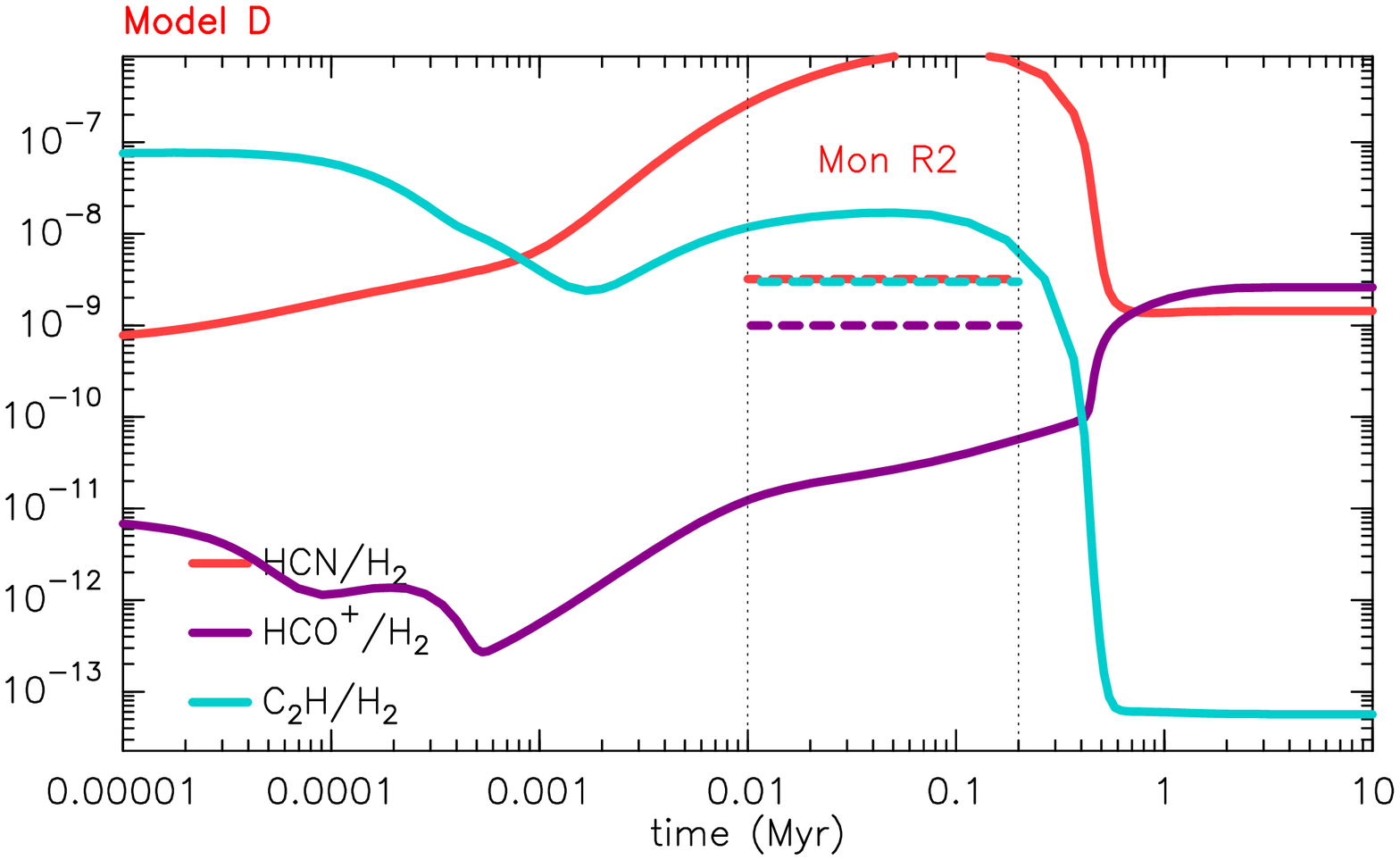} \\     
\caption{Molecular abundances relative to H$_{2}$ predicted by models A to D. The input parameters of these models are shown in Table 4. The red line represents the [HCN]/[H$_{2}$] ratio, the purple line shows the [HCO$^{+}$]/[H$_{2}$] ratio, and the blue line shows the [C$_{2}$H]/[H$_{2}$] ratio. The dashed lines shows the ratios of observational results toward MonR2 for the 10.5 \kms\ component at the IF position.}
\label{models}
\end{figure*}

\end{appendix}

\end{document}